\documentclass{aims}

\usepackage{txfonts}
\usepackage{booktabs}
\usepackage{colortbl}
\def\typeofarticle{Regular article}
\def\currentvolume{}
\def\currentissue{x}
\def\currentyear{}

\def\ppages{Accepted}
\def\DOI{ to be inserted}
\def\Received{to be inserted}
\def\Accepted{to be inserted}
\def\Published{to be inserted}

\numberwithin{equation}{section}

\makeatletter
\renewcommand{\@biblabel}[1]{#1\hfill \hspace{-0.2cm}}
\makeatother

\begin{document}

\title{ Computational efficiency of numerical integration methods for
  the tangent dynamics of many-body Hamiltonian systems in one and two
  spatial dimensions }

\author{%
  Carlo Danieli\affil{1},
  Bertin Many Manda\affil{2},
  Thudiyangal Mithun\affil{1}
  and
  Charalampos Skokos\affil{2,3,}\corrauth
}

% \shortauthors is used in copyright information in the end of the paper
\shortauthors{the Author(s)}

\address{%
  \addr{\affilnum{1}}{Center for Theoretical Physics of Complex
    Systems, Institute for Basic Sciences, Daejeon, Korea}
  \addr{\affilnum{2}}{Department of Mathematics and Applied
    Mathematics, University of Cape Town, Rondebosch, 7701, Cape Town,
    South Africa} \addr{\affilnum{3}}{Max Planck Institute for the
    Physics of Complex Systems, N\"othnitzer Str.~38, 01187 Dresden,
    Germany}}

% corresponding author
\corraddr{haris.skokos@uct.ac.za; Tel: +27-021-650-5074; Fax: +27-021-650-2334.
}

\begin{abstract}
  We investigate the computational performance of various numerical
  methods for the integration of the equations of motion and the
  variational equations for some typical classical many-body models of
  condensed matter physics: the Fermi-Pasta-Ulam-Tsingou (FPUT) chain
  and the one- and two-dimensional disordered, discrete nonlinear
  Schr\"odinger equations (DDNLS). In our analysis we consider methods
  based on Taylor series expansion, Runge-Kutta discretization and
  symplectic transformations.  The latter have the ability to exactly preserve the symplectic structure of
  Hamiltonian systems, which results in keeping bounded the error of the system's computed total energy. We perform extensive numerical
  simulations for several initial conditions of the studied models and
  compare the numerical efficiency of the used integrators by testing
  their ability to accurately reproduce characteristics of the
  systems' dynamics and quantify their chaoticity through the
  computation of the maximum Lyapunov exponent. We also report the
  expressions of the implemented symplectic schemes and provide the
  explicit forms of the used differential operators.  Among the tested
  numerical schemes the symplectic integrators $ABA864$ and
  $SRKN^a_{14}$ exhibit the best performance, respectively for moderate
  and high accuracy levels in the case of the FPUT chain, while for
  the DDNLS models $s9\mathcal{ABC}6$ and $s11\mathcal{ABC}6$ (moderate accuracy), along with
  $s17\mathcal{ABC}8$ and $s19\mathcal{ABC}8$ (high accuracy) proved to be the most efficient
  schemes.
\end{abstract}


\maketitle

%-----------------------------------------------------
\section{Introduction}
\label{sec:intro}

A huge number of important problems in physics, astronomy, chemistry,
etc.~are modeled via sets of ordinary differential equations (ODEs)
governed by the Hamiltonian formalism.  Due to non-integrability, the
investigation of the time evolution of these problems, and generally
their properties, often rely solely on numerical techniques.  As
modern research requires numerical simulations to be pushed to their
very limit (e.g.~large integration times, macroscopic limits), a
methodical assessment of the properties of different numerical methods
becomes a fundamental issue.  Such studies allow to highlight the most
suitable scheme for both general purposes and specific tasks,
according to criteria of stability, accuracy, simplicity and
efficiency.

The beginning of the era of computational physics is considered to be
the computer experiment performed in the 1950s by Fermi, Pasta, Ulam
and Tsingou (FPUT)
\cite{fermi1955studies,ford1992fermi,campbell2005introduction} to
observe energy equipartition due to weak-anharmonic coupling in a set
of harmonic oscillators. Indeed, the breaking of integrability in
Hamiltonian systems is often performed with the introduction of
nonlinear terms in the equations of motion.  These additional
nonlinear terms describe new physical processes, and led to important
questions and significant advancements in condensed matter physics.
For instance, the FPUT problem has been used to answer questions
related to ergodicity, dynamical thermalization and chaos occurrence
(see
e.g.~\cite{lepri2003universality,lepri2005studies,antonopoulos2006chaotic}
and references therein) and led to the observation of solitons
\cite{Zabusky1965soliton,zabuskydeem}, and progress in Hamiltonian
chaos \cite{izrailevchirikov}.
The interested reader can find a concise review of numerical results concerning the FPUT system in \cite{Paleari2007methods}.

Another important example concerns disordered media.  In 1958,
Anderson \cite{anderson1958absence} theoretically showed the
appearance of energy localization in one-dimensional lattices with
uncorrelated disorder (a phenomenon which is now referred to as {\it
  Anderson localization}). This phenomenon was  later investigated also
for two-dimensional lattices \cite{Abraham1979}. An important question
which has attracted wide attention in theory, numerical simulations
and experiments is what happens when nonlinearity (which appears
naturally in real world problems) is introduced in a disordered system
\cite{shepelyansky1993dl,molina1998transport,
  clement2005suppression,fort2005effect,
  schwartz2007transport,lahini2008anderson, billy2008direct,
  roati2008g, fallani2008bose, flach2009universal,vicencio2009control,
  skokos2009delocalization, skokos2010spreading,
  laptyeva2010crossover,  modugno2010anderson, bodyfelt2011wave, bodyfelt2011nonlinear,
  laptyeva2012subdiffusion}.

Two basic Hamiltonian models are at the center of these
investigations: the disordered Klein-Gordon (DKG) chain and the
disordered, discrete nonlinear Schr\"odinger equation (DDNLS). In
Refs.~\cite{
  flach2009universal,vicencio2009control,skokos2009delocalization,
  skokos2010spreading,laptyeva2010crossover,bodyfelt2011wave,
  bodyfelt2011nonlinear, laptyeva2012subdiffusion,
  pikovsky2008destruction, skokos2013nonequilibrium,
  senyange2018characteristics,KYF19} it was shown that Anderson localization
is destroyed by the presence of nonlinearity,  resulting to the
subdiffusive spreading of energy due to deterministic chaos. Such
results rely on the accurate, long time integration of the systems'
equations of motion and variational equations. We note that the
variational equations are a set of ODEs describing the time evolution
of small deviations from a given trajectory, something which is needed
for the computation of chaos indicators like the maximum Lyapunov
exponent (mLE) \cite{benettin1980lyapunov2, benettin1980lyapunov1,
  skokos2010lyapunov}. The numerical integration of these sets of ODEs
can be performed by any general purpose integrator. For example
Runge-Kutta family schemes are still used~\cite{mulansky2009dynamical,
  sales2018sub}. Another category of integrators is the so-called
symplectic integrators (SIs), which were particularly designed for
Hamiltonian systems (see e.g.~\cite{hairer2002geometric,
  mclachlan2002splitting,mclachlan2006geometric,
  forest2006geometric,blanes2008splitting} and references
therein).
The main characteristic of SIs is that they conserve the
symplectic structure of Hamiltonian systems, so that the numerical approximation they
provide corresponds to the exact solution of a system which can be
considered as a perturbation of the original one.
Consequently
the error in the computed value of the Hamiltonian function (usually refereed to as the system's energy) is kept almost constant over long integration times.
This almost constant error can be used as an indicator of the accuracy of the used
integration scheme.

SIs have been successfully implemented for the
long time integration of multidimensional Hamiltonian systems like
the $\alpha$- and $\beta$-FPUT systems and FPUT-like models, the DKG, the DDNLS and systems of coupled rotors (see
e.g.~\cite{
  flach2009universal,vicencio2009control,skokos2009delocalization,skokos2010spreading,
  laptyeva2010crossover,
  bodyfelt2011wave,bodyfelt2011nonlinear,laptyeva2012subdiffusion,
  skokos2013nonequilibrium, senyange2018characteristics, Benettin2001like,
  antonopoulos2014complex,antonopoulos2017analyzing,tieleman2014chaoticity,
  skokos2014high,
  gerlach2016symplectic,danieli2017intermittent,thudiyangal2017gross,
  mithun2018:rotor, danieli2018:dynamical}).  In these studies SIs belonging to the so-called
SABA family~\cite{laskar2001high} were mostly used, since the FPUT,
the DKG and systems of coupled rotors can be split into two integrable
parts (the system's kinetic and the potential energy), while in some
more recent studies~\cite{senyange2018characteristics,
senyange2017symplectic} the $ABA864$~\cite{blanes2013new} SI was
implemented as it displayed an even better performance. As the DDNLS
model is not explicitly decomposed into two integrable parts, the
implementation of SABA schemes requires a special treatment involving
the application of fast Fourier transforms which are
computationally time
consuming~\cite{bodyfelt2011nonlinear}. In~\cite{skokos2014high,
  gerlach2016symplectic} it was shown that for the DDNLS model, the
split of the Hamiltonian function into three integrable parts is
possible, and this approach proved to be the best choice for the
model's integration with respect to both speed and accuracy. It is
worth noting here that the numerical integration of the variational
equations by SIs is done by the so-called {\it Tangent Map
  Method}~\cite{skokos2010numerical, gerlach2011comparing,
  gerlach2012efficient}.

The intention of this work is to present several efficient (symplectic
and non-symplectic) integration techniques for the time evolution of
both the equations of motion and the variational equations of
multidimensional Hamiltonian systems.  In particular, in our study we
use these techniques to investigate the chaotic dynamics of the
one-dimensional (1D) FPUT system, as well as one- and two-dimensional
(2D) DDNLS models.  We carry out this task by considering particular
initial conditions for both the systems and for the deviation vectors,
which we evolve in time requiring a predefined level of energy
accuracy. Then, we record the needed CPU time to perform these
numerical simulations and highlight those methods that ensure the
smallest CPU times for obtaining accurate results. A similar analysis
for the 1D and 2D DKG systems was performed in
\cite{senyange2017symplectic}.

The paper is organized as follows. In Sec.~\ref{sec:hamiltonians} we
introduce the models considered in our study.  In
Sec.~\ref{sec:integration_schemes} we introduce in  detail the
symplectic and non-symplectic schemes implemented in our
investigations.  In Sec.~\ref{sec:tangent_map} we present our
numerical results and report on the efficiency of each studied scheme.
In Sec.~\ref{sec:conclusion} we review our findings and discuss the
outcomes of our study.  Finally, in Appendix~\ref{sec:app_A} we
provide the explicit forms of the used tangent map
operators. Moreover, we present there the explicit expressions of the
tangent map operators for some additional important many-body systems,
the so-called $\beta$-FPUT chain, the KG system and the classical XY
model (a Josephson junction chain - JJC) in order to facilitate the
potential application of SIs for these systems from the interested
reader, although we do not present numerical results for these models.

%-----------------------------------------------------
\section{Models and Hamiltonian functions}\label{sec:hamiltonians}

In this work we focus our analysis on the 1D FPUT system and the 1D
and 2D DDNLS models. In what follows we briefly present the
Hamiltonian functions of these systems.

%-----------------------------------------------------
\subsection{The $\alpha$-Fermi-Pasta-Ulam-Tsingou model}
\label{sec:FPUT_ham}

The 1D FPUT model \cite{fermi1955studies,ford1992fermi,
  campbell2005introduction} was first introduced in 1955 to study the
road toward equilibrium of a chain of harmonic oscillators in the
presence of weak anharmonic interactions. Since then, this system has
been widely used as a toy model for investigating energy equipartition
and chaos in nonlinear lattices. In the literature there exist two
types of the FPUT model the so-called $\alpha-$ and $\beta-$FPUT
systems.
In our study we consider the $\alpha-$FPUT model whose
Hamiltonian function reads
\begin{equation}
  H_\text{1F} = \sum_{ i=0}^N\left[ \frac{p_{i}^2}{2} + \frac{1}{2}(q_{i+1} - q_{ i})^2  + \frac{\alpha}{3}(q_{i+1} - q_{ i})^3 \right].
\label{eq:alphaFPUT}
\end{equation}
In Eq.(\ref{eq:alphaFPUT}), $q_i$ and $p_i$ are respectively the generalized
coordinates and momenta of the $i$ lattice site and $\alpha$ is a real
positive parameter. In our study we consider fixed boundary
conditions $q_0 = p_0 = p_{N+1} = q_{N+1} = 0$. We also note that this
model conserves the value of the total energy $H_{\text{1F}}$.

In contrast to the $\alpha$-FPUT system, the $\beta$-FPUT model  is characterized by a quartic nonlinear term [see Eq.~(\ref{eq:beta}) in Appendix \ref{sec:app_other}]. The fact that the value of the  Hamiltonian function of the $\beta$-FPUT model is bounded from below leads to differences between the two models. For example, phenomena of chopping time which occur in the $\alpha$- model do not appear in the $\beta$-FPUT system~\cite{Carati2018chopping}.
Further discussion of the differences between the  $\alpha$- and the $\beta$- models can be found in \cite{Flach2005qbreather,Flach2008periodic,Flach2008fermi}.

%-----------------------------------------------------
\subsection{The  1D and 2D disordered discrete nonlinear Schr\"{o}dinger equations}
\label{sec:DDNLS_ham}

The DDNLS model describes anharmonic interactions between oscillators
in disordered media and has been used to study the propagation of
light in optical media or Bose-Einstein condensates through the famous
Gross-Pitaevskii equation~\cite{thudiyangal2017gross}, as well as
investigate, at a first approximation, several physical processes
(e.g.~electron tight binding systems~\cite{sales2018sub}).  The
Hamiltonian function of the 1D DDNLS model reads
\begin{equation}
  H_{1D} = \sum_{i = 1}^{N} \left[ \frac{\epsilon _i}{2}(q_i ^2 + p_i ^2) + \frac{\beta}{8}(q_i ^2 + p_i ^2)^2 - p_{i+1}p_i - q_{i+1}q_i  \right],
\label{eq:DNLS_1D_real}
\end{equation}
where $q_i$ and $p_i$ are respectively the generalized coordinates and
momenta of the $i$ lattice site and the onsite energy terms
$\epsilon _i$ are uncorrelated numbers uniformly distributed in the
interval $\left[ -\frac{W}{2}, \frac{W}{2} \right]$.  The real,
positive numbers $W$ and $\beta$ denote the disorder and the
nonlinearity strength respectively. We also consider here fixed
boundary conditions i.e.~$q_0 = p_0 = p_{N+1} = q_{N+1} = 0$.

The two-dimensional version of the DDNLS model was considered
in~\cite{laptyeva2012subdiffusion,garcia2009delocalization}.  Its
Hamiltonian function is
\begin{equation}
\label{eq:DNLS_2D_real}
H_{2D} = \sum _{i = 1}^{N} \sum _{j=1}^{M} \bigg\{ \frac{\epsilon _{i, j} }{2} \left[q_{i, j} ^2 + p_{i, j}^2\right] + \frac{\beta}{8} \left[q_{i, j} ^2 + p_{i, j}^2\right] ^2 - \left[ q_{i, j+1} q _{i, j} + q_{i + 1, j} q _{i, j} +  p _{i, j+1} p _{i, j} + p _{i + 1, j}p _{i, j}\right] \bigg\},
\end{equation}
where $q_{i, j}$ and $p_{i, j}$ are respectively the generalized
positions and momenta at site $(i, j)$ and $\epsilon _{i, j}$ are the
disorder parameters uniformly chosen in the interval
$\left[-\frac{W}{2},\frac{W}{2}\right]$. We again consider fixed
boundary conditions i.e.~$q_{0, j} = p_{0, j} = q_{N+1, j}= p_{N+1,
  j}=0$ for $1\leq j\leq M$ and $q_{i, 0}= p_{i, 0}= q_{i, M+1} =
p_{i, M+1} = 0$ for $1\leq i\leq N$.

Additionally to the energies $H_{1D}$ [Eq.~\eqref{eq:DNLS_1D_real}]
and $H_{2D}$ [Eq.~\eqref{eq:DNLS_2D_real}] both systems conserve their
 respective norms $S_{\text{1D}} $ and $S_{\text{2D}}$:
\begin{equation}
S_{\text{1D}} = \frac{1}{2}\sum _{i=1}^{N} \left( q_i^2 + p_i ^2\right)\ ;
\qquad \quad
S_{\text{2D}} =\frac{1}{2}\sum _{i=1}^{N} \sum _{j=1}^{M} \left(q_{i, j}^2 + p_{i, j}^2 \right)\ .
\label{eq:normDDNLS}
\end{equation}

%-----------------------------------------------------
\section{Numerical integration schemes}\label{sec:integration_schemes}

The Hamilton equations of motion
\begin{equation}
\label{eq:ham-em}
\frac{d\boldsymbol{ q}}{dt} = \frac{\partial H}{\partial \boldsymbol{ p}}\ ,\qquad \frac{d\boldsymbol{ p}}{dt} = -\frac{\partial H}{\partial \boldsymbol{ q}},
\end{equation}
of the $N$ degree of freedom (dof) Hamiltonian $H=H(\boldsymbol{ q},
\boldsymbol{ p})$, with $\boldsymbol{ q}= (q_1,q_2,\ldots,q_N)$ and
$\boldsymbol{ p}= (p_1,p_2,\ldots,p_N)$ being respectively the
system's generalized positions and momenta, can be expressed in the
general setting of first order ODEs as
\begin{equation}
\label{eq:ham-em_ode}
\frac{d\boldsymbol{ x}}{dt} =\dot{\boldsymbol{ x}}=\boldsymbol{ J}_{2N} \cdot \boldsymbol{ D}_H(\boldsymbol{ x}(t)),
\end{equation}
where $\boldsymbol{ x} = \left(\boldsymbol{ q}, \boldsymbol{ p}
\right) = (x_1, x_2, \ldots, x_{N}, x_{N+1}, \ldots, x_{2N} ) =
(q_1,q_2,\ldots,q_N,p_1,p_2,\ldots,p_N)$ is a vector representing the
position of the system in its phase space and $(\,\dot{}\,)$ denotes
differentiation with respect to time $t$. In Eq.~(\ref{eq:ham-em_ode})
\begin{equation}
\label{eq:symplmat}
\boldsymbol{ J}_{2N} =
\begin{bmatrix}
    \boldsymbol{O}_N       &  \boldsymbol{I}_N \\
   - \boldsymbol{I}_N       &  \boldsymbol{O}_N
\end{bmatrix},
\end{equation}
is the symplectic matrix with $\boldsymbol{I}_N$ and
$\boldsymbol{O}_N$ being the $N\times N$ identity and the null
matrices respectively, and
\begin{equation}
\label{eq:DH}
\boldsymbol{ D}_H =\left[ \frac{\partial H}{\partial q_1}, \ldots, \frac{\partial H}{\partial q_N},\frac{\partial H}{\partial p_1}, \ldots, \frac{\partial H}{\partial p_N} \right]^T,
\end{equation}
with $(^T)$ denoting the transpose matrix.

The variational equations (see for
example~\cite{skokos2010lyapunov,skokos2010numerical}) govern the time
evolution of a small perturbation $\boldsymbol{ w}(t)$ to the trajectory
$\boldsymbol{ x}(t)$ with $\boldsymbol{w}(t)= (\delta \boldsymbol{q}(t),
\delta \boldsymbol{p}(t))=\left(\delta q_1 (t), \delta q_2 (t), \ldots,
  \delta q_N (t), \delta p_1(t), \delta p_2(t), \ldots, \delta p_N (t)
\right)$ (which can also be written as $\boldsymbol{w}(t)=\delta
\boldsymbol{x}(t)= \left(\delta x_1 (t), \ldots, \delta x_N (t),
  \delta x_{N+1}(t), \ldots, \delta x_{2N} (t) \right)$) and have the
following form
\begin{equation}
\label{eq:var_eq}
\dot{\boldsymbol{ w}}(t) =\big[  \boldsymbol{ J}_{2N} \cdot \boldsymbol{ D}_{ H}^2 (\boldsymbol{ x}(t) ) \big] \cdot \boldsymbol{ w} (t),
\end{equation}
where
\begin{equation}
\label{eq:hessian}
\left[\boldsymbol{ D}_{H}^2 (\boldsymbol{ x}(t)) \right]_{i, j} =
\frac{\partial^2 H}{\partial x_i \partial x_j}\bigg|_{\boldsymbol{ x}(t)}, \,\,\, i,j=1,2,\ldots,N,
\end{equation}
are the $2N\times 2N$ elements of the Hessian matrix $\boldsymbol{
  D}_{ H}^2 (\boldsymbol{ x}(t))$ of the Hamiltonian function $H$
computed on the phase space trajectory $\boldsymbol{
  x}(t)$. Eq.~\eqref{eq:var_eq} is linear in $\boldsymbol{w}(t)$,
with coefficients depending on the system's trajectory $\boldsymbol{
  x}(t)$. Therefore, one has to integrate the variational
equations~\eqref{eq:var_eq} along with the equations of motion
(\ref{eq:ham-em_ode}), which means to evolve in time the general
vector $ \boldsymbol{ X}(t) =(\boldsymbol{ x}(t),\delta \boldsymbol{
  x}(t))$ by solving the system
\begin{equation}
\label{eq:genaralX}
\dot{\boldsymbol{ X}} = (\dot{\boldsymbol{ x}}(t),\dot{\delta \boldsymbol{ x}}(t))=\boldsymbol{ f}(\boldsymbol{ X})=
  \begin{bmatrix}
\boldsymbol{ J}_{2N} \cdot \boldsymbol{ D}_H(\boldsymbol{ x}(t))         \\
   \big[  \boldsymbol{ J}_{2N} \cdot \boldsymbol{ D}_{ H}^2 (\boldsymbol{ x}(t) ) \big] \cdot \delta \boldsymbol{ x}(t)
\end{bmatrix}.
\end{equation}
In what follows we will briefly describe several numerical schemes for
integrating the set of equations (\ref{eq:genaralX}).

%-----------------------------------------------------
\subsection{Non-symplectic integration schemes}
\label{subsec:nonsymplectic}

Here we present the non-symplectic schemes we will use in this
work: the Taylor series method and the Runge-Kutta discretization
scheme. These methods are referred to be \textit{non-symplectic}
because they do not preserve the geometrical properties of Hamiltonian
equations of motion (e.g.~their symplectic structure). The immediate consequence of that is that they do not
conserve constants of motion (e.g.~the system's energy).

%-----------------------------------------------------
\subsubsection{Taylor series method - $TIDES$}
\label{subsec:TIDES}

The Taylor series method consists in expanding the solution
$\boldsymbol{ X}(t) $ at time $t_0 + \tau$ in a Taylor series of
$\boldsymbol{X}(t)$ at $t = t_0$
\begin{equation}
  \boldsymbol{X} (t_0 + \tau) = \boldsymbol{ X}(t_0) + \frac{\tau^1}{1!}\frac{d \boldsymbol{ X} (t_0)}{dt} + \frac{\tau ^2}{2!}\frac{d^2 \boldsymbol{ X} (t_0)}{dt^2}+ \ldots + \frac{\tau ^{n}}{n!}\frac{d ^n \boldsymbol{ X} (t_0)}{dt^{n}} + \mathcal{O}\left(\frac{\tau^{n + 1}}{(n + 1)!}\frac{d^{n + 1} \boldsymbol{ X} (t_0)}{dt^{n+1}} \right).
\label{eq:taylorseriesmethod}
\end{equation}

Once the solution $\boldsymbol{X}$ at time $ t_0 + \tau $ is
approximated, $\boldsymbol{ X} (t_0 + \tau) $ is considered as the  new
initial condition and the procedure is repeated to approximate
$\boldsymbol{X} (t_0 + 2 \tau) $ and so on so forth. Further
information on this integrator can be found in
~\cite[Sec. I.8]{hairer1993solving}. If we consider in
Eq.~\eqref{eq:taylorseriesmethod} only the first $n + 1$ terms we then
account the resulting scheme to be of order $n$. In addition, in order
to explicitly express this numerical scheme, one has to perform $n-1$
successive differentiations of the field vector $\boldsymbol{
  f}$. This task becomes very elaborate for complex structured vector
fields. One can therefore rely on successive differentiation to
automate the whole process (see e.g.~\cite{barrio2005performance}).
For the simulations reported in this paper we used the software
package called $TIDES$~\cite{barrio2005performance, abad2012algorithm}
which is freely available \cite{freelyTIDES}. We particularly focused
on the implementation of the $TIDES$ package as it has been already used
in studies of lattice dynamics \cite{gerlach2012efficient}. The $TIDES$ package
comes as a \textit{Mathematica} notebook in which the user
provides the Hamiltonian function, the potential energy or the set of
ODEs themselves. It then produces FORTRAN (or C) codes which can be
compiled directly by any available compiler producing the appropriate
executable programs.  In addition, the  $TIDES$ package allows us to
choose both the integration time step $\tau$ and the desired
`one-step' precision of the integrator $\delta$. In practice, the
integration time step is accepted if the estimated local truncation
error is smaller than $\delta$.

It is worth noting that there exists an equivalent numerical scheme to
the Taylor series method, derived from Lie
series~\cite{grobner1967lie} (for more details see e.g.~the appendix of~\cite{blanes2016concise}). Indeed, Eq.~\eqref{eq:genaralX} can be
expressed as
\begin{equation}
\frac{d\boldsymbol{X}}{dt} = L_{HZ} \boldsymbol{X},
\label{eq:formal-em}
\end{equation}
where $L_{HZ}$ is the Lie operator~\cite{hanslmeier1984numerical}
defined as
\begin{equation}
  L_{HZ}  = \sum _{i=1}^{2N} \left(\frac{dx_i}{dt}\frac{\partial }{\partial x_i}  + \frac{d\delta x_i}{dt}\frac{\partial }{\partial \delta x_i}\right).
\end{equation}
The formal solution of Eq.~\eqref{eq:formal-em} reads
 \begin{equation}
   \boldsymbol{X} (t_0 + \tau ) = e^{\tau L_{HZ} } \boldsymbol{X} (t_0),
\label{eq:formal_sol}
 \end{equation}
and can be expanded as
\begin{equation}
  \boldsymbol{X}(t_0 + \tau )= L_{HZ}^{0}\boldsymbol{X}(t_0) + \frac{\tau^1}{1!}L_{HZ}^{1}\boldsymbol{X}(t_0)  + \frac{\tau^2}{2!}L_{HZ}^{2}\boldsymbol{X}(t_0)  + \ldots + \frac{\tau^n}{n!}L_{HZ}^{n}\boldsymbol{X}(t_0)  + \mathcal{O}\left(\frac{\tau^{n+1}}{(n+1)!}L_{HZ}^{n+1}\boldsymbol{X}(t_0)  \right).
\label{eq:lieseriesoperator}
\end{equation}
This corresponds to a Lie series integrator of order $n$.  Similarly
to the Taylor series method, one has to find the analytical expression
of the successive action of the operator $L_{HZ}$ onto the vector
$\boldsymbol{X} (t_0)$.  For further details concerning Lie series one
can refer to~\cite{hanslmeier1984numerical, eggl2010introduction}.
The equivalence between the Lie and Taylor series approaches can be
seen in the following way: for each element $x_i$ and $\delta x_i$ of
the phase space vector $\boldsymbol{X}$ we compute
\begin{equation}
\begin{array}{lcll}
  \displaystyle L_{HZ}^0 x_i  &=& \displaystyle  Id\cdot x_i  = x_i (t_0), & \displaystyle 1\leq i\leq 2N,\\
  \displaystyle L_{HZ}^1 x_i &=&\displaystyle  \sum _{j=1}^{2N} \frac{d x_j}{dt}  \frac{\partial x_i}{\partial x_j} = \frac{d x_i}{dt} = f_i,
  &\displaystyle 1\leq i\leq 2N,    \\
  \displaystyle L_{HZ}^2 x_i &=& \displaystyle \sum _{j=1}^{2N} \frac{d x_j}{dt}  \frac{\partial f_i}{\partial x_j} = \frac{d f_i}{dt} = \frac{d^2 x_i}{dt^2} , &\displaystyle 1\leq i\leq 2N, \\
  \qquad  & & \qquad \qquad \vdots \nonumber \qquad \qquad & \\
  \displaystyle L_{HZ}^0 \delta x_i  &=& \displaystyle  Id\cdot \delta x_i  = \delta x_i (t_0) , & \displaystyle 1\leq i\leq 2N,\\
  \displaystyle L_{HZ}^1 \delta x_i &=& \displaystyle  \sum _{j=1}^{2N} \frac{d \delta x_j}{dt}  \frac{\partial \delta x_i}{\partial \delta x_j} = \frac{d \delta x_i}{dt} = f_{2N + i} , & \displaystyle  1\leq i\leq 2N,\\
  \displaystyle L_{HZ}^2 \delta x_i &=& \displaystyle  \sum _{j=1}^{2N} \frac{d x_j}{dt}  \frac{\partial  f_{2N+i}}{\partial  x_j} + \frac{d \delta x_j}{dt}  \frac{\partial f_{2N+i}}{\partial \delta x_j}  = \frac{d f_{2N + i} }{dt} = \frac{d^2 \delta x_i}{dt^2} , &\displaystyle  1\leq i\leq 2N,\\
  \qquad  & & \qquad \qquad \vdots \nonumber \qquad \qquad &
 \end{array}
 \label{eq:lie_one}
\end{equation}
Therefore $L_{HZ}^0 \boldsymbol{X} = \boldsymbol{X}$, $L_{HZ}^1
\boldsymbol{X} = \frac{d \boldsymbol{X}}{dt}$, $L_{HZ}^2\boldsymbol{X}
= \frac{d^2 \boldsymbol{X}}{dt^2}$ etc.

%-----------------------------------------------------
\subsubsection{A Runge-Kutta family scheme $-$ $DOP853$}
\label{susec:RK8-DOP}

A huge hurdle concerning the applicability of the Taylor and Lie series methods can be the explicit derivation of the differential operators (see~\cite{eggl2010introduction} and references therein).
Over the years other methods have been developed in order to overcome such issues and to efficiently and accurately
approximate Eq.~\eqref{eq:taylorseriesmethod}  up to a certain order in $n$.
One way to perform this task was
through the use of the well-known `Runge-Kutta family' of algorithms
(see e.g.~\cite{hairer2002geometric, eggl2010introduction}) which
nowadays  is one of the most popular general purpose schemes for
numerical computations. This is a sufficient motivation for us to
introduce a $s-$stage Runge-Kutta method of the form
\begin{equation}
  \boldsymbol{X} (t_0+\tau ) = \boldsymbol{X} (t_0) + \tau \sum_{i=1}^{ s}b_i\boldsymbol{k}_i,~\text{with}~\boldsymbol{k}_i = \boldsymbol{f} \left(t_0 + c_i \tau , \boldsymbol{X} (t_0) + \tau \sum_{j=1}^{ i-1} a_{i, j} \boldsymbol{k}_j \right)~\text{and}~c_i =\sum_{j=1}^{ i-1} a_{i,j},
\label{eq:rungekuttageneralexplicit}
\end{equation}
where the real coefficients $a_{i,j}$, $b_i$ with $i, j=1, \ldots, s$
are appropriately chosen in order to obtain the desired accuracy
(see e.g.~\cite[Sec.~II.1]{hairer1993solving}).
Eq.~\eqref{eq:rungekuttageneralexplicit} can be understood in the following way: in order to propagate the phase space vector $\boldsymbol{X}$ from time $t=t_0$ to $t=t_0+\tau$, we compute $s$ intermediate values $\boldsymbol{X}_1, \boldsymbol{X}_2, \ldots, \boldsymbol{X}_s$ and $s$ intermediate derivatives $\boldsymbol{k}_1, \boldsymbol{k}_2, \ldots, \boldsymbol{k}_s$, such that $\boldsymbol{k}_i = \boldsymbol{f} ( \boldsymbol{\boldsymbol{X}_i})$, at  times $t_i = t_0 + \tau \sum _{j = 1}^{i - 1} a_{ij}$. Then each $\boldsymbol{X}_i$ is found through  a linear combination of the known $\boldsymbol{k}_i$ values, which are added to  the initial condition $\boldsymbol{X}(t_0)$.

In this work we use a $12$-stage explicit Runge-Kutta algorithm of order $8$, called $DOP853$~\cite[Sec.~II.5]{hairer1993solving}. This method is the most precise scheme among the Runge-Kutta algorithms presented in in~\cite{hairer1993solving} (see Fig.~$4.2$ of \cite{hairer1993solving}). A free access Fortran77 implementation  of $DOP853$ is available in~\cite{freelyDOP853} (see also the appendix of~\cite{hairer1993solving}).
As in the case of the  $TIDES$ package, apart from the integration time
step $\tau$, $DOP853$ also admits a `one-step' precision $\delta$ based on embedded formulas of orders $5$ and $3$ (see~\cite[Sec.~II.10]{hairer1993solving}, or \cite{gerlach2012efficient} and references therein for
more details).

%-----------------------------------------------------
\subsection{Symplectic integration schemes}\label{sec:symp_integration_schemes}

Hamiltonian systems are characterized by a symplectic structure (see e.g.~\cite[Chap. VI]{hairer2002geometric} and
references therein) and they may also possess integrals of motion, like for example the
energy, the angular momentum, etc.
SIs are appropriate for the numerical integration of Hamiltonian systems as they keep  the computed  energy (i.e.~the value of the Hamiltonian) of the system almost constant over the integration in time. Let us remark that, in general, SIs do not preserve any additional conserved quantities of the system (for a specific exception see~\cite{Boreux2010high}). SIs
have already been
extensively used in fields such as celestial
mechanics~\cite{laskar2003chaos}, molecular
dynamics~\cite{leimkuhler2004simulating}, accelerator
physics~\cite{forest2006geometric}, condensed matter
physics~\cite{flach2009universal,bodyfelt2011nonlinear,senyange2018characteristics}, etc.

Several methods can be used to build SIs~\cite{hairer2002geometric}.
For instance, it has been proved that under certain circumstances the Runge-Kutta algorithm can preserve the
symplectic structure (see e.g.~\cite{lasagni1988canonical, sanz1988runge}).
However, the most common way to construct SIs is through the \textit{split}
of the Hamiltonian function into the sum of two or more integrable parts.
For example, many Hamiltonian functions are expressed as the sum of the
\textit{kinetic} and \textit{potential} energy terms, with each one of them corresponding to an integrable system (see Appendices \ref{sec:FPUT_app} and \ref{sec:app_other}).
Let us remark that, in general, even if each component of the total Hamiltonian is integrable the corresponding analytical solution might be unknown. Nevertheless,  we will not consider such cases in this work.
In our study we also want to track the evolution of
the deviation vector $\boldsymbol{ w}(t)$ by solving
Eq.~\eqref{eq:genaralX}.  Indeed this can be done by using SIs since
upon splitting the Hamiltonian into integrable parts we know
analytically for each part the exact mapping $\boldsymbol{ x} (t)
\rightarrow \boldsymbol{ x}(t+\tau )$, along with the mapping $\delta
\boldsymbol{ x} (t) \rightarrow \delta \boldsymbol{ x}(t+\tau )$.

Let us outline the whole process by considering a general autonomous
Hamiltonian function $H\left(\boldsymbol{ q}, \boldsymbol{ p}\right)$,
which can be written as sum of $I$ integrable intermediate Hamiltonian
functions $A_i$, i.e.~$H = \sum_{i=1}^I A_i$. This decomposition
implies that the  operator $e^{ \tau L_{A_iZ}}$ in the formal
solution in Eq.~\eqref{eq:formal_sol} of each intermediate
Hamiltonian function $A_i$ is known.  A symplectic integration scheme
to integrate the Hamilton equations of motion from $t_0$ to $t_0+\tau$
consists in approximating the action of  $e^{\tau L_{HZ}}$ in Eq.~\eqref{eq:formal_sol}
by a product of the operators $e^{\gamma_i \tau L_{A_iZ}}$ for a set
of properly chosen coefficients $\gamma_i$.  In our analysis we will
call as number of steps of a particular SI the total number of
successive application of each individual operator $ e^{ \tau
  L_{A_iZ}}$.
  Further details about this class of integrators can be
found in ~\cite{yoshida1990construction, suzuki1990fractal,
  yoshida1993recent} and references therein.  In what follows, we
consider the most common cases where the Hamiltonian function can be
split into two ($I=2$) or three ($I=3$) integrable parts.
Let us remark here that, in general, the split $H = \sum_{i=1}^I A_i$ is not necessarily unique (see Appendix \ref{sec:FPUT_app}).
Studying  the efficiency and the stability of different SIs upon different choices of splitting the Hamiltonian is an interesting topic by itself, which, nevertheless, is beyond the scope of our work.

%-----------------------------------------------------
\subsubsection{Two part split}\label{sec:2steps}

Let us consider that the Hamiltonian $H\left(\boldsymbol{ q},
  \boldsymbol{ p}\right)$ can be separated into two integrable parts,
namely $H = A(\boldsymbol{ q}, \boldsymbol{ p}) + B(\boldsymbol{ q},
\boldsymbol{ p})$. Then we can approximate the action of the operator
$e^{\tau L_{HZ}} = e^{\tau( L_{AZ} + L_{BZ})}$ by the successive
actions of products of the operators $e^{\tau L_{AZ}}$ and $e^{\tau
  L_{BZ}}$ \cite{yoshida1990construction,
  suzuki1990fractal,yoshida1993recent, ruth1983canonical}
\begin{equation}
  e^{\tau L_{HZ}} = \prod_{j=1}^{ p} e^{c_j\tau L_{AZ}} e^{d_j\tau L_{BZ}} + \mathcal{O}\left( \tau ^{n + 1}\right),
\label{eq:general_approx_elh_2ps}
\end{equation}
for appropriate choices of the real coefficients
$c_j, d_j$ with $j=1,\dots,p$.
Different choices of $p$ and coefficients $c_j$, $d_j$ lead to schemes
of different accuracy.  In Eq.~\eqref{eq:general_approx_elh_2ps} the integer $n$ is called the {\it order} of a symplectic integrator.

The Hamiltonian function of the 1D $\alpha$-FPUT system
[Eq.~\eqref{eq:alphaFPUT}] can be split into two integrable parts
$H_{\text{1F}} = A\left(\boldsymbol{ p}\right) + B\left(\boldsymbol{
    q}\right)$, with each part possessing $N$ cyclic coordinates. The
kinetic part
\begin{equation}
  A\left(\boldsymbol{ p}\right) = \sum _{i}\frac{p_i^2}{2},
\end{equation}
depends only on the generalized momenta, whilst the potential part
 \begin{equation}
   B\left(\boldsymbol{ q}\right) = \sum _{i} \frac{1}{2}(q_{i+1} - q_{ i})^2  + \frac{\alpha}{3}(q_{i+1} - q_{ i})^3,
 \end{equation}
depends only on the generalized positions. This type of split is
 the most commonly used in the literature, therefore a large number of
 SIs have been developed for such splits. Below we briefly
 review the SIs used in our analysis, based mainly on results
 presented in \cite{senyange2017symplectic}.

%-----------------------------------------------------
\paragraph{Symplectic integrators of order two.}
These integrators constitute the most basic schemes we can develop
from Eq.~\eqref{eq:general_approx_elh_2ps}

%-----------------------------------------------------
\subparagraph{$LF$:} The simplest example of
Eq.~\eqref{eq:general_approx_elh_2ps} is the so-called   {\it St\"ormer-Verlet} or {\it leap-frog}
scheme (e.g.~see \cite[Sect. I.3.1]{hairer2002geometric} and
\cite{ruth1983canonical}) having $3$ individual steps
\begin{equation}
LF(\tau ) = e^{a_1 \tau L_{AZ}} e^{b_1 \tau L_{BZ}}e^{a_1 \tau L_{AZ}},
\label{eq:leapfrog}
\end{equation}
where $a_1 = \frac{1}{2}$ and $b_1 = 1$.

%-----------------------------------------------------
\subparagraph{$SABA_2/SBAB_2$:}\label{sec:sub_saba-sbab}
We consider the $SABA_2$ and the $SBAB_2$ SIs with $5$ individual steps
 \begin{equation}
   SABA_2(\tau )= e^{a_1 \tau L_{AZ}}e^{b_1 \tau L_{BZ}}e^{a_2 \tau L_{AZ}}e^{b_1 \tau L_{BZ}}e^{a_1 \tau L_{AZ}},
 \label{eq:saba2}
 \end{equation}
 where $a_1 = \frac{1}{2} - \frac{1}{2\sqrt{3}}$, $a_2 =
 \frac{1}{\sqrt{3}}$ and $b_1 = \frac{1}{2}$, and
\begin{equation}
  SBAB_2(\tau) = e^{b_1 \tau L_{BZ}}e^{a_1 \tau L_{AZ}}e^{b_2 \tau L_{BZ}}e^{a_1 \tau L_{AZ}}e^{b_1 \tau L_{BZ}},
\label{eq:sbab2}
\end{equation}
with $a_1 = \frac{1}{2}$, $b_1 = \frac{1}{6}$ and $b_2 = \frac{2}{3}$.
These schemes were presented in \cite{mclachlan1995composition}, where they were named the (4,2) methods, and also used in \cite{laskar2001high}. We note  that  the $SABA_2$ and  $SBAB_2$ SIs (as well as other two part split SI schemes) have been introduced for Hamiltonian systems of the form $H = A + \varepsilon B$, with $\varepsilon$ being a small parameter. Both the   $SABA_2$ and  $SBAB_2$  integrators have only positive time steps and  are characterized by an accuracy of order
$\mathcal{O}(\tau^4\varepsilon + \tau^2\varepsilon^2)$ \cite{laskar2001high}.
Although these integrators are particularly efficient for small perturbations ($\varepsilon \ll1$), they have also shown a very good performance in cases of $\varepsilon = 1$ (see e.g.~\cite{skokos2009delocalization}).

%-----------------------------------------------------
\subparagraph{$ABA82$:} In addition, we use in our analysis the SI
\begin{equation}
  ABA82(\tau ) = e^{a_1 \tau L_{AZ}}e^{b_1 \tau L_{BZ}}e^{a_2 \tau L_{AZ}}e^{b_2 \tau L_{BZ}}e^{a_3 \tau L_{AZ}}e^{b_2 \tau L_{BZ}}e^{a_2 \tau L_{AZ}}e^{b_1 \tau L_{BZ}}e^{a_1 \tau L_{AZ}},
\label{eq:aba82}
\end{equation}
with $9$ individual
steps~\cite{mclachlan1995composition,farres2013high}, where the
constants $a_i, b_i$ with $i=1, 2, 3$ can be found in Table 2
of~\cite{farres2013high}.
We note that the $ABA82$ method is
called $SABA_4$ in~\cite{laskar2001high}.

%-----------------------------------------------------
\paragraph{Symplectic integrators of order four.}
The order of symmetric SIs can be increased by using a composition
technique presented in \cite{yoshida1990construction}. According to
that approach starting from a symmetric SI $S_{2n} (\tau)$ of order
$2n~(n\geq 1)$, we can construct a SI $S_{2n+2} (\tau)$ of order
$2n+2$ as \footnote{We adopt the notation of Iserles and
  Quispel~\cite{iserles2018geometric}.}
\begin{equation}
  S_{2n+2} (\tau) = S_{2n} ((1 + d) \tau)\, S_{2n} (-(1 + 2d) \tau)\, S_{2n} ((1 + d)\tau),\quad\text{where}\quad d = \frac{2^{1/(2n+1)} - 1}{ 2 - 2^{1/(2n+1)}}.
\label{eq:yoshida_composition_technique}
\end{equation}

%-----------------------------------------------------
\subparagraph{$FR4$:} Using the composition given in
Eq.~\eqref{eq:yoshida_composition_technique} for the $LF$ SI of
Eq.~\eqref{eq:leapfrog} we construct a SI which we name
$FR4$~\cite{yoshida1990construction, forest1990fourth} having $7$
individual steps
\begin{equation}
  FR4(\tau ) = e^{a_1 \tau L_{AZ}}e^{b_1 \tau L_{BZ}}e^{a_2 \tau L_{AZ}}e^{b_2 \tau L_{BZ}}e^{a_2 \tau L_{AZ}}e^{b_1 \tau L_{BZ}}e^{a_1 \tau L_{AZ}},
\label{eq:fr4}
\end{equation}
with coefficients $a_1 = \frac{1}{2(2 - 2^{1/3})}$, $a_2 = \frac{1 -
  2^{1/3}}{2(2 - 2^{1/3})}$, $b_1 = \frac{1}{2 - 2^{1/3}}$ and $b_2 =
-\frac{2^{1/3}}{2 - 2^{1/3}}$.

%-----------------------------------------------------
\subparagraph{$SABA_2Y4/SBAB_2Y4$:} Applying the composition given in
Eq.~\eqref{eq:yoshida_composition_technique} to the $SABA_2$
[Eq.~\eqref{eq:saba2}] and the $SBAB_2$ [Eq.~\eqref{eq:sbab2}]
integrators we obtain the fourth order SIs $SABA_2Y4$ and $SBAB_2Y4$
having $13$ individual steps. In particular, we get
\begin{align}
  \nonumber SABA_2Y4(\tau ) = e^{d_1a_1 \tau L_{AZ}}e^{d_1b_1 \tau
    L_{BZ}} & e^{d_1a_2 \tau L_{AZ}}e^{d_1b_1 \tau L_{BZ}}e^{a_0 \tau
    L_{AZ}}
  e^{d_0b_1 \tau L_{BZ}}e^{d_0a_2 \tau L_{AZ}}e^{d_0b_1 \tau L_{BZ}} \\
  & \times e^{a_0 \tau L_{AZ}}e^{d_1b_1 \tau L_{BZ}}e^{d_1a_2 \tau
    L_{AZ}}e^{d_1b_1 \tau L_{BZ}}e^{d_1a_1 \tau L_{AZ}},
\label{eq:saba2y4}
\end{align}
with coefficients
$d_0 = - \frac{2^{1/3}}{2 - 2^{1/3} }$,  $d_1 = \frac{1}{ 2 - 2^{1/3} }$,
$a_1 = \frac{1}{2} - \frac{1}{2\sqrt{3}}$, $a_2 = \frac{1}{\sqrt{3}}$, $b_1 = \frac{1}{2}$,
and $a_0 = d_1a_1 + d_0 a_1$, and
\begin{align}
  \nonumber
  SBAB_2Y4 (\tau ) = e^{d_1b_1 \tau L_{BZ}}e^{d_1a_1 \tau L_{AZ}} &  e^{d_1b_2 \tau L_{BZ}}e^{d_1a_1 \tau L_{AZ}}  e^{b_0 \tau L_{BZ}}  e^{d_0a_1 \tau L_{AZ}}e^{d_0b_2 \tau L_{BZ}}e^{d_0a_1 \tau L_{AZ}} \\
  & \times e^{b_0 \tau L_{BZ}}e^{d_1a_1 \tau L_{AZ}}e^{d_1b_2 \tau
    L_{BZ}}e^{d_1a_1 \tau L_{AZ}}e^{d_1b_1 \tau L_{BZ}},
\label{eq:sbab2y4}
\end{align}
with coefficients
$d_0 = -\frac{2^{1/3}}{ 2 - 2^{1/3} }$, $d_1 = \frac{1}{ 2 - 2^{1/3} }$,
$a_1 = \frac{1}{2}$, $b_1 =\frac{1}{6}$, $b_2 = \frac{2}{3}$ and $b_0 = d_1 b_1 + d_0 b_1$.

%-----------------------------------------------------
\subparagraph{$ABA82Y4$:} Using the composition given in
Eq.~\eqref{eq:yoshida_composition_technique} for the second order
scheme $ABA82$ of Eq.~\eqref{eq:aba82} we obtain an integrator with
$25$ individual steps having the form
\begin{align}
  \nonumber ABA82Y4 (\tau ) = & e^{d_1 a_1 \tau L_{AZ}}e^{d_1b_1 \tau
    L_{BZ}}e^{d_1a_2 \tau L_{AZ}}e^{d_1b_2 \tau L_{BZ}}e^{d_1a_3 \tau
    L_{AZ}}e^{d_1b_2 \tau L_{BZ}}e^{d_1a_2 \tau L_{AZ}}e^{d_1b_1 \tau
    L_{BZ}}e^{a_0 \tau L_{AZ}} \\ \nonumber & \times
  e^{d_0b_1 \tau L_{BZ}}e^{d_0a_2 \tau L_{AZ}}e^{d_0b_2 \tau L_{BZ}}e^{d_0a_3 \tau L_{AZ}}e^{d_0b_2 \tau L_{BZ}}e^{d_0a_2 \tau L_{AZ}}e^{d_0b_1 \tau L_{BZ}}e^{a_0 \tau L_{AZ}} \\
  & \times e^{d_1b_1 \tau L_{BZ}}e^{d_1a_2 \tau L_{AZ}}e^{d_1b_2 \tau
    L_{BZ}}e^{d_1a_3 \tau L_{AZ}}e^{d_1b_2 \tau L_{BZ}}e^{d_1a_2 \tau
    L_{AZ}}e^{d_1b_1 \tau L_{BZ}}e^{d_1a_1 \tau L_{AZ}} ,
\label{eq:aba82y4}
\end{align}
where   $d_0 = -\frac{2^{1/3}}{ 2 - 2^{1/3} }$, $d_1 = \frac{1}{ 2 - 2^{1/3} }$,
while $a_i, b_i$ with $i=1, 2, 3$  can be found
in Table 2 of~\cite{farres2013high}. Here $a_0 = d_1a_1 + d_0 a_1$.

%-----------------------------------------------------
\subparagraph{$SABA_2K/SBAB_2K$:} As was explained in
\cite{laskar2001high} the accuracy of the $SABA_n$ (and the $SBAB_n$)
class of SIs
can be improved by a {\it corrector} term $K = \{
B,\{B,A\}\}$, defined by two successive applications of Poisson brackets
($\{\cdot,\cdot\}$), if $K$ corresponds to a solvable Hamiltonian
function.  In that case, the second order integration schemes can be
improved by the addition of two extra operators
with negative time steps
in the following way
\begin{equation}
  SABA_nK(\tau) \equiv e^{-\frac{g \tau^2}{2} L_{KZ}} SABA_n(\tau) e^{-\frac{g \tau^2}{2} L_{KZ}},
\label{eq:2st_int_corre}
\end{equation}
with analogous result holding for the $SBAB_n$ scheme.
By following
this approach for the $SABA_2$ and $SBAB_2$ SIs
[which are of the order
$\mathcal{O}(\tau^4\varepsilon + \tau^4\varepsilon^2)$]
we produce the fourth
order SIs $SABA_2K$ and $SBAB_2K$, with $g = (2 - \sqrt{3})/24$ and
$g=1/72$ respectively.
These new integration schemes are of the order $\mathcal{O}(\tau^4\varepsilon + \tau^4\varepsilon^2)$ \cite{laskar2001high}.

%-----------------------------------------------------
\subparagraph{$ABA864/ABAH864$:} The fourth order SIs $ABA864$ and
$ABAH864$ were proposed in~\cite{blanes2013new, farres2013high}. They
have respectively $15$ and $17$ individual steps and have the form
\begin{align}
  \nonumber
  ABA864(\tau ) = e^{a_1 \tau L_{AZ}}e^{b_1 \tau L_{BZ}}e^{a_2 \tau L_{AZ}}e^{b_2 \tau L_{BZ}} & e^{a_3 \tau L_{AZ}}e^{b_3 \tau L_{BZ}}e^{a_4 \tau L_{AZ}}e^{b_4 \tau L_{BZ}}e^{a_4 \tau L_{AZ}}e^{b_3 \tau L_{BZ}} \\
  & \times e^{a_3 \tau L_{AZ}}e^{b_2 \tau L_{BZ}}e^{a_2 \tau
    L_{AZ}}e^{b_1 \tau L_{BZ}}e^{a_1 \tau L_{AZ}},
\label{eq:aba864}
\end{align}
with coefficients $a_i, b_i~i=1, 2, 3, 4$ taken from Table 3 of
\cite{blanes2013new}, and
\begin{align}
  \nonumber
  ABAH864(\tau ) = e^{a_1 \tau L_{AZ}}e^{b_1 \tau L_{BZ}}e^{a_2 \tau L_{AZ}}e^{b_2 \tau L_{BZ}} & e^{a_3 \tau L_{AZ}}e^{b_3 \tau L_{BZ}}e^{a_4 \tau L_{AZ}}e^{b_4 \tau L_{BZ}}e^{a_5 \tau L_{AZ}}e^{b_4 \tau L_{BZ}}e^{a_4 \tau L_{AZ}}e^{b_3 \tau L_{BZ}} \\
  & \times e^{a_3 \tau L_{AZ}}e^{b_2 \tau L_{BZ}}e^{a_2 \tau
    L_{AZ}}e^{b_1 \tau L_{BZ}}e^{a_1 \tau L_{AZ}},
\label{eq:abah864}
\end{align}
with coefficients $a_i, b _i, ~i =1,2,\dots, 5$ found in Table 4
of~\cite{blanes2013new}.
We note that both schemes were designed for near-integrable systems of the form  $H = A + \varepsilon B$, with $\varepsilon$ being a small parameter, but the construction of $ABAH864$ was based on the assumption that the integration of the $B$ part cannot be done explicitly, but can be approximated by the action of some second order SI, since $B$ is expressed as the sum of two explicitly integrable parts, i.e.~$B=B_1+B_2$. The $ABA864$ and $ABAH864$ SIs are of order four, but their construction satisfy several other conditions at higher orders, improving in this way their performance.

%-----------------------------------------------------
\paragraph{Symplectic integrators of order six.}

Applying the composition technique of
Eq.~\eqref{eq:yoshida_composition_technique} to the fourth order SIs
$FR4$ [Eq.~\eqref{eq:fr4}], $SABA_2Y4$ [Eq.~\eqref{eq:saba2y4}],
$SBAB_2Y4$ [Eq.~\eqref{eq:sbab2y4}], $ABA82Y4$
[Eq.~\eqref{eq:aba82y4}], $SABA_2K$ and $ABA864$
[Eq.~\eqref{eq:aba864}], we construct the sixth order SIs $FR4Y6$,
$SABA_2Y4Y6$, $SBAB_2Y4Y6$, $ABA82Y4Y6$, $SABA_2KY6$ and $ABA864Y6$
with $19$, $37$, $37$, $73$, 19 and $43$ individual
steps.

In~\cite{yoshida1990construction} a composition technique using fewer
individual steps than the one obtained by the repeated application of
Eq.~\eqref{eq:yoshida_composition_technique} to SIs of order two was
proposed, having the form
\begin{equation}
  S_6(\tau) = S_2(w_3 \tau)S_2(w_2 \tau)S_2(w_1 \tau)S_2(w_0 \tau)S_2(w_1 \tau)S_2(w_2 \tau)S_2(w_3 \tau),
\label{eq:yoshida_composition_6_order}
\end{equation}
whose coefficients $w_i, ~i=0, 1, 2, 3$ are given in Table 1 in
\cite{yoshida1990construction} for the case of the so-called `solution
A' of that table. Here $S_2$ and $S_6$ respectively represent a second and a sixth
order symmetric SI.  Note that
Eq.~\eqref{eq:yoshida_composition_6_order} corresponds to the
composition scheme $s6odr6$ of \cite{kahan1997composition}. Applying
the composition given in Eq.~\eqref{eq:yoshida_composition_6_order} to the
$SABA_2$ [Eq.~\eqref{eq:saba2}], the $SBAB_2$ [Eq.~\eqref{eq:sbab2}]
and the $ABA82$ [Eq.~\eqref{eq:aba82}] SIs we generate the order six
schemes $SABA_2Y6$, $SBAB_2Y6$ and $ABA82Y6$ having $29$, $29$ and
$57$ individual steps respectively.

We also consider in our study the composition scheme $s9odr6b$ of
\cite{kahan1997composition} which is based on $9$ successive
applications of $S_2$
\begin{equation}
  s9odr6b(\tau ) = S_2(\delta _1 \tau)S_2(\delta _2 \tau)S_2(\delta _3 \tau)S_2(\delta _4 \tau)S_2(\delta _5 \tau)S_2(\delta _4 \tau)S_2(\delta _3 \tau)S_2(\delta _2 \tau)S_2(\delta _1 \tau).
\label{eq:s9odr6b_general}
\end{equation}
The values of $\delta _i,~i=1, 2, \dots, 5$ in
Eq.~\eqref{eq:s9odr6b_general} can be found in the Appendix
of~\cite{kahan1997composition}. Furthermore, we also implement the
composition method
\begin{equation}
  s11odr6(\tau) = S_2 (\gamma _1\tau)S_2 (\gamma _2\tau)S_2 (\gamma _3\tau)S_2 (\gamma _4\tau)S_2 (\gamma _5\tau)S_2 (\gamma _6\tau)S_2 (\gamma _5\tau) S_2 (\gamma _4\tau)S_2 (\gamma _3\tau)S_2 (\gamma _2\tau)S_2 (\gamma _1\tau)
\label{eq:s11odr6_general}
\end{equation}
of~\cite{sofroniou2005derivation}, which involves 11 applications of a
second order SI $S_2$, whose coefficients $\gamma_i$, $i=1, 2, \dots,
6$ are reported in Section 4.2 of~\cite{sofroniou2005derivation}.
Using the $SABA_2$ of Eq.~\eqref{eq:saba2} as $S_2$ in
Eqs.~\eqref{eq:s9odr6b_general} and \eqref{eq:s11odr6_general}, we
respectively build two SIs of order six, namely the $s9SABA_26$ and
the $s11SABA_26$ SIs with $37$ and $45$ individual steps.  In
addition, using the $ABA82$ integrator of Eq.~\eqref{eq:aba82} as
$S_2$ in Eqs.~\eqref{eq:s9odr6b_general}
and~\eqref{eq:s11odr6_general} we construct two other order six SIs
namely the $s9ABA82\_6$ and $s11ABA82\_6$ schemes with $73$ and $89$
individual steps respectively.

\subparagraph{Runge-Kutta-Nystr\"{o}m methods:}
In addition, we consider in our analysis two  SIs of order six,  belonging in the category of the so-called Runge-Kutta-Nystr\"{o}m (RKN) methods
(see e.g.~\cite{hairer2002geometric,blanes2002practical,blanes2008splitting, blanes2016concise}
and references therein), which respectively have $21$ and $29$
individual steps
\begin{align}
\nonumber
SRKN_{11}^b(\tau) = e^{b_1 \tau L_{BZ}}e^{a_1 \tau L_{AZ}} & e^{b_2 \tau L_{BZ}}  e^{a_2 \tau L_{AZ}}e^{b_3 \tau L_{BZ}}e^{a_3 \tau L_{AZ}} e^{b_4 \tau L_{BZ}}e^{a_4 \tau L_{AZ}}e^{b_5 \tau L_{BZ}}e^{a_5 \tau L_{AZ}}e^{b_6 \tau L_{BZ}}
e^{a_5 \tau L_{BZ}} \\
& \times
e^{b_5 \tau L_{AZ}}
e^{a_4 \tau L_{BZ}}e^{b_4\tau L_{AZ}}
e^{a_3 \tau L_{BZ}}e^{b_3 \tau L_{AZ}}
e^{a_2 \tau L_{AZ}}e^{b_2 \tau L_{AZ}} e^{a_1 \tau L_{AZ}}
e^{b_1 \tau L_{AZ}},
\label{eq:SRKN_11_b}
\end{align}
and
\begin{align}
\nonumber
SRKN_{14}^a(\tau) = e^{a_1 \tau L_{AZ}}
e^{b_1 \tau L_{AZ}}e^{a_2 \tau L_{AZ}}&
e^{b_2 \tau L_{AZ}} \times \ldots \times e^{a_7 \tau L_{AZ}}
e^{b_7 \tau L_{AZ}} e^{a_8 \tau L_{AZ}}e^{b_7 \tau L_{AZ}} e^{a_7 \tau L_{AZ}}  \\
& \times \ldots \times e^{b_2 \tau L_{AZ}}e^{a_2 \tau L_{AZ}}
e^{b_1 \tau L_{AZ}}e^{a_1 \tau L_{AZ}}.
\label{eq:SRKN_14_a}
\end{align}
The values of the coefficients appearing in Eqs.~(\ref{eq:SRKN_11_b}) and (\ref{eq:SRKN_14_a})  can be found in Table 3 of \cite{blanes2002practical}.
This class of integrators has, for example,  been successfully implemented in a recent investigation of the chaotic behavior of the DNA molecule \cite{hillebrand2018chaos}.

%-----------------------------------------------------
\paragraph{Symplectic integrators of order eight.}\label{sec: 2ps SI order 8}

Following \cite{yoshida1990construction} we can construct an eighth
order SI $S_8$, starting from a second order one $S_2$, by using the
composition
\begin{align}
  \nonumber
  S_8(\tau) =S_2(w_7 \tau)S_2(w_6 \tau)S_2(w_5 \tau) & S_2(w_4 \tau)S_2(w_3 \tau)S_2(w_2 \tau)S_2(w_1 \tau)S_2(w_0 \tau)S_2(w_1 \tau)S_2(w_2 \tau) \\
  & \times S_2(w_3 \tau)S_4(w_4 \tau)S_2(w_5 \tau)S_2(w_6 \tau)S_2(w_7
  \tau).
\label{eq:yoshida_composition_8_order}
\end{align}
In our study we consider two sets of coefficients $w_i$, $i=1, \ldots,
7$, and in particular the ones corresponding to the so-called
`solution A' and `solution D' in Table 2
of~\cite{yoshida1990construction}.  Using in
Eq.~\eqref{eq:yoshida_composition_8_order} the $SABA_2$
[Eq.~\eqref{eq:saba2}] SI as $S_2$ we construct the eighth order SIs
$SABA_2Y8\_A$ (corresponding to `solution A') and $SABA_2Y8\_D$
(corresponding to `solution D') with $61$ individual steps each.  In a
similar way the use of $ABA82$ [Eq.~\eqref{eq:aba82}] in
Eq.~\eqref{eq:yoshida_composition_8_order} generates the SIs
$ABA82Y8\_A$ and $ABA82Y8\_D$ with $121$ individual steps each.

In addition, considering the composition scheme $s15odr8$
of~\cite{kahan1997composition}, having 15 applications of $S_2$, we
construct the eighth order SI $s15SABA_28$ [$s15ABA82\_8$] having $61$
[$121$] individual steps when $SABA_2$ of Eq.~\eqref{eq:saba2}
[$ABA82$ of Eq.~\eqref{eq:aba82}] is used in the place of $S_2$.

Furthermore, implementing the composition technique $s19odr8$
presented in~\cite{sofroniou2005derivation}, which uses 19
applications of a second order SI $S_2$, we construct the SI
$s19SABA_28$ [$s19ABA82\_8$] with $77$ [$153$] individual steps, when
$SABA_2$ of Eq.~\eqref{eq:saba2} [$ABA82$ of Eq.~\eqref{eq:aba82}] is
used in the place of $S_2$.

The SIs of this section  will be implemented to numerically integrate
the $\alpha$-FPUT model [Eq.~\eqref{eq:alphaFPUT}] since this
Hamiltonian system can be split into two integrable parts
$A(\boldsymbol{ p})$ and $B(\boldsymbol{ q})$. In
Sec.~\ref{sec:FPUT_app} of the Appendix the explicit forms of the
operators $e^{\tau L_{AZ} }$ and $e^{\tau L_{BZ} }$ are given, along
with the operator $e^{\tau L_{KZ} }$ of the corrector term used by the
$SABA_2K$ and $SBAB_2K$ SIs. In addition, in Sec.~\ref{sec:app_other}
of the Appendix the explicit forms of the tangent map method operators
of some commonly used lattice systems, whose Hamiltonians can be split
in two integrable parts, are also reported.

%-----------------------------------------------------
\subsubsection{Three part split}\label{sec:3steps}

Let us now consider the case of a Hamiltonian function
$H\left(\boldsymbol{ q}, \boldsymbol{p}\right)$ which can be separated
into three integrable parts, namely $H\left(\boldsymbol{ q},
  \boldsymbol{p}\right) = \mathcal{A}\left(\boldsymbol{ q},
  \boldsymbol{p}\right) + \mathcal{B}\left(\boldsymbol{ q},
  \boldsymbol{ p}\right) + \mathcal{C}\left(\boldsymbol{ q},
  \boldsymbol{ p}\right)$.
This for example could happen because the
Hamiltonian function may not be split into two integrable parts,
or to simplify the solution of one of the two
components, $A$ or $B$, of the two part split schemes discussed in Sec.~\ref{sec:2steps}.
In such cases we approximate the action of the operator $e^{\tau
  L_{HZ}}$ of Eq.~\eqref{eq:formal_sol} by the successive application
of operators $e^{\tau L_{\mathcal{A}Z}}$, $e^{\tau L_{\mathcal{B}Z}}$
and $e^{\tau L_{\mathcal{C}Z}}$ i.e.
\begin{equation}
  e^{\tau L_{HZ}} = \prod_{j=1}^{p } e^{c_j\tau L_{\mathcal{A}Z}} e^{d_j\tau L_{\mathcal{B}Z}}e^{e_j\tau L_{\mathcal{C}Z}} + \mathcal{O}\left(\tau ^{n + 1} \right),
  \label{eq:general_approx_elh_3ps}
\end{equation}
for appropriate choices of the real coefficients  $c_j$, $d_j$ and $e_j$ with $j=1,\dots,p$.
As in Eq.~\eqref{eq:general_approx_elh_2ps}, in Eq.~\eqref{eq:general_approx_elh_3ps} the integer $n$ is the order of a symplectic integrator.

As examples of Hamiltonians which can be split in three integrable
parts we mention the Hamiltonian function of a free rigid
body~\cite{laskar2018dedicated} and the Hamiltonian functions of the
1D [Eq.~\eqref{eq:DNLS_1D_real}] and 2D [Eq.~\eqref{eq:DNLS_2D_real}]
DDNLS models we consider in this work. For example the 1D DDNLS
Hamiltonian of Eq.~\eqref{eq:DNLS_1D_real} can be split in the
following three integrable Hamiltonians: a system of $N$ independent
oscillators
\begin{equation}
  \mathcal{A}_1 = \sum _{i=1}^{N} \epsilon _i J_i + \frac{\beta}{2} J_i^2,
\label{eq:a_part_1DDNLS}
\end{equation}
where $J_i = (q_i^2 + p_i^2)/2$, $i = 1, \ldots, N$ are $N$ constants
of motion, and the Hamiltonian functions of the $\boldsymbol{q}-$ and
$\boldsymbol{p}-$hoppings
\begin{equation}
  \mathcal{B}_1 = -\sum _{i=1}^{N} p_{i}p_{i+1},\qquad \text{and} \qquad \mathcal{C}_1 = -\sum _{i=1}^{N} q_{i}q_{i+1},
\label{eq:bc_part_1DDNLS}
\end{equation}
with each one of them having $N$ cyclic coordinates.  The three part
split of the 2D DDNLS of Eq.~\eqref{eq:DNLS_2D_real} can be found
in Sec.~\ref{sec:2D_DNLS_app} of the Appendix
[Eq.~\eqref{eq:2D_DNLS_3split_app}].

We note that a rather thorough survey on three part split SIs can
be found in~\cite{skokos2014high, gerlach2016symplectic}. We decided
to include in our study a smaller number of schemes than the one presented in these works, focusing on the more efficient SIs. We briefly
present these integrators below.

%-----------------------------------------------------
\paragraph{Symplectic integrators of order two.}
We first present the basic three part split scheme obtained by the application of the St\"ormer-Verlet/leap-frog method to three-part separable Hamiltonians. This scheme has $5$ individual steps and we call it
$\mathcal{ABC}2$ \cite{koseleff1996exhaustiv}
\begin{equation}
  \mathcal{ABC}2 (\tau ) = e^{a_1 \tau L_{\mathcal{A}Z}}e^{b_1 \tau L_{\mathcal{B}Z}}e^{c_1 \tau L_{\mathcal{C}Z}}e^{b_1 \tau L_{\mathcal{B}Z}}e^{a_1 \tau L_{\mathcal{A}Z}},
\label{eq:abc2}
\end{equation}
where $a_1 = \frac{1}{2}$, $b_1 = \frac{1}{2}$ and $c_1 = 1$.

%-----------------------------------------------------
\paragraph{Symplectic integrators of order four.}
In order to built higher order three part split SIs we apply some
composition techniques on the basic $\mathcal{ABC}2$ SI of
Eq.~\eqref{eq:abc2}.

%-----------------------------------------------------
\subparagraph{$\mathcal{ABC}Y4$:} Using the composition given in
Eq.~\eqref{eq:yoshida_composition_technique} for $n=1$, we construct
\begin{equation}
  \mathcal{ABC}Y4(\tau ) = \mathcal{ABC}2(d_1\tau) \mathcal{ABC}2(d_0 \tau)\mathcal{ABC}2(d_1 \tau),
\label{eq:abcy4}
\end{equation}
with $d_0 = \frac{- 2^{1/3}}{2 - 2^{1/3}}$ and $d_1 = \frac{1}{2 -
  2^{1/3}}$.
This integrator has $13$ individual steps,  it has been explicitly introduced in \cite{koseleff1996exhaustiv} and implemented in \cite{gerlach2016symplectic}, where it was called
$\mathcal{ABC}^{4}_{[Y]}$.

%-----------------------------------------------------
\subparagraph{$\mathcal{ABC}S4$:} Implementing a composition scheme
which was introduced in~\cite{suzuki1990fractal} and studied in
~\cite{kahan1997composition} (where it was named $s5odr4$) we obtain
the SI
\begin{equation}
  \mathcal{ABC}S4(\tau ) = \mathcal{ABC}2(p_2\tau) \mathcal{ABC}2(p_2\tau)\mathcal{ABC}2((1- 4p_2)\tau)\mathcal{ABC}2(p_2\tau) \mathcal{ABC}2(p_2\tau),
\label{eq:abcs4}
\end{equation}
where $p_2 = \frac{1}{4 -4^{1/3}}$ and $1 - 4p_2 = \frac{-4^{1/3}}{4 -
  4^{1/3}}$, having $21$ individual steps. This integrator was denoted
as $\mathcal{ABC}^{4}_{[S]}$ in ~\cite{gerlach2016symplectic}.

%-----------------------------------------------------
\subparagraph{$SS864S$:} Using the $ABAH864$ integrator of Eq.~(\ref{eq:abah864}) where $B$ is considered to be the sum of functions $\mathcal{B}_1$ and $\mathcal{C}_1$ of Eq.~(\ref{eq:bc_part_1DDNLS}), i.e.~$B=\mathcal{B}_1+\mathcal{C}_1$, and its solution is approximated by the second order $SABA_2$ SI of Eq.~(\ref{eq:saba2}), we construct a SI with 49 steps, which we call $SS864S$. This integrator has been implemented for the integration of the equations of motion of the 1D DDNLS system [Eq.~(\ref{eq:DNLS_1D_real})] in \cite{skokos2014high}, where it was called $SS_{864}^4$.

%-----------------------------------------------------
\paragraph{Symplectic integrators of order six.}

%-----------------------------------------------------
\subparagraph{$\mathcal{ABC}Y4Y6/\mathcal{ABC}S4Y6$:} Applying the
composition technique of Eq.~\eqref{eq:yoshida_composition_technique}
to the fourth order SIs $\mathcal{ABC}Y4$ [Eq.~\eqref{eq:abcy4}] and
$\mathcal{ABC}S4$ [Eq.~\eqref{eq:abcs4}], we respectively construct the schemes
$\mathcal{ABC}Y4Y6$ and $\mathcal{ABC}S4Y6$ with $37$ and $49$
individual steps.

%-----------------------------------------------------
\subparagraph{$\mathcal{ABC}Y6\_A$:} Using the composition given in
Eq.~\eqref{eq:yoshida_composition_6_order} we build a sixth order SI
with $29$ individual steps, considering the integrator
$\mathcal{ABC}2$ in the place of $S_2$
\begin{equation}
  \mathcal{ABC}Y6\_A (\tau)= \mathcal{ABC}2(w_3 \tau)\mathcal{ABC}2(w_2 \tau)\mathcal{ABC}2(w_1 \tau)\mathcal{ABC}2(w_0 \tau)\mathcal{ABC}2(w_1 \tau)\mathcal{ABC}2(w_2 \tau)\mathcal{ABC}2(w_3 \tau).
\label{eq:abc6y}
\end{equation}
In particular, we consider in this construction the coefficients
$w_i$, $i=0,1,2,3$, corresponding to the `solution A' of Table 1
in~\cite{yoshida1990construction}.  Note that this SI has already been
implemented in~\cite{skokos2014high,gerlach2016symplectic}, where it
was denoted as $\mathcal{ABC}^{6}_{[Y]}$.

%-----------------------------------------------------
\subparagraph{$s9\mathcal{ABC}6$:} Implementing the composition given in
Eq.~\eqref{eq:s9odr6b_general}, with $\mathcal{ABC}2$ in the place of
$S_2$, we get
\begin{align}
  \nonumber
  s9\mathcal{ABC}6(\tau ) = \mathcal{ABC}2(\delta _1 \tau) & \mathcal{ABC}2(\delta _2 \tau)  \mathcal{ABC}2(\delta _1 \tau)\mathcal{ABC}2(\delta _4 \tau) \\
  & \times \mathcal{ABC}2(\delta _5 \tau)\mathcal{ABC}2(\delta _4
  \tau)\mathcal{ABC}2(\delta _3 \tau)\mathcal{ABC}2(\delta _2
  \tau)\mathcal{ABC}2(\delta _1 \tau),
\label{eq:abc6kl}
\end{align}
which was referred to as $\mathcal{ABC}^{6}_{[KL]}$
in~\cite{skokos2014high,gerlach2016symplectic}.

%-----------------------------------------------------
\subparagraph{$s11\mathcal{ABC}6$:} From the composition given in
Eq.~\eqref{eq:s11odr6_general} we create the SI scheme
\begin{align}
  \nonumber
  s11\mathcal{ABC}6(\tau) = \mathcal{ABC}2 (\gamma _1\tau) & \mathcal{ABC}2 (\gamma _2\tau)\times \dots \times \mathcal{ABC}2 (\gamma _5\tau)\mathcal{ABC}2 (\gamma _6\tau)\mathcal{ABC}2 (\gamma _5\tau) \\
  & \times \dots \times \mathcal{ABC}2 (\gamma
  _2\tau)\mathcal{ABC}2(\gamma _1\tau),
\label{eq: abc6ss}
\end{align}
which has 45 individual steps. This integrator was referred to as
$\mathcal{ABC}^{6}_{[SS]}$
in~\cite{skokos2014high,gerlach2016symplectic}.

%-----------------------------------------------------
\paragraph{Symplectic integrators of order eight.}

%-----------------------------------------------------
\subparagraph{$\mathcal{ABC}Y8\_A/\mathcal{ABC}Y8\_D$:} Based on the composition given in
Eq.~\eqref{eq:yoshida_composition_8_order} we construct the SI
\begin{align}
  \nonumber \mathcal{ABC}Y8(\tau ) =\mathcal{ABC}2(w _7\tau)\mathcal{ABC}2 (w
  _6\tau) & \times \dots \times \mathcal{ABC}2 (w
  _1\tau)\mathcal{ABC}2 (w _0\tau)\mathcal{ABC}2 (w _1\tau) \\& \times
  \dots \times \mathcal{ABC}2 (w _6\tau)\mathcal{ABC}2(w _7\tau),
\label{eq: abc8y}
\end{align}
setting $\mathcal{ABC}2$ in the place of $S_2$. This SI has $61$
individual steps. Considering the `solution A' of Table 2
in~\cite{yoshida1990construction} for the coefficients $w_i$, $0\leq i
\leq 7$, we obtain the $\mathcal{ABC}Y8\_A$ SI, while the use of
`solution D' of the same table leads to the construction of the SI
$\mathcal{ABC}Y8\_D$.

%-----------------------------------------------------
\subparagraph{$s17\mathcal{ABC}8$:} Consider the composition method
$s17odr8b$ of \cite{kahan1997composition} we build the SI (referred to
as $\mathcal{ABC}^{8}_{[KL]}$ in
\cite{skokos2014high,gerlach2016symplectic}) $s17\mathcal{ABC}8$
having $69$ individual steps
\begin{align}
\nonumber
s17\mathcal{ABC}8 (\tau) = \mathcal{ABC}2 (\delta _1\tau)\mathcal{ABC}2 (\delta _2\tau) & \times \dots \times \mathcal{ABC}2 (\delta _8\tau)\mathcal{ABC}2 (\delta _9\tau)\mathcal{ABC}2 (\delta _8\tau) \\
& \times \dots \times \mathcal{ABC}2 (\delta _2\tau)\mathcal{ABC}2(\delta _1\tau).
\label{eq: abc8kl}
\end{align}

%-----------------------------------------------------
\subparagraph{$s19\mathcal{ABC}8$:} Finally, we also implement the
composition  $s19odr8b$ reported in~\cite[Eq.~(13)]{sofroniou2005derivation} and construct the SI
\begin{align}
  \nonumber
  s19\mathcal{ABC}8(\tau) = \mathcal{ABC}2 (\gamma _1\tau)\mathcal{ABC}2(\gamma _2\tau) & \times \dots \times \mathcal{ABC}2 (\gamma _8\tau)\mathcal{ABC}2 (\gamma _9\tau)\mathcal{ABC}2 (\gamma _8\tau) \\
  & \times \dots \times \mathcal{ABC}2 (\gamma
  _2\tau)\mathcal{ABC}2(\gamma _1\tau),
\label{eq: abc8ss}
\end{align}
which has $72$ individual steps. We note that this scheme corresponds
to the SI $\mathcal{ABC}^{8}_{[SS]}$ considered in
\cite{skokos2014high,gerlach2016symplectic}.

The explicit forms of the operators related to the three part split of
the DDNLS Hamiltonians [Eqs.~\eqref{eq:DNLS_1D_real} and
\eqref{eq:DNLS_2D_real}] are given in Secs.~\ref{sec:1D_DNLS_app} and
\ref{sec:2D_DNLS_app} of the Appendix.

%------------------------------------
\section{Numerical results}\label{sec:tangent_map}

We test the efficiency of the integrators presented in Sec.~\ref{sec:integration_schemes} by
using them to follow the dynamical evolution of the $\alpha$-FPUT model [Eq.~\eqref{eq:alphaFPUT}], the 1D DDNLS system [Eq.~\eqref{eq:DNLS_1D_real}] and the 2D DDNLS Hamiltonian [Eq.~\eqref{eq:DNLS_2D_real}]. For each model, given an initial condition $\boldsymbol{X} (t_0) = \left( \boldsymbol{ x} (t_0), \delta\boldsymbol{ x} (t_0)\right)$ we compute the trajectory $\{\boldsymbol{x}(t_n)\}_{n\in \mathbb{N}}$ with $\boldsymbol{ x} (t) = \left(q_1 (t), q_2 (t), \ldots, q_N (t), p_1(t), p_2(t), \ldots, p_N (t) \right) $ and check the integrators' efficiency through their ability to correctly reproduce certain observables of the dynamics.
We also follow the evolution of a small initial perturbation to that trajectory  $\boldsymbol{ w}(t_0)= \delta \boldsymbol{x} (t_0) =\left(\delta q_1 (t_0), \delta q_2 (t_0), \ldots, \delta q_N (t_0), \delta p_1(t_0), \delta p_2(t_0), \ldots, \delta p_N (t_0) \right)$ and use it to compute the time evolution of the finite time mLE~\cite{benettin1980lyapunov2, benettin1980lyapunov1,skokos2010lyapunov}
\begin{equation}
X_1(t) = \frac{1}{t}\ln \left[ \frac{\|\boldsymbol{ w}( t_0 + \tau)\|}{\|\boldsymbol{ w}(t_0)\|} \right],
\label{eq:finitetimeLE}
\end{equation}
in order to characterize the regular or chaotic nature of the trajectory through the estimation of the most commonly used chaos indicator, the mLE $\chi$, which is defined as $\chi = \lim_{t\rightarrow +\infty} X_1(t)$.
In Eq.~\eqref{eq:finitetimeLE}  $\|\cdot \|$ is the  usual Euclidian norm, while $\boldsymbol{ w}(t_0)$ and $\boldsymbol{ w}(t_0 + \tau)$ are respectively the deviation vectors at $t=t_0$ and $t_0 + \tau> t_0$. In the case of regular  trajectories $X_1(t)$ tends to zero following the power law~\cite{skokos2010lyapunov, benettin1976g}
\begin{equation}
X_1(t) \propto t^{-1},
\label{eq:regularmle}
\end{equation}
whilst it takes positive values for chaotic ones.

%------------------------------------
\subsection{The $\alpha$-Fermi-Pasta-Ulam-Tsingou model}\label{sec:NR_FPUT}

We present here  results on the computational efficiency of the symplectic and non-symplectic schemes of Sec.~\ref{sec:integration_schemes} for the case of the  $\alpha$-FPUT chain [Eq.~\eqref{eq:alphaFPUT}].
As this system can be split into two integrable parts we will use for its study the two part split  SIs of Sec.~\ref{sec:2steps}. In our investigation we consider a lattice of $N=2^{10}$ sites with $\alpha=0.25$
and integrate up to the final time $t_f = 10^6$  two sets of initial conditions:
\begin{itemize}
\item Case $\text{I}_{\text{F}}$: We  excite all lattice sites by attributing to their position and momentum coordinates a  randomly chosen value from a uniform distribution in the interval
$[-1,1]$. These values are rescaled to achieve a particular energy density, namely  $H_{1F}/N=0.1$.
\item Case $\text{II}_{\text{F}}$: Same as in case $\text{I}_{\text{F}}$, but for $H_{1F}/N=0.05$.
\end{itemize}
We consider these two initial conditions in an attempt to investigate the potential dependence of the performance of the tested integrators on initial conditions~\cite[Sec. 8.3]{eggl2010introduction}. Since we have chosen non-localized initial conditions, we also use an initial normalized deviation vector $\boldsymbol{ w} (t)$  whose components are  randomly selected from a uniform distribution in the interval $[-1,1]$.

To evaluate the performance of each integrator  we investigate  how accurately it  follows the considered trajectories   by checking the numerical constancy of the energy integral of motion, i.e.~the value of $H_\text{1F}$ [Eq.~\eqref{eq:alphaFPUT}]. This is done by registering the time evolution of the  \textit{relative energy  error}
\begin{equation}
E_r (t) = \left| \frac{H_{\text{1F}} (t)- H_{\text{1F}} (0)}{ H_{\text{1F}} (0)} \right|,
 \label{eq:absoluterelenergyerror}
\end{equation}
at each time step.
In our analysis  we consider  two  energy error thresholds $E_r \approx 10^{-5}$ and $E_{r} \approx 10^{-9}$.
The former, $E_r \approx 10^{-5}$, is typically considered to be a good accuracy in many studies in the field of lattice dynamics, like for example in investigations of the  DKG and the  DDNLS models, as well as in systems of coupled rotors (see for example \cite{skokos2009delocalization,  skokos2010spreading,  laptyeva2010crossover,  bodyfelt2011wave,
bodyfelt2011nonlinear,  laptyeva2012subdiffusion}). In some cases, e.g.~for very small values of conserved quantities, one may desire more accurate computations. Then $E_r \approx 10^{-9}$ is a more appropriate accuracy level. In addition, in order to check  whether the variational equations are properly evolved, we compute the finite time mLE $X_1(t)$ [Eq.~\eqref{eq:finitetimeLE}].

\begin{figure}
\centering
\includegraphics[scale=0.85]{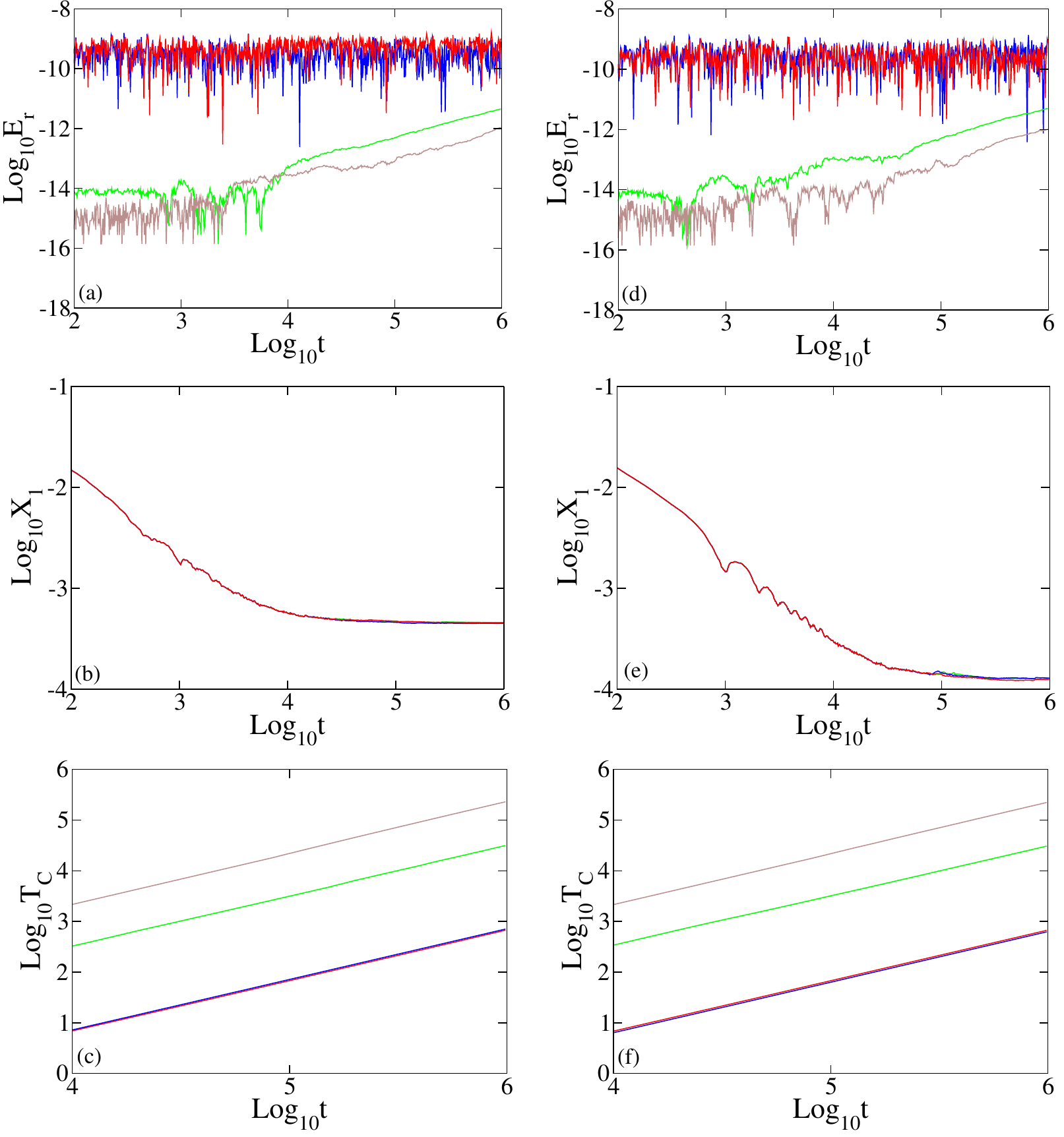}\caption{Results for the integration of the equations of motion and
  the variational equations of the $\alpha$-FPUT Hamiltonian
  [Eq.~\eqref{eq:alphaFPUT}] for cases (see text for details) $\text{I}_{\text{F}}$ [panels
  (a), (b) and (c)] and $\text{II}_{\text{F}}$ [panels (d), (e) and
  (f)] by the SIs $ABA864$ (blue curves) and $SABA_2Y6$ (red
  curves), and the non-symplectic schemes $DOP853$ (green curves) and $TIDES$ (brown curves): the
  time evolution of, (a) and (d) the  relative energy error
  $E_r(t)$ [Eq.~\eqref{eq:absoluterelenergyerror}], (b) and (e) the
  finite time mLE $X_1(t)$ [Eq.~\eqref{eq:finitetimeLE}], (c) and (f)
  the required CPU time $T_C$. All curves in panels (b) and (e),
  as well as the blue and red curves in panels (c) and (f) overlap.}
\label{fig:FPUT}
\end{figure}

In Fig.~\ref{fig:FPUT} we show the time evolution of the relative energy error $E_r(t)$ [panels (a) and (d)], the finite time mLE $X_1(t)$ [panels (b) and (e)], and the required CPU time $T_C$ [panels (c) and (f)],  for  cases $\text{I}_{\text{F}}$ and $\text{II}_{\text{F}}$ respectively, when the following four integrators were used:  the fourth order SI $ABA864$ (blue curves), the sixth order SI $SABA_2Y6$ (red curves),  the $DOP853$ scheme (green curves) and the   $TIDES$ package (brown curves). These results are indicative of our analysis as in our study we considered in total 37 different integrators (see Tables \ref{table:performance_FPUT_Case1} and  \ref{table:performance_FPUT_Case2}). In Fig.~\ref{fig:FPUT} the integration time steps $\tau$ of the SIs (reported in Tables \ref{table:performance_FPUT_Case1} and  \ref{table:performance_FPUT_Case2}) were appropriately chosen in order to achieve  $E_r \approx 10^{-9}$, while for the $DOP853$ algorithm and the  $TIDES$ package $E_r(t)$ eventually grows in time as a power law [Figs.~\ref{fig:FPUT}(a) and (d)].
Nevertheless, all schemes succeed in capturing correctly the chaotic nature of the dynamics as they do not present any noticeable difference in the computation of the finite time mLE $X_1$ in Figs.~\ref{fig:FPUT}(b) and (e). For both sets of initial conditions  $X_1$ eventually saturates to a constant positive value indicating that both trajectories are chaotic. The  CPU time $T_C$ needed for the integration of the equations of motion and the variational equations are reported in Figs.~\ref{fig:FPUT}(c) and (f). From these plots we see that the SIs need less computational time to perform the simulations than the $DOP853$ and $TIDES$ schemes.
%%%%%%%%%%%%%%%%%%%%%%%%%%%%%
%\begin{figure}[!hhtb]
%%%%%%%%%%%%%%%%%%%%%%%%%%%%%
%\begin{table}[!htbp]
\begin{table}
\centering
\caption{Information on the performance of the numerical schemes used for the integration of the equations of motion and the variational equations of the $\alpha$-FPUT system [Eq.~\eqref{eq:alphaFPUT}] up to the final time $t_f = 10^6$ for case $\text{I}_{\text{F}}$ (see text for details). The order $n$ and the number of steps of each SI, along with the integration time step $\tau$ used to reach a  relative energy error $E_r \approx 10^{-5}$ and $E_r \approx 10^{-9}$, as well as the required CPU time $T_C$ in seconds are reported. $\delta$ is the one-step precision of the non-symplectic schemes. Results are presented in increasing $T_C$ values.
See \cite{sim_details_fput} for practical information on the simulations.
}
  \begin{tabular}{lcclr|lcclr}
  \toprule
  \multicolumn{5}{c}{$E_r \approx 10^{-5}$} &
  \multicolumn{5}{c}{$E_r \approx 10^{-9}$} \\
    \toprule
    Integrator & $n$ & Steps & $\tau$ & $T_C$ & Integrator & $n$ & Steps & $\tau$ & $T_C$\\
  \midrule
  $ABA864$ &4 & 15   & $ 0.6$ & 88
  & $ SRKN_{14}^a$ &6 & 29   & $ 0.45$ & 160        \\
   $ABAH864$ &4  & 17 &  $  0.55$ & 115
  & $ SRKN_{11}^b$ &6 &23  & $0.35$ &  177       \\
  $SABA_2Y6$ &6 &  29 & $ 0.575 $  &167
  & $s11SABA_26$ &6 &45  & $0.3$ &  536          \\
 $ABA864Y6$ &6 & 43 &  $ 0.625$ &  202
   & $s19SABA_28$ &8 & 77 &  $ 0.45 $ & 594   \\
   $s9SABA_26$ &6 & 37 &  $ 0.575$ & 205
  &   $s9ABA82\_6$ &6 & 73 &  $ 0.35$ &   607      \\
   $FR4$ &4 &7 &  $0.14$  & 228
 & $s15SABA_28$ &8 & 61  &  $ 0.35 $ & 611
    \\
   $SBAB_2Y6$ &6 & 29  &  $0.5 $  &  233
   &  $SABA_2Y6$ &6 &  29 &  $ 0.14 $& 683
      \\
    $SABA_2K$ &4 & 9 & $0.3 $ & 234
   &  $ABA864$ &4 & 15   &   $ 0.08$ & 717
       \\
    $ABA82Y4$ &4 &25&  $ 0.375$ & 240
    & $s19ABA82\_8$ &8 &  153 &   $0.65$ & 773
     \\
   $s11SABA_26$ &6 &45  & $ 0.65$  & 247
    & $s9SABA_26$ &6 & 37 &  $ 0.16$ &  779
        \\
    $SABA_2Y4$ &4 &13 & $0.18$  & 265
    &  $ABA864Y6$ &6 & 43 &   $ 0.16$ & 791
       \\
  $ABA82$ & 2 & 5  & $ 0.125$  &  278
   & $s15ABA82\_8$ &8 & 121 & $ 0.475$ &  838
      \\
  $ABA82Y6$  &6 & 57 & $ 0.675$  &  283
     & $ABA82Y6$  &6 & 57 &  $ 0.2$ & 841
        \\
   $s15SABA_28$ &8 & 61  & $  0.65$ & 339
  &  $s11ABA82\_6$ &6 & 89 &    $ 0.275$   &  941
       \\
   $SABA_2$ & 2 & 5  & $ 0.07$ & 347
  &  $SBAB_2Y6$ &6 & 29  &   $0.12$& 965
     \\
  $s19SABA_28$ &8 & 77 &  $ 0.775$ & 356
   & $ABAH864$ &4  & 17 &    $0.055 $ & 1013
        \\
    $SBAB_2Y4$ &4 & 13 &  $0.18$ & 358
    &  $SABA_2Y8\_D$ &8 &61&   $ 0.175$ &  1223
       \\
     $FR4Y6$ &6 & 19 &  $ 0.21$ & 366
  &  $ABA82Y8\_D$ &8 & 121 &   $ 0.25$& 1575
       \\
   $s9ABA82\_6$ &6 & 73 &  $ 0.575$ &  369
    &    $SABA_2Y4Y6$ &6 & 37&   $0.07 $ &  1701
        \\
   $s11ABA82\_6$ &6 & 89 & $ 0.675$ & 382
     &   $FR4Y6$ &6 & 19 & $0.45$ &   1787
         \\
    $SBAB_2$ & 2  & 5 & $ 0.07$  & 387
    & $ABA82Y4Y6$ &6 & 73 &  $ 0.125 $ & 1932
       \\
  $SABA_2Y4Y6$ &6 & 37&  $0.3$  &  394
    & $ABA82Y4$ &4 &25& $0.0375$ &  2156
      \\
   $ABA82Y4Y6$ &6 & 73 &   $ 0.525$  & 405
     &  $SBAB_2Y4Y6$ &6 & 37 &  $ 0.065$ & 2239
         \\
    $SABA_2Y8\_D$ &8 &61& $ 0.525$ &  408
  & $SABA_2K$ &4 & 9 &  $ 0.03$ &  2344
     \\
   $SBAB_2K$ &4 & 9& $ 0.2$ & 416
   &  $SABA_2KY6$ &6 & $19$ &  $ 0.09 $ & 2465
         \\
  $s19ABA82\_8$ &8 &  153 & $ 1.15$ & 439
    &  $FR4$ &4 &7 &   $0.01$  & 2597
        \\
   $SABA_2KY6$&6 & $19$ &  0.4  &  535
   &  $SABA_2Y4$ &4 &13 &  $0.018$ & 2654
         \\
   $SBAB_2Y4Y6$ &6 & 37 &  $ 0.275$ & 553
    & $SBAB_2Y4$ &4 & 13 &  $0.018$& 3156
      \\
   $s15ABA82\_8$ &8 & 121 &  $ 0.775 $ & 618
   & $SABA_2Y8\_A$ &8 & 61 & $ 0.06 $ &  3570
      \\
  $ABA82Y8\_D$ &8 & 121 & $ 0.6$  & 656
   &  $SBAB_2K$ &4 & 9& $0.02$ & 4167
       \\
  $SABA_2Y8\_A$ &8 & 61 &  $ 0.225 $ & 1090
    & $ABA82Y8\_A$ &8 & 121 &  $ 0.07$ & 5624
       \\
   $LF$ & 2  & 3 &   $0.018 $   & 1198
  & $ABA82$ & 2 & 5  & $0.00125$ & 27796
     \\
   $ABA82Y8\_A$ &8 & 121 & $ 0.225$ &  1749
  &  $DOP853$ & 8 & $\delta=10^{-16}$     &0.05  &31409
         \\
 & & & &
   & $SABA_2$ & 2 & 5  & $0.0007 $ & 34595
        \\
 & & & &
     & $SBAB_2$ & 2  & 5 & $ 0.0007  $   & 39004
           \\
 & & & & &  $LF$ & 2  & 3 & $ 0.0002 $   & 95096    \\
 & & & & &  $TIDES$ & - & $\delta=10^{-16}$      & 0.05 & $232785$   \\
    \bottomrule
  \end{tabular}
  \label{table:performance_FPUT_Case1}
\end{table}
%%%%%%%%%%%%%%%%%%%%%%%%%%%%%

%%%%%%%%%%%%%%%%%%%%%%%%%%%%%
%\begin{table}[!htbp]
\begin{table}
\centering
\caption{
Similar to Table \ref{table:performance_FPUT_Case1} but for  case $\text{II}_{\text{F}}$ (see text for details) of the $\alpha$-FPUT  system of Eq.~\eqref{eq:alphaFPUT}.
See \cite{sim_details_fput} for practical information on the simulations}.
  \begin{tabular}{lcclr|lcclr}
  \toprule
  \multicolumn{5}{c}{$E_r \approx 10^{-5}$} &
  \multicolumn{5}{c}{$E_r \approx 10^{-9}$} \\
    \toprule
    Integrator & $n$ & Steps & $\tau$ & $T_C$ & Integrator & $n$ & Steps & $\tau$ & $T_C$\\
  \midrule
    \midrule
$ABA864$ &4 & 15   & $ 0.6$    & 77
    & $ SRKN_{14}^a$ &6 & 29   & $ 0.475$ & 156       \\
$ABAH864$ &4  & 17 &  $  0.55$   &  101
  & $ SRKN_{11}^b$ &6 &23  & $0.35$ &  179       \\
 $SABA_2Y6$ &6 &  29 & $ 0.575 $    &  183
&$s11SABA_26$ &6 &45  &  $0.3$    &  480   \\
    $ABA864Y6$ &6 & 43 &  $ 0.625$  & 195
&$s19SABA_28$ &8 & 77 &  $ 0.45 $  & 596\\
$s9SABA_26$ &6 & 37 &  $ 0.575$ & 214
 &  $s9ABA82\_6$ &6 & 73 &   $ 0.35$  &  611    \\
 $ABA82Y4$ &4 &25&  $ 0.375$    & 224
  & $ABA864$ &4 & 15   &   $ 0.08$  &   613   \\
$s11SABA_26$ &6 &45  & $ 0.65$   & 227
& $s15SABA_28$ &8 & 61  &  $ 0.35 $  & 618  \\
$FR4$ &4 &7 &  $0.14$    & 231
   &  $SABA_2Y6$ &6 &  29 &  $ 0.14 $  &  676    \\
 $SABA_2K$ &4 & 9 & $0.3 $  &  241
& $s19ABA82\_8$ &8 &  153 &$0.65$   & 760 \\
$SBAB_2Y6$ &6 & 29  &  $0.5 $    &  255
  & $ABA864Y6$ &6 & 43 &   $ 0.16$  & 811    \\
$SABA_2Y4$ &4 &13 & $0.18$    & 270
&$s15ABA82\_8$ &8 & 121 & $ 0.475$  & 828 \\
 $ABA82$ & 2 & 5  & $ 0.125$  &   280
  &$s9SABA_26$ &6 & 37 &   $ 0.16$  &  838    \\
$ABA82Y6$  &6 & 57 & $ 0.675$   & 285
&$ABA82Y6$  &6 & 57 &   $ 0.2$  &  937 \\
 $SBAB_2Y4$ &4 & 13 &  $0.18$  &  316
   & $SBAB_2Y6$ &6 & 29  &   $0.12$ &  964   \\
$s15SABA_28$ &8 & 61  & $  0.65$ & 329
  & $ABAH864$ &4  & 17 &    $0.055 $  &   1023   \\
$s19SABA_28$ &8 & 77 &  $ 0.775$   & 336
  &$s11ABA82\_6$ &6 & 89 &   $ 0.275$     &  1062   \\
  $FR4Y6$ &6 & 19 &  $ 0.21$   & 337
&$SABA_2Y8\_D$ &8 &61& $ 0.175$  & 1230 \\
 $SBAB_2K$ &4 & 9& $ 0.2$   & 366
   & $FR4Y6$ &6 & 19 & $0.45$  & 1577    \\
$s9ABA82\_6$ &6 & 73 &  $ 0.575$  & 373
&$ABA82Y8\_D$ &8 & 121 &   $ 0.25$ &  1613 \\
$SABA_2$ & 2 & 5  & $ 0.07$ &  392
& $ABA82Y4Y6$ &6 & 73 &   $ 0.125 $  &  1702     \\
$SBAB_2$ & 2  & 5 & $ 0.07$  &  393
&   $ABA82Y4$ &4 &25&   $0.0375$  &  2113  \\
$ABA82Y4Y6$ &6 & 73 &   $ 0.525$    & 398
&  $SABA_2Y4Y6$ &6 & 37& $0.07 $ &  2159   \\
$SABA_2Y8\_D$ &8 &61& $ 0.525$   & 407
&   $SBAB_2Y4Y6$ &6 & 37 & $ 0.065$  &  2310\\
$s11ABA82\_6$ &6 & 89 & $ 0.675$   &  415
  & $SABA_2K Y6$ &6 & 19 &  $ 0.09 $ &  2380   \\
$s19ABA82\_8$ &8 &  153 & $ 1.15$   & 431
   &     $FR4$ &4 &7 &   $0.01$   &  2615 \\
  $SABA_2Y4Y6$ &6 & 37&  $0.3$   &  444
  & $SABA_2K$ &4 & 9 &  $ 0.03$   &   2651    \\
  $SBAB_2Y4Y6$ &6 & 37 &  $ 0.275$   & 533
&   $SABA_2Y4$ &4 &13 & $0.018$  &  3016 \\
$SABA_2K Y6$ &6 & 19 &  0.4  &  540
  &   $SBAB_2Y4$ &4 & 13 & $0.018$ &  3585   \\
$s15ABA82\_8$ &8 & 121 &  $ 0.775 $  & 598
 & $SABA_2Y8\_A$ &8 & 61 & $ 0.06 $  & 3647 \\
$ABA82Y8\_D$ &8 & 121 & $ 0.6$    & 629
 &     $SBAB_2K$ &4 & 9& $0.02$  & 3663     \\
$SABA_2Y8\_A$ &8 & 61 &  $ 0.225 $  & 952
& $ABA82Y8\_A$ &8 & 121 &  $ 0.07$ & 5691  \\
$LF$ & 2  & 3 &   $0.018 $   &   1059
&   $ABA82$ & 2 & 5  &  $0.00125$ & 31576      \\
$ABA82Y8\_A$ &8 & 121 & $ 0.225$    & 1787
& $DOP853$ & 8 & $\delta=10^{-16}$    &0.05  & 31709 \\
 & & & &
 &  $SABA_2$ & 2 & 5  & $0.0007 $  & 39333 \\
 & & & &
 &  $SBAB_2$ & 2  & 5 & $ 0.0007  $      &  45690      \\
 & & & &  & $LF$ & 2  & 3 &  $ 0.0002 $     & 106653     \\
 & & & & & $TIDES$ & - & $\delta=10^{-16}$    & 0.05  &  $225565$\\
    \bottomrule
  \end{tabular}
  \label{table:performance_FPUT_Case2}
\end{table}
%%%%%%%%%%%%%%%%%%%%%%%%%%%%%

In  Table \ref{table:performance_FPUT_Case1} (Table \ref{table:performance_FPUT_Case2}) we present information on the performance of all considered integration schemes for the initial condition of case $\text{I}_{\text{F}}$ (case $\text{II}_{\text{F}}$).
From the results of these tables we see that the performance and ranking (according to $T_C$) of the integrators do not practically depend on the considered initial condition. It is worth noting that although the non-symplectic schemes manage to achieve better accuracies than the symplectic ones, as their $E_r$ values are smaller [Figs.~\ref{fig:FPUT}(a) and (d)], their implementation is not recommended for the long time evolution of the Hamiltonian system, because they require more CPU time and eventually their $E_r$ values will increase above the bounded $E_r$ values obtained by the symplectic schemes.

From the results of Tables \ref{table:performance_FPUT_Case1} and \ref{table:performance_FPUT_Case2} we see that the best performing integrators are the fourth order SIs  $ABA864$ and $ABAH864$ for  $E_r\approx 10^{-5}$, and
the sixth   order SIs $SRKN_{14}^a$  and $SRKN_{11}^b$  for  $E_r \approx 10^{-9}$.
We note that the best  SI for $E_r\approx 10^{-5}$, the $ABA864$ scheme, shows a quite good behavior also for $E_r \approx 10^{-9}$, making this integrator a  valuable numerical tool for dynamical studies of multidimensional Hamiltonian systems. We remark  that the eighth order SIs we implemented to achieve the moderate accuracy level  $E_r\approx 10^{-5}$ exhibited an unstable behavior failing to keep their $E_r$ values bounded.
A similar behavior was also observed for the two RKN schemes $SRKN_{14}^a$, $SRKN_{11}^b$ when they were used to obtain  $E_r\approx 10^{-5}$. Thus, the  higher order SIs are best suited for more accurate computations. It is also worth mentioning here that the ranking presented in Tables \ref{table:performance_FPUT_Case1} and \ref{table:performance_FPUT_Case2} is indicative of the performance of the various SIs in the sense that small changes in the implementation (e.g.~a change in the last digit of the used $\tau$ value) of integrators with similar behaviors (i.e.~similar $T_C$ values) could interchange their ranking positions without any noticeable difference in the produced results.

%-----------------------------------------------------
\subsection{The 1D  disordered discrete nonlinear Schr\"{o}dinger equation system}\label{sec:DNLS_1D}

We investigate the performance of various integrators of Sec.~\ref{sec:integration_schemes} for the 1D DDNLS system [{Eq.~\eqref{eq:DNLS_1D_real}] by considering a lattice of $N=2^{10}$ sites and integrating two sets of initial conditions (for the same reason we did that for the $\alpha$-FPUT system) up to the final time $t_f = 10^6$. We note that, as was already mentioned in Sec.~\ref{sec:3steps}, this model can be split into  three integrable parts, so we will implement the SIs presented in that section. In particular, we consider the following two cases of initial conditions:
\begin{itemize}
\item Case $\text{I}_{\text{1D}}$: We initially excite $21$ central sites by attributing to each one of them  the same constant norm $s_j = (q_j^2 + p_j^2)/2=1$, $1\leq i \leq N$, for $W=3.5$ and $\beta = 0.62$. This choice sets the total norm
$S_{\text{1D}}=21$. The random disorder parameters $\epsilon _i$, $1\leq i \leq N$, are chosen so that the total energy is $H_{1D} \approx 0.0212$.
\item Case $\text{II}_{\text{1D}}$: Similar set of initial conditions as in  case $\text{I}_{\text{1D}}$ but for $W=3$, $\beta = 0.03$. The random disorder parameters $\epsilon _i$, $1\leq i \leq N$, are chosen such that $H_{\text{1D}}\approx 3.4444$.
\end{itemize}
We note that  cases $\text{I}_{\text{1D}}$ and $\text{II}_{\text{1D}}$ have been studied in~\cite{senyange2018characteristics} and  respectively correspond  to the so-called `strong chaos' and `weak chaos' dynamical regimes of this model.
As initial normalized deviation vector we use  a vector having non-zero coordinates only at the  central site of the lattice,  while its remaining  elements  are set to  zero.

To evaluate the performance of each implemented integrator we check if the obtained  trajectory correctly captures the statistical behavior of the normalized norm density distribution $\zeta _j =s_j/S_{\text{1D}}$, $1\leq j\leq N$, by computing the distribution's  second moment
\begin{equation}
 m_2 = \sum_{j=1}^N  \left( j - \overline{j} \right)^2 \zeta _j,
\label{eq:1D_DNLS_obs}
\end{equation}
where $\overline{j} = \sum_{j=1}^N j \zeta _j$ is the position of the center of the distribution \cite{flach2009universal, skokos2009delocalization,laptyeva2010crossover,  bodyfelt2011nonlinear,senyange2018characteristics,  skokos2014high, gerlach2016symplectic}.
We also check how accurately the values of the system's two conserved quantities, i.e.~its total energy  $H_{\text{1D}}$ [Eq.~\eqref{eq:DNLS_1D_real}] and norm $S_{\text{1D}}$ [Eq.~\eqref{eq:normDDNLS}], are kept constant throughout the integration by  evaluating the  relative energy error $E_r (t)$ [similarly to Eq.~\eqref{eq:absoluterelenergyerror}] and the  \textit{relative norm error}
 \begin{equation}
 S_r (t) = \left| \frac{S_{\text{1D}} (t)- S_{\text{1D}} (0)}{ S_{\text{1D}} (0)} \right|.
 \label{eq:absoluterelnormerror}
 \end{equation}
In addition, we compute the finite time mLE $X_1(t)$ [Eq.~\eqref{eq:finitetimeLE}] in order to characterize the system's chaoticity and check the proper integration of the variational equations.

We consider  several three part split SIs, which we divide into two groups: \textit{(i)} those integrators with order $n \leq 6$, which we implement in order to achieve  an accuracy  of $E_r \approx 10^{-5}$, and \textit{(ii)} SIs of order eight used for $E_r \approx 10^{-9}$. In addition, the two best performing integrators of the first group are also included in the  second group.
We do not use higher order SIs for obtaining the accuracy level of $E_r \approx 10^{-5}$ because, as was shown in \cite{skokos2014high} and also discussed in Sec.~\ref{sec:NR_FPUT}, usually this task   requires large integration time steps, which typically make the integrators unstable.
Moreover, increasing the order $n$ of SIs beyond eight does not improve significantly the performance of the symplectic schemes for  high precision ($E_r\approx 10^{-9}$) simulations \cite{skokos2014high}. Therefore we do not consider such integrators in our study.

In Fig.~\ref{fig:figE9S19S11DOPTIDES} we show the time evolution of the relative energy error $E_r(t)$ [panel (a)], the  relative norm error $S_r(t)$ [panel (b)], the second moment $m_2(t)$ [panel (d)], as well as the norm density distribution $\zeta_j$ at time $t_f\approx 10^6$ [panel (c)] for  case $\text{I}_{\text{1D}}$  (we note that analogous results were also obtained for case $\text{II}_{\text{1D}}$, although we do not report them here). These results are obtained by the implementation of the
the second order SI $\mathcal{ABC}2$ (red curves), the  fourth order SI $\mathcal{ABC}Y4$ (blue curves), and the non-symplectic schemes $DOP853$ (green curves) and $TIDES$ (brown curves).
The results of Fig.~\ref{fig:figE9S19S11DOPTIDES} are indicative of the results obtained by the integrators listed in Table~\ref{table:performance1DDNLSstrong}. The integration time step $\tau$ of the SIs was selected so that the relative energy error is kept  at $E_r \approx 10^{-5}$ [Fig~\ref{fig:figE9S19S11DOPTIDES}(a)].
From the results of Fig.~\ref{fig:figE9S19S11DOPTIDES}(b) we see that the SIs do not keep  $S_r$ constant.
Nevertheless, our results show that we lose no more than two orders  of precision (in the worst case of the  $\mathcal{ABC}2$ scheme) during the whole integration. On the other hand,  both the  relative energy  [$E_r(t)$] and  norm [$S_r(t)$] errors of the
$TIDES$ and $DOP853$ integrators increase in time, with the  $TIDES$ scheme behaving better than the  $DOP853$ one.
Figs.~\ref{fig:figE9S19S11DOPTIDES}(c) and (d) show that all integrators correctly reproduce the dynamics of the system, as all of them practically produce the same norm density distribution  at  $t_f = 10^6$ [Fig.~\ref{fig:figE9S19S11DOPTIDES}(c)] and the same evolution of the $m_2(t)$ [Fig.~\ref{fig:figE9S19S11DOPTIDES}(d)]. We note that $m_2$ increases by following a power law $m_2 \propto t^\alpha$ with $\alpha = 1/2$, as was expected for the strong chaos dynamical regime (see for example \cite{senyange2018characteristics} and references therein).
%############# FIGURE (1) ################
\begin{figure}[!hhtb]
%\begin{figure}
\centering
\includegraphics[scale=0.55]{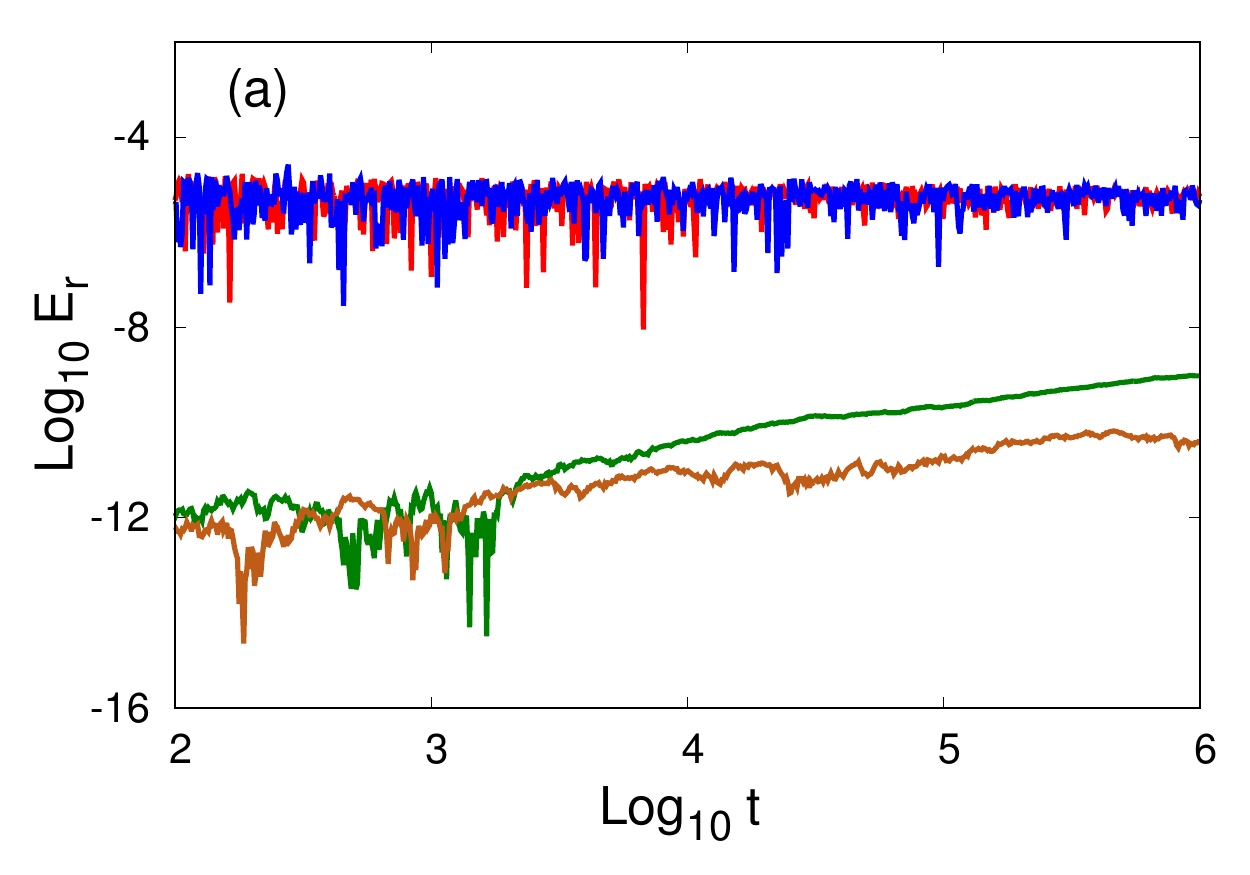}
\includegraphics[scale=0.55]{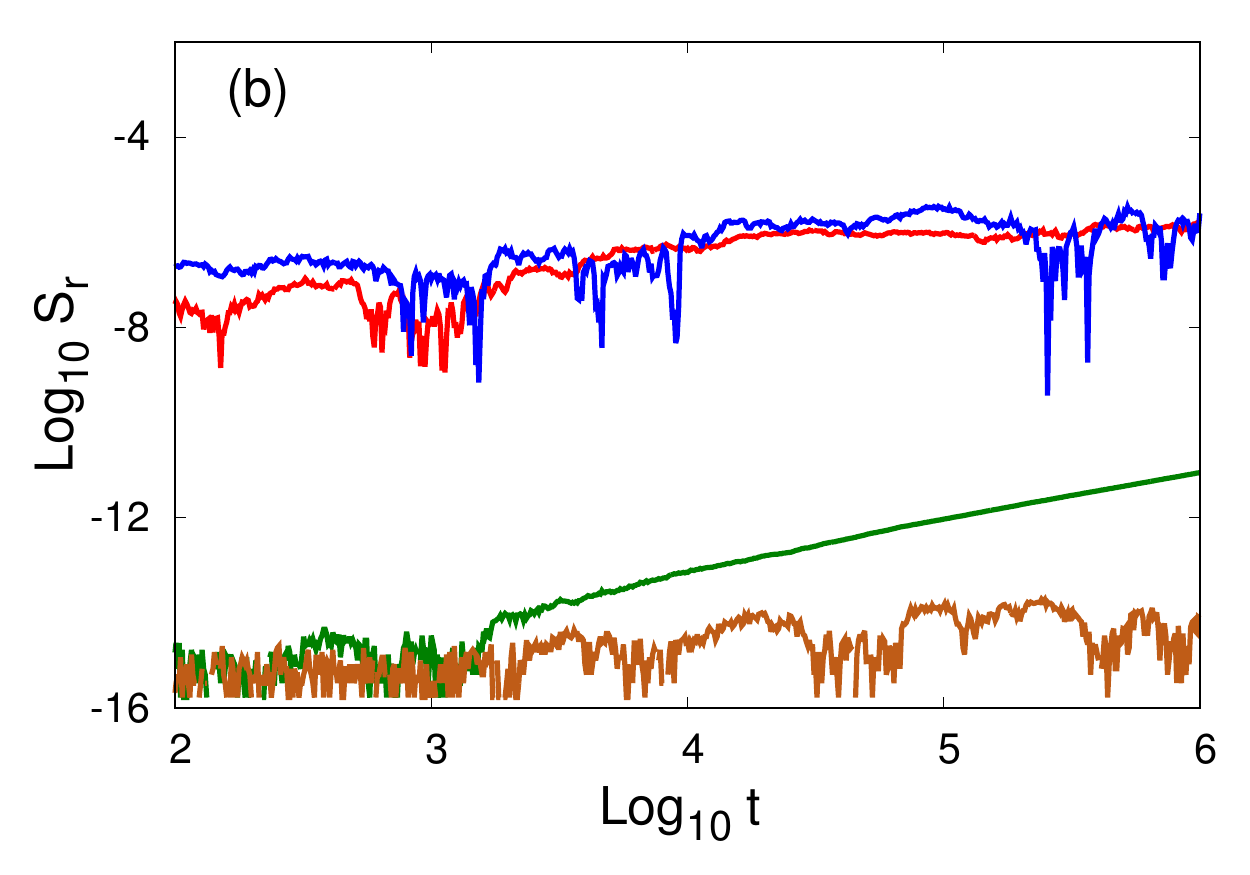}
\includegraphics[scale=0.55]{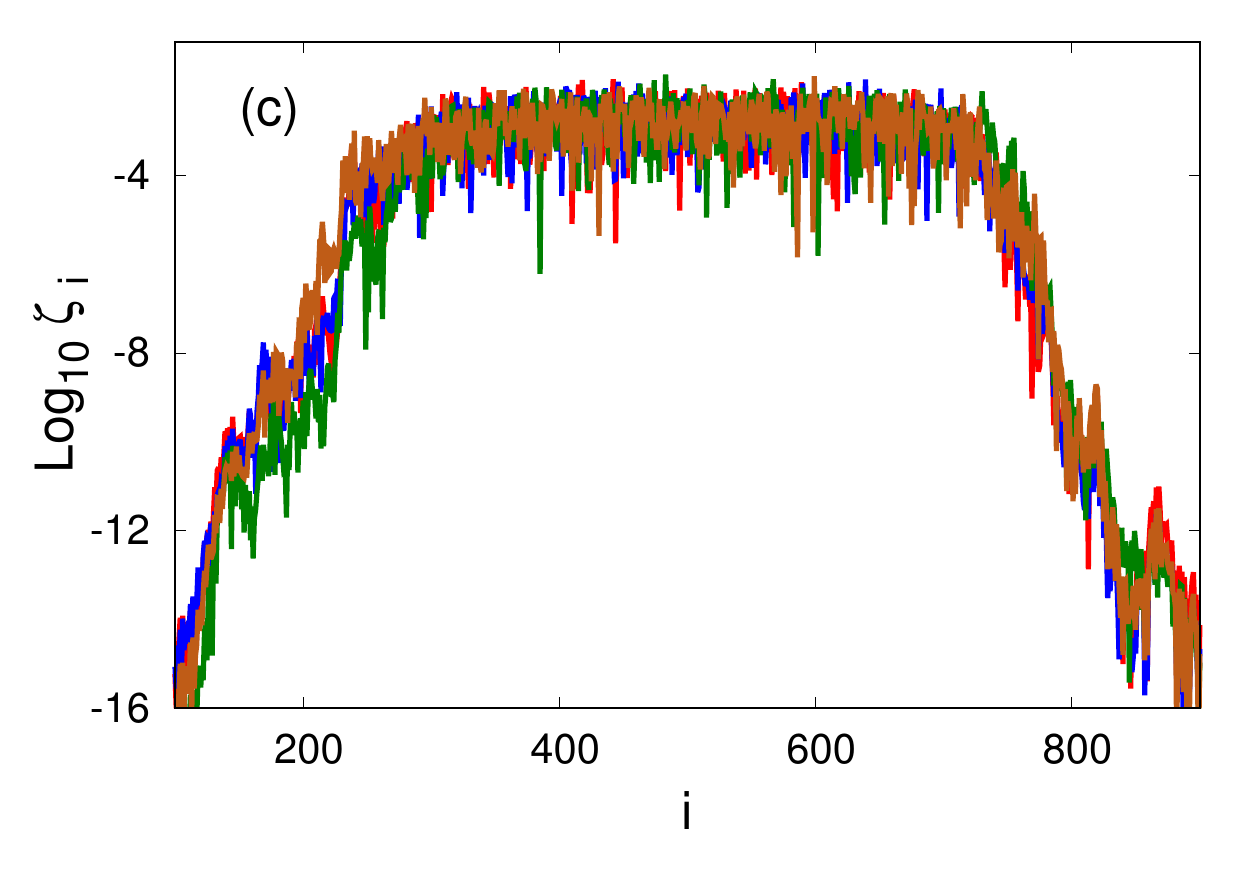}
\includegraphics[scale=0.55]{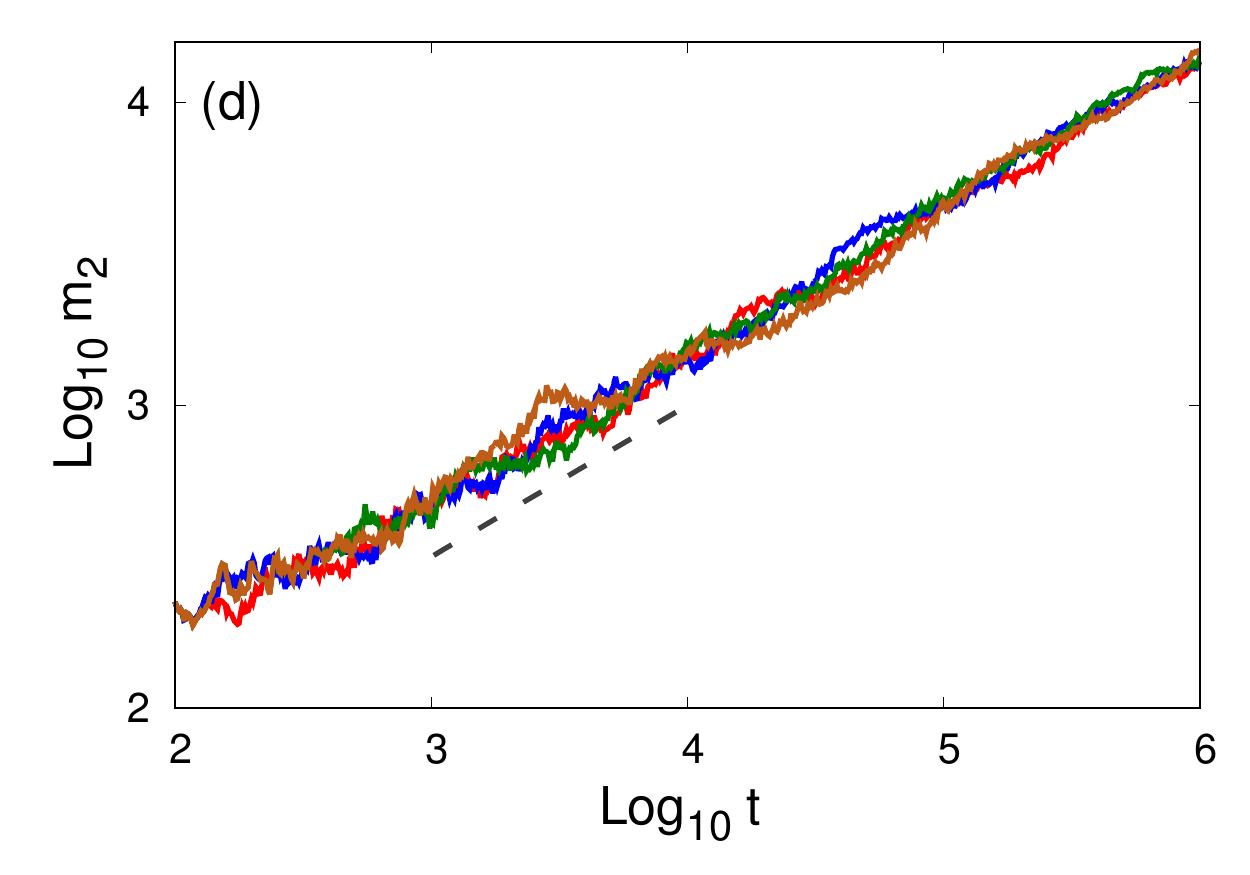}
\caption{Results for the integration of case $\text{I}_{\text{1D}}$
  (see text for details) of the 1D DDNLS model
  [Eq.~\eqref{eq:DNLS_1D_real}] by the second order SI
  $\mathcal{ABC}2$ for $\tau = 0.0002$ (red curves), the fourth order
  SI $\mathcal{ABC}Y4$ for $\tau = 0.0125$ (blue curves) and the
  non-symplectic schemes $DOP853$ (green curves) and $TIDES$ (brown
  curves): time evolution of (a) the  relative energy error
  $E_r(t)$, (b) the  relative norm error $S_r(t)$ and (d) the
  second moment $m_2(t)$. In (c) the norm density distribution at time $t_f
  =10^{6}$ is shown. The dashed line in (d) guides the eye for slope
  $1/2$.  }
\label{fig:figE9S19S11DOPTIDES}
\end{figure}
%#######################################

%########### TABLE (1) ################
%\begin{table}[!hhhtb]
\begin{table}
\centering
\caption{Data similar to the ones presented in Tables \ref{table:performance_FPUT_Case1} and \ref{table:performance_FPUT_Case2} but for the performance of the numerical schemes  used for the integration of the equations of motion and the variational equations of the 1D DDNLS model [Eq.~\eqref{eq:DNLS_1D_real}] up to the final time $t_f = 10^6$ for case $\text{I}_{\text{1D}}$ (see text for details).
See \cite{sim_details_dnls} for practical information on the simulations.
}
	\begin{tabular}{lcclr|lcclr}
		\toprule
		\multicolumn{5}{c}{$E_r \approx 10^{-5}$} &
		\multicolumn{5}{c}{$E_r \approx 10^{-9}$} \\
		\toprule
		Integrator           & $n$ & Steps & $\tau$    & $T_{\text{C}}$ & Integrator           & $n$ & Steps               & $\tau$  & $T_{\text{C}}$ \\
		\midrule
		$s11\mathcal{ABC}6$  & $6$ & $45$  & $0.115$   & $3395$         & $s19\mathcal{ABC}8$  & $8$ & $77$                & $0.09$  & $7242$         \\
		$s9\mathcal{ABC}6$   & $6$ & $37$  & $0.095$   & $3425$         & $s17\mathcal{ABC}8$  & $8$ & $69$                & $0.08$  & $7301$         \\
		$\mathcal{ABC}Y6\_A$ & $6$ & $29$  & $0.07$    & $3720$         & $s11\mathcal{ABC}6$  & $6$ & $45$                & $0.025$ & $15692$        \\
		 $SS864S$             & $4$ & $17$  &  $0.05$    & $6432$         & $s9\mathcal{ABC}6$   & $6$ & $37$                & $0.02$  & $16098$        \\
		$\mathcal{ABC}Y4$    & $4$ & $13$  & $0.0125$  & $10317$        & $DOP853$             & $8$ & $\delta=10^{-16}$   & $0.05$  & $18408$        \\
		$\mathcal{ABC}S4Y6$  & $6$ & $49$  & $0.015$   & $35417$        & $\mathcal{ABC}Y8\_D$ & $8$ & $61$                & $0.002$ & $258891$       \\
		$\mathcal{ABC}Y4Y6$  & $6$ & $37$  & $0.008$   & $40109$        & $TIDES$              & $-$ & $\delta = 10^{-16}$ & $0.05$  & $419958$       \\
		$\mathcal{ABC}S4$    & $4$ & $21$  & $0.00085$ & $267911$       &                      &     &                     &         &                \\
		$\mathcal{ABC}2$     & $2$ & $5$   & $0.0002$  & $320581$       &                      &     &                     &         &                \\
		\bottomrule
	\end{tabular}
  \label{table:performance1DDNLSstrong}
\end{table}
%#######################################

In Fig.~\ref{fig:figE5S19S11DOPTIDES} we show the evolution of the finite time mLE $X_1$ [panels (a) and (c)] and the required CPU time  $T_C$ [panels (b) and (d)] for the integration of the Hamilton equations of motion and the variational equations for  cases $\text{I}_{\text{1D}}$ [panels (a) and (b)] and $\text{II}_{\text{1D}}$ [panels (c) and (d)] obtained by using  the same  integrators of Fig.~\ref{fig:figE9S19S11DOPTIDES}. Again the results obtained by these integrators are  practically the same for both sets of initial conditions, reproducing the tendency of the finite time mLE to asymptotically decrease according to the power law $X_1(t) \propto t^{\alpha_L}$ with
$\alpha _L \approx -0.3$ (case $\text{I}_{1D}$) and $\alpha _L \approx -0.25$ (case $\text{II}_{1D}$), in accordance to the results  of \cite{ skokos2013nonequilibrium, senyange2018characteristics}.
%############# FIGURE (2) ################
\begin{figure}[!hhtb]
\centering
\includegraphics[scale=0.55]{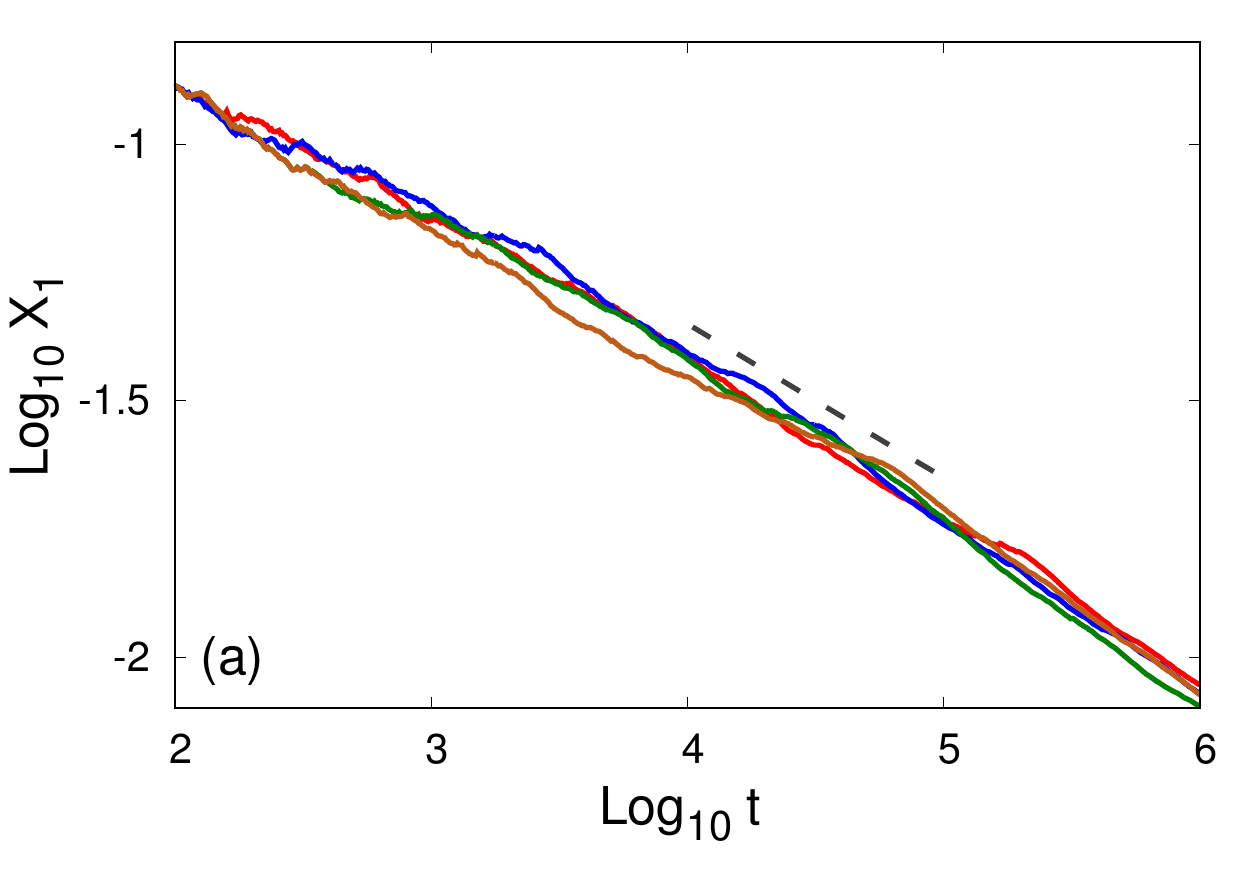}
\includegraphics[scale=0.55]{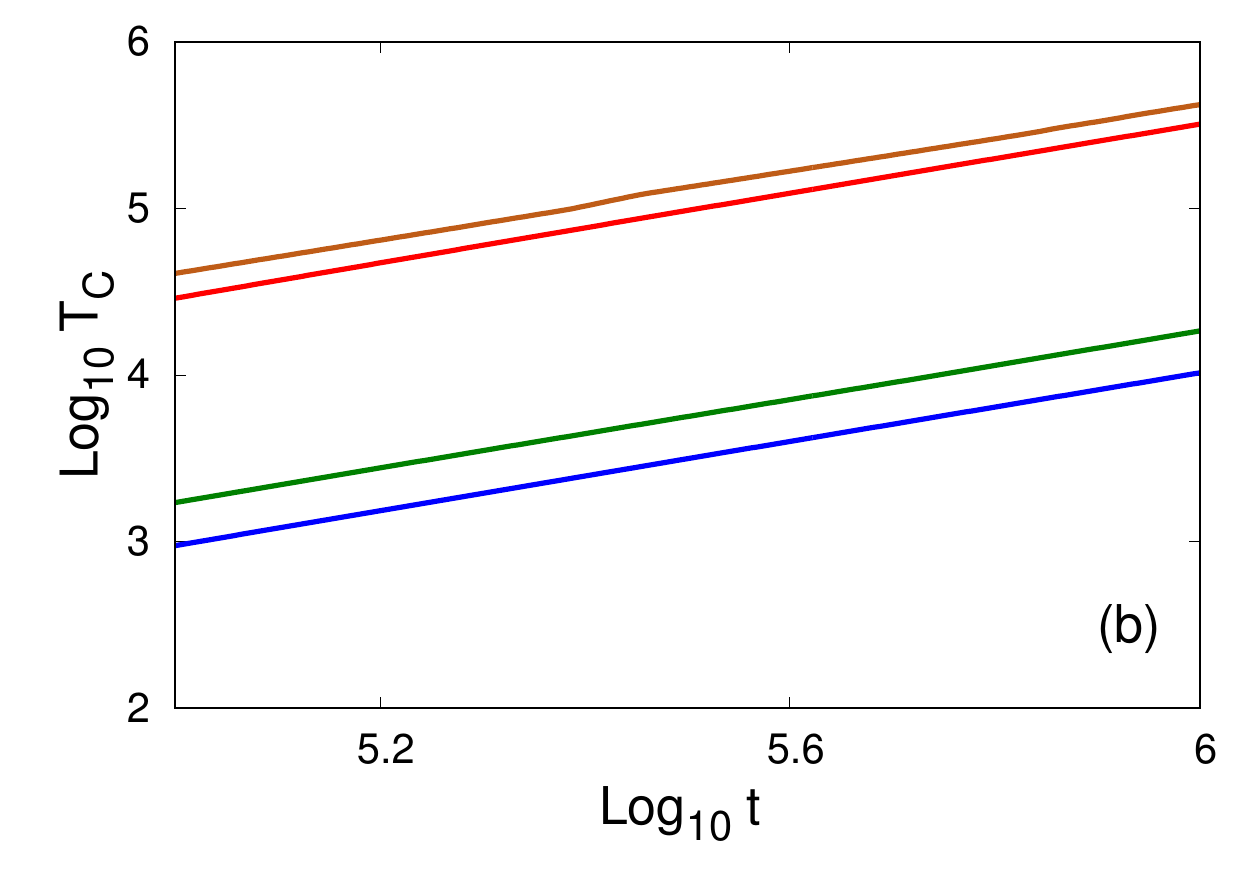}
\includegraphics[scale=0.55]{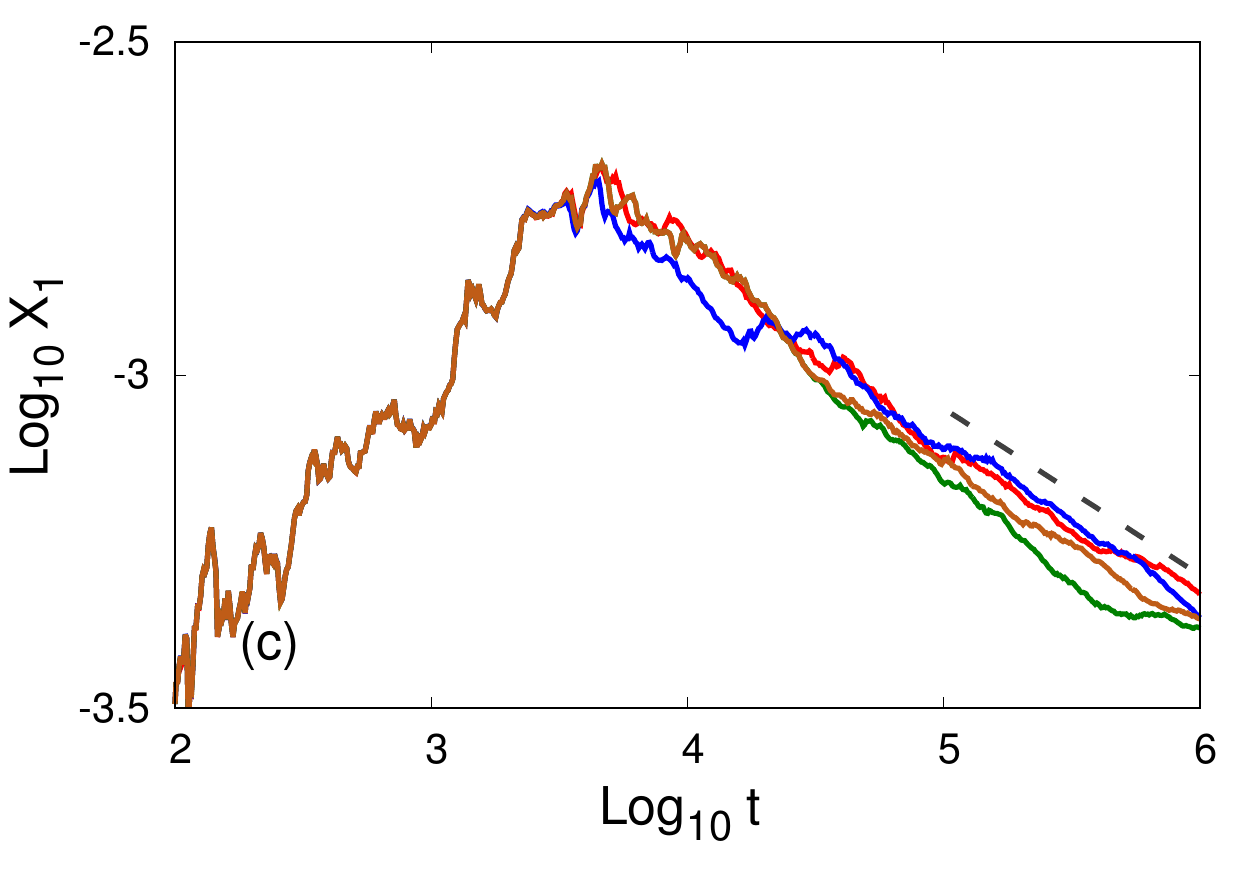}
\includegraphics[scale=0.55]{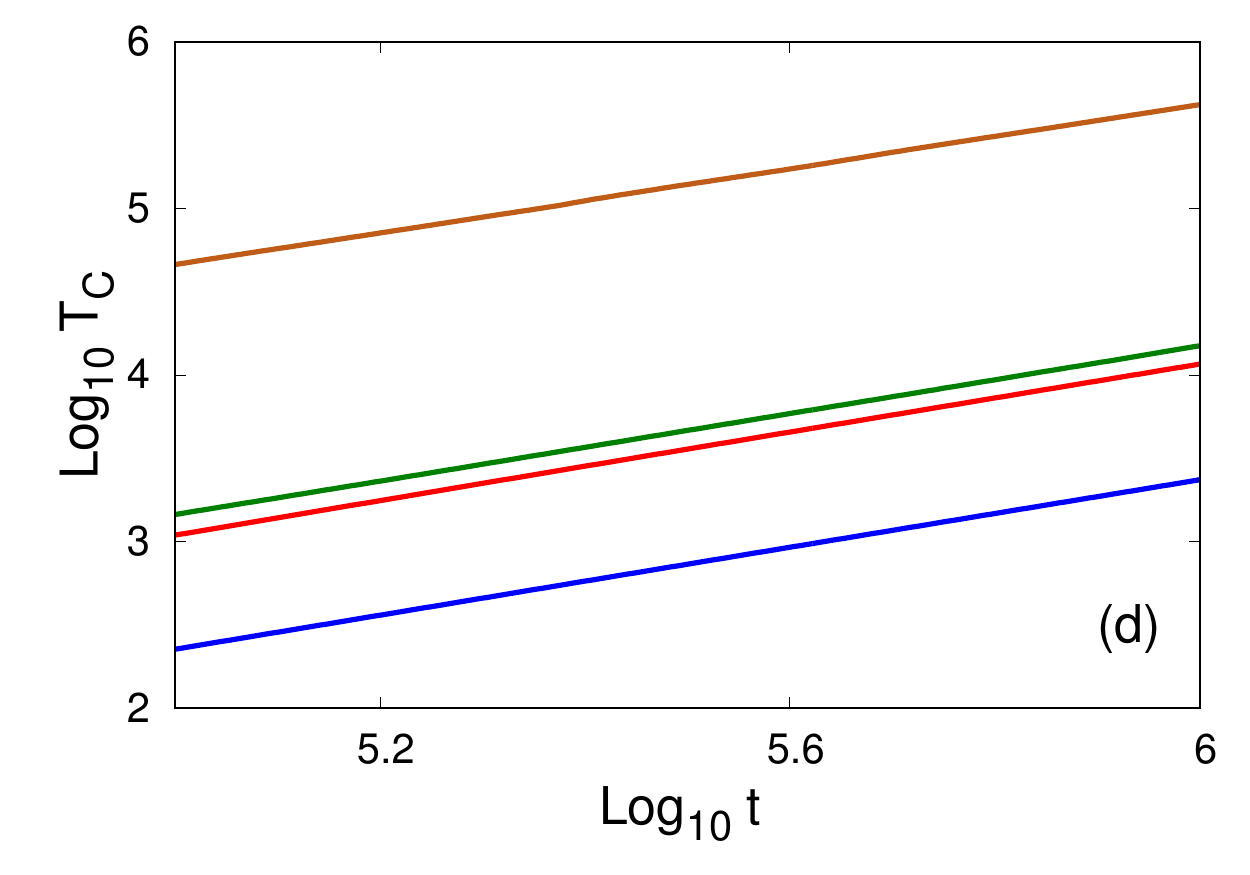}
\caption{Results obtained by the integration of the variational
  equations of the 1D DDNLS Hamiltonian [Eq.~\eqref{eq:DNLS_1D_real}]
  for the initial conditions described in cases (see text for details)
  $\text{I}_{\text{1D}}$ [panels (a) and (b)] and
  $\text{II}_{\text{1D}}$ [panels (c) and (d)]: time evolution of, (a)
  and (c) the finite time mLE $X_1(t)$ [Eq.~\eqref{eq:finitetimeLE}],
  and (b) and (d) the required CPU time $T_C$ in seconds. The dashed
  lines in (a) and (c) guide the eye for slopes $-0.3$ and $-0.25$
  respectively. The integrators and the curve colors are the ones used
  in Fig.~\ref{fig:figE9S19S11DOPTIDES}.  }
\label{fig:figE5S19S11DOPTIDES}
\end{figure}
%#######################################

We now check the efficiency of  the used symplectic and non-symplectic methods by comparing the CPU time $T_C$ they require to carry out the simulations. These results  are
reported in Table \ref{table:performance1DDNLSstrong} for case $\text{I}_{\text{1D}}$ and in Table \ref{table:performance1DDNLSweak} for case $\text{II}_{\text{1D}}$.
These tables show that the comparative performance of the integrators does not depend on the chosen initial condition, as the ranking of the schemes is practically the same in both tables. As in the case of the $\alpha$-FPUT model, the $DOP853$ and $TIDES$ integrators required, in general,  more CPU time than the SIs, although they produced more accurate results (smaller $E_r$ and $S_r$ values) at least up to $t_f=10^{6}$, with $TIDES$ being more precise.  The integrators exhibiting the best performance for $E_r \approx 10^{-5}$ are the sixth order SIs $s11\mathcal{ABC}6$ and $s9\mathcal{ABC}6$, while for  $E_r \approx 10^{-9}$ we have the  eighth order SIs $s19\mathcal{ABC}8$ and $s17\mathcal{ABC}8$, with the  $s11\mathcal{ABC}6$ scheme performing quite well also in this case.
%########### TABLE (2) ################
%\begin{table}[!hhhtb]
\begin{table}[!hhhtb]
\centering
\caption{Similar to Table \ref{table:performance1DDNLSstrong} but for  case $\text{II}_{\text{1D}}$ (see text for details) of  the 1D DDNLS model [Eq.~\eqref{eq:DNLS_1D_real}].
See \cite{sim_details_dnls} for practical information on the simulations.}
	\begin{tabular}{lcclr|lcclr}
		\toprule
		\multicolumn{5}{c}{$E_r \approx 10^{-5}$} &
		\multicolumn{5}{c}{$E_r \approx 10^{-9}$} \\
		\toprule
		Integrator           & $n$ & Steps & $\tau$  & $T_{\text{C}}$ & Integrator           & $n$ & Steps               & $\tau$  & $T_{\text{C}}$ \\
		\midrule
		$s11\mathcal{ABC}6$  & $6$ & $45$  & $0.4$   & $1132$         & $s19\mathcal{ABC}8$  & $8$ & $77$                & $0.3$   & $2184$         \\
		$s9\mathcal{ABC}6$   & $6$ & $37$  & $0.285$ & $1147$         & $s17\mathcal{ABC}8$  & $8$ & $69$                & $0.225$ & $2632$         \\
		$\mathcal{ABC}Y6\_A$ & $6$ & $29$  & $0.2$   & $1308$         & $s11\mathcal{ABC}6$  & $6$ & $45$                & $0.1$   & $4137$         \\
		$SS864S$             & $4$ & $17$  & $0.265$ & $1365$         & $s9\mathcal{ABC}6$   & $6$ & $37$                & $0.075$ & $4462$         \\
		$\mathcal{ABC}Y4$    & $4$ & $13$  & $0.055$ & $2354$         & $\mathcal{ABC}Y8\_D$ & $8$ & $61$                & $0.065$ & $8528$         \\
		$\mathcal{ABC}S4Y6$  & $6$ & $49$  & $0.105$ & $4965$         & $DOP853$             & $8$ & $\delta=10^{-16}$   & $0.05$  & $14998$        \\
		$\mathcal{ABC}Y4Y6$  & $6$ & $37$  & $0.04$  & $8091$         & $TIDES$              & $-$ & $\delta = 10^{-16}$ & $0.05$  & $420050$       \\
		$\mathcal{ABC}S4$&$4$&$21$&$0.02$&$9774$&  & & &	\\
		$\mathcal{ABC}2$&$2$&$5$&$0.0055$&$11700$&  & & &	\\
		\bottomrule
	\end{tabular}
  \label{table:performance1DDNLSweak}
\end{table}
%#######################################

%-----------------------------------------------------
\subsection{The 2D  disordered discrete nonlinear Schr\"{o}dinger equation system}

We now investigate the performance of the integrators used in Sec.~\ref{sec:DNLS_1D} for the computationally much more difficult case of the 2D DDNLS lattice of  Eq.~\eqref{eq:DNLS_2D_real}, as its Hamiltonian function can also be split into three integrable parts. In order to  test the performance of the various schemes we consider a lattice with $N\times M=200\times 200$ sites, resulting to a system  of $4\times 40\,000=160\,000$ ODEs (equations of motion and variational equations). The numerical integration of this huge number of ODEs is a very demanding computational task. For this reason we integrate this model only up to a final time $t_f =10^5$, instead of the $t_f =10^6$ used for the $\alpha$-FPUT and the 1D DDNLS systems. It is worth noting that due to the computational difficulty of the problem very few numerical results for the 2D DDNLS system exist in the literature (e.g.~\cite{sales2018sub, garcia2009delocalization}). We consider again two sets of initial conditions:
\begin{itemize}
\item Case $\text{I}_{\text{2D}}$: We initially excite $7\times 7$ central sites attributing to each one of them the same norm  $s_{i, j} = (q_{i, j} ^2 + p_{i, j}^2)/2 = 1/6$ so that the total norm is $S_{\text{2D}} =49/6$, for $W=15$ and $\beta =6$. The disorder parameters  $\epsilon_{i, j}$, $1\leq i \leq N$, $1\leq j \leq M$, are chosen so that the initial total energy is $H_{\text{2D}} \approx 1.96$.
\item Case $\text{II}_{\text{2D}}$: We initially excite a single central site of the lattice with a total norm $S_{\text{2D}} = 1$, i.e.~$\zeta _{100, 100} = 1$, for $W=16$,  $\beta = 1.25$ and $H_{\text{2D}} = 0.625$.
\end{itemize}
The initial normalized deviation vector considered in our simulations  has random non-zero values only at the $7\times 7$ initially excited sites for case $\text{I}_{\text{2D}}$, and only at site $i=100$, $j=100$  for case $\text{II}_{\text{2D}}$. In both cases, all  others elements of the vectors are initially set to zero. Both considered cases  belong to a Gibbsian regime
where the thermalization processes are well defined by  Gibbs ensembles~\cite{thudiyangal2017gross, rasmussen2000statistical}. Therefore, we expect a subdiffusive spreading of the initial excitations to take place for both cases, although their initial conditions are significantly different.

As was done in the case of the 1D DDNLS system (Sec.~\ref{sec:DNLS_1D}), in order to evaluate the performance of the used integrators we follow the time evolution of the normalized norm density distribution $\zeta _{i, j}=s_{i,j} /S_{\text{2D}}$, $1\leq i \leq N$, $1\leq j \leq M$ and  compute  the related second moment $m_2$ and  participation number $P$
\begin{equation}
m_2 = \sum _{i=1}^N \sum_{ j=1}^M \left\|
(i,j)^T
-
(\overline{i} ,
\overline{j})^T
\right\|^2 \zeta_{i,j}
, \qquad \displaystyle P = \frac{1}{ \sum_{i=1}^{N}\sum _{j=1}^{M} \zeta _{i, j}^2},
\label{eq:2D_DNLS_obs}
\end{equation}
where $(\overline{i} ,\overline{j})^T = \sum _{i, j} (i,j)^T  \zeta _{i, j}$ is the mean position of the  norm density distribution. We also evaluate the relative energy [$ E_r (t) $] and norm [$ S_r (t) $] errors  and  compute the finite time mLE $X_1(t)$.

In Fig.~\ref{fig:weakchaos2Dequationsmotion} we present  results obtained for case $\text{I}_{\text{2D}}$ by  the four best performing SIs (see Table \ref{table:performance2DDNLSstrong_1new}), the  $s11\mathcal{ABC}6$ (red curves), $s9\mathcal{ABC}6$ (blue curves), $\mathcal{ABC}Y6\_A$ (green curves) and $\mathcal{ABC}Y4$ (brown curves) schemes, along with the   $DOP853$ integrator (grey curves). The integration time  steps $\tau$ of the SIs were adjusted in order to obtain an accuracy of $E_r\approx 10^{-5}$ [Fig.~\ref{fig:weakchaos2Dequationsmotion}(a)], while results for  the conservation of the second integral of motion, i.e.~the system's total norm, are shown in Fig.~\ref{fig:weakchaos2Dequationsmotion}(b). We see that for all  SIs the $S_r$ values increase slowly, remaining always below $S_r \approx 10^{-4}$, which indicates  a good conservation  of the system's  norm.
As in the case of the 1D DDNLS system, the $E_r$ and $S_r$ values obtained by the $DOP853$ integrator increase, although the choice of  $\delta = 10^{-16}$ again  ensures high precision computations.
%%%%%%%%%%%%%%%%%%%%%%%%%%%%%
%\begin{figure}[!hhtb]
\begin{figure}[!hhtb]
\centering
\includegraphics[scale=0.5]{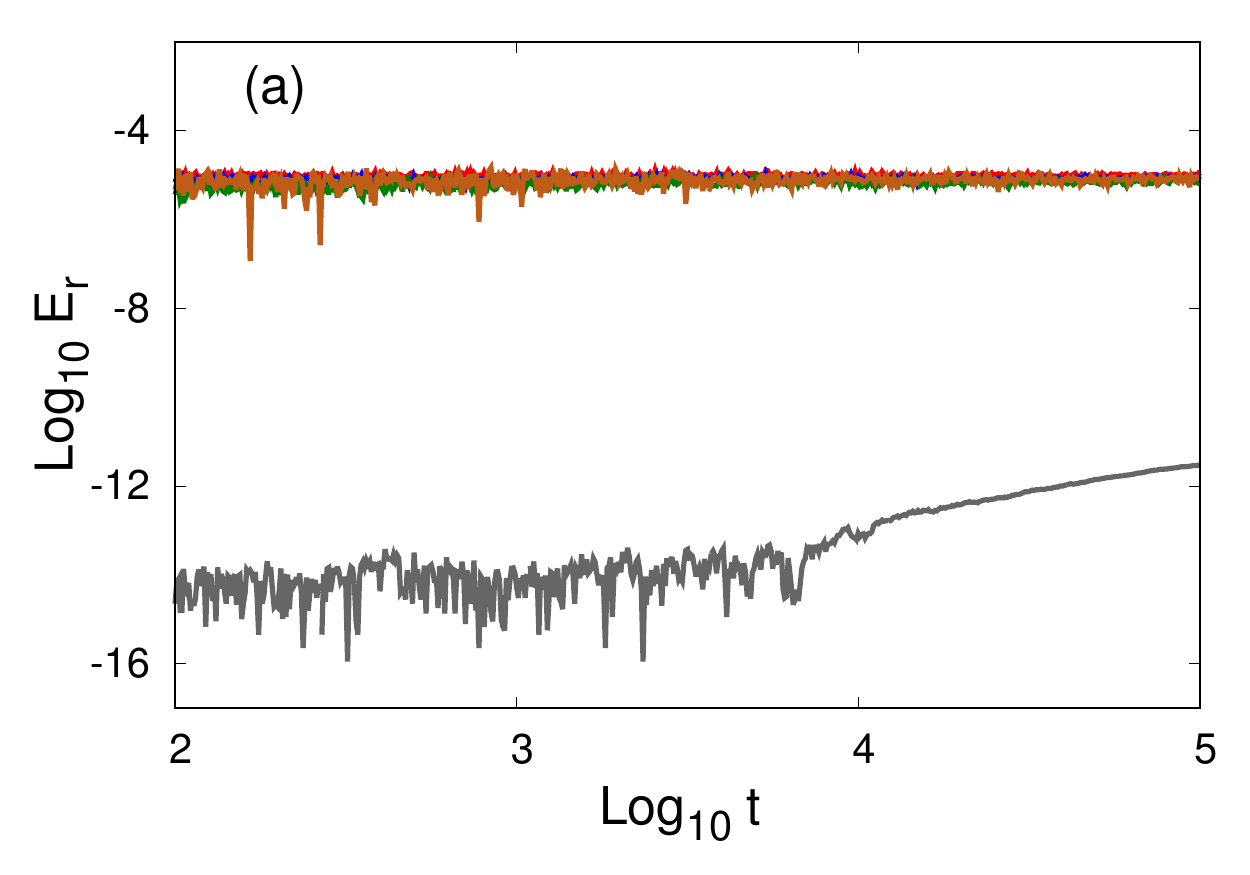}
\includegraphics[scale=0.5]{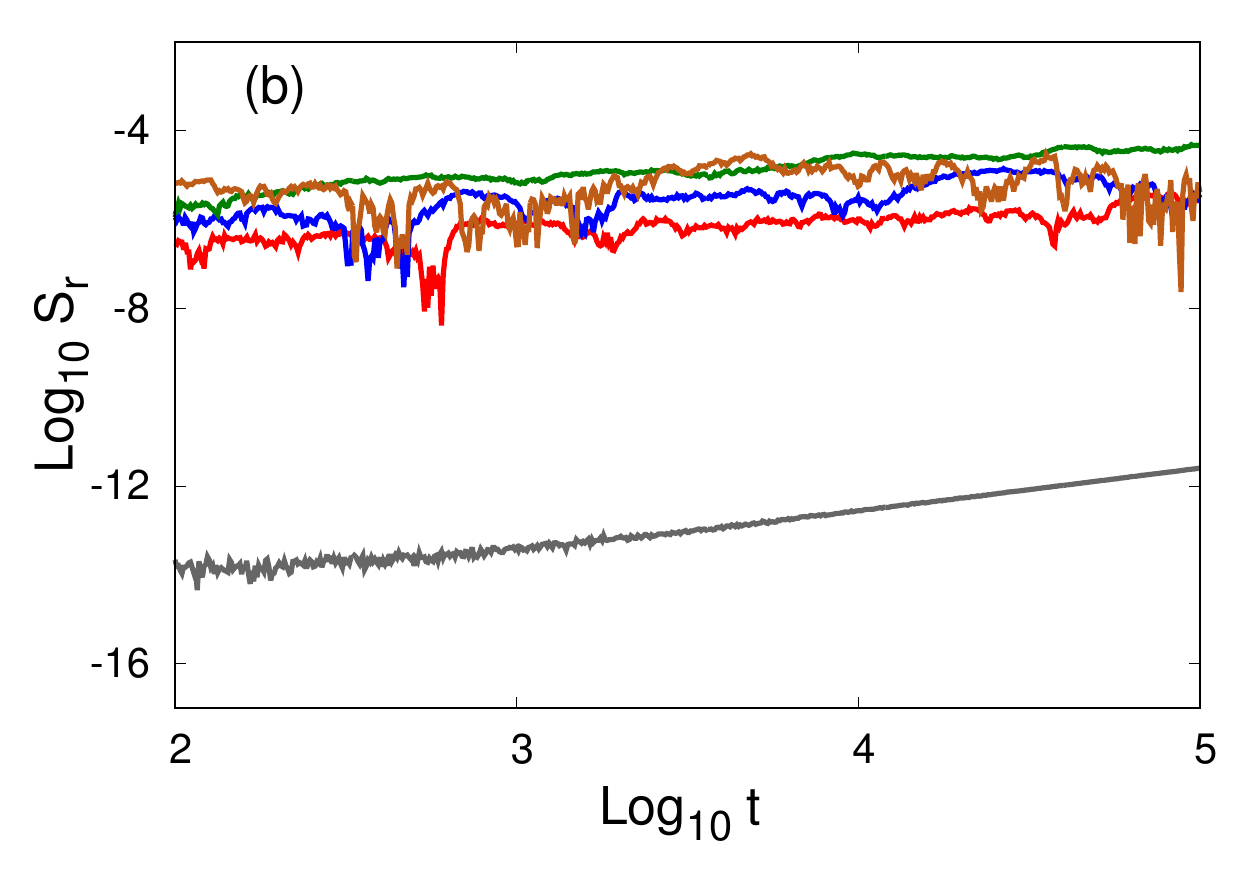}
\includegraphics[scale=0.5]{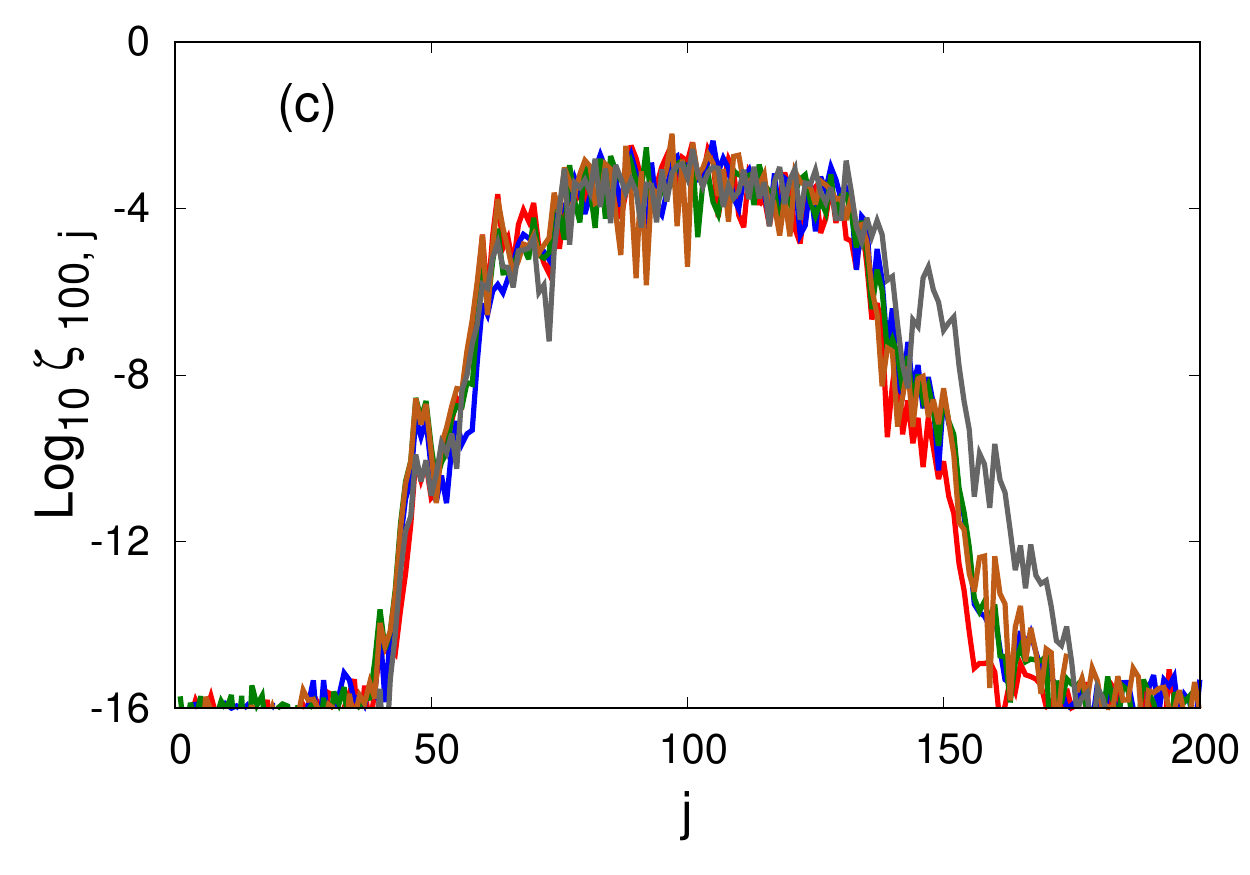}
\includegraphics[scale=0.5]{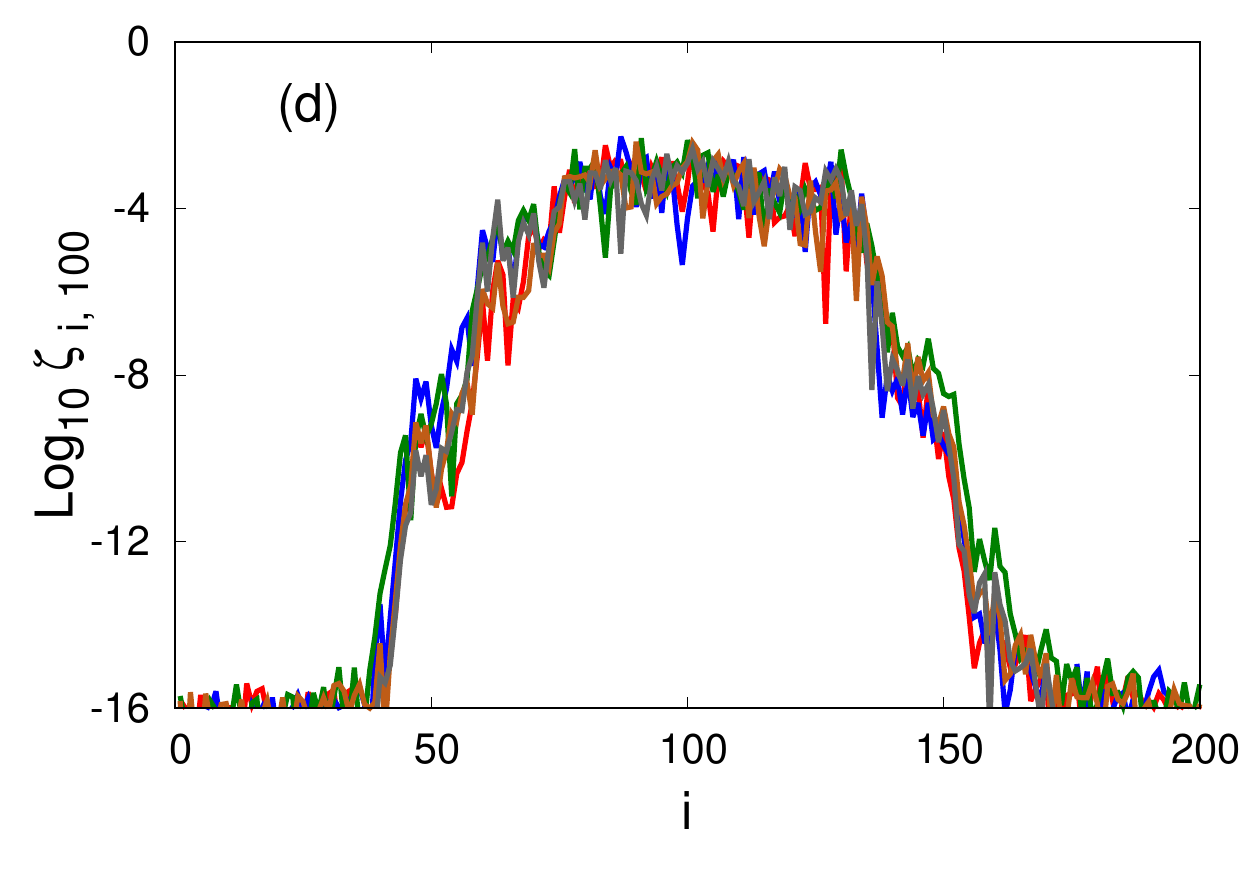}
\includegraphics[scale=0.5]{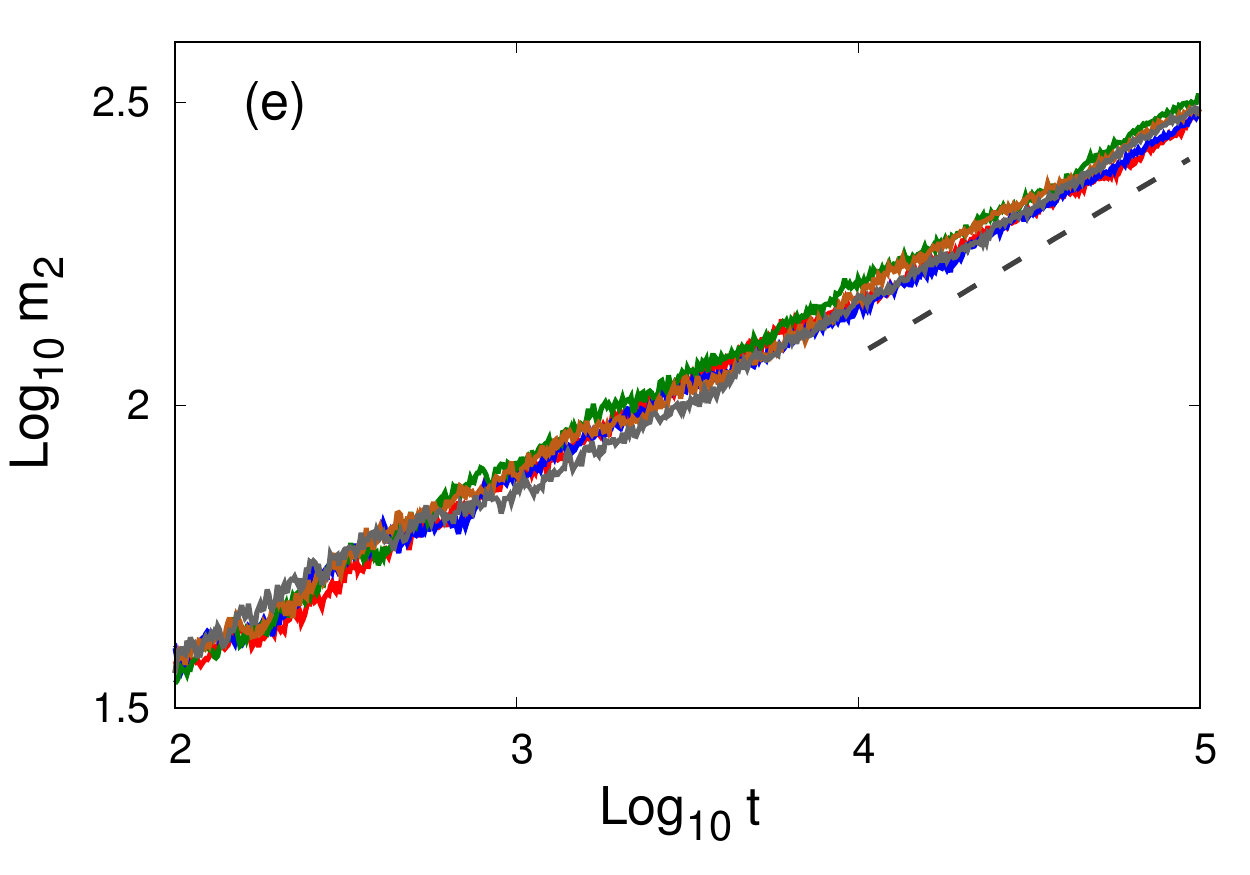}
\includegraphics[scale=0.5]{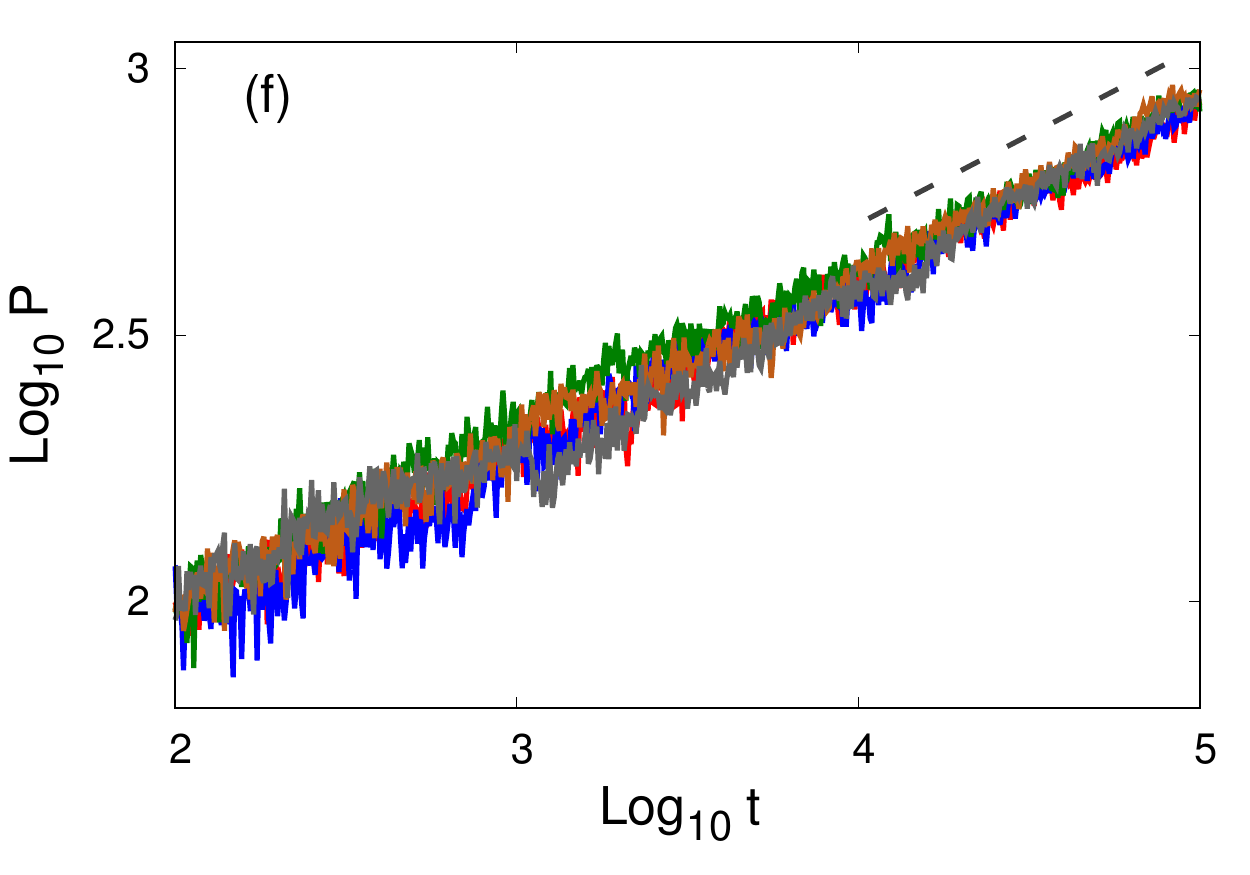}
\includegraphics[scale=0.5]{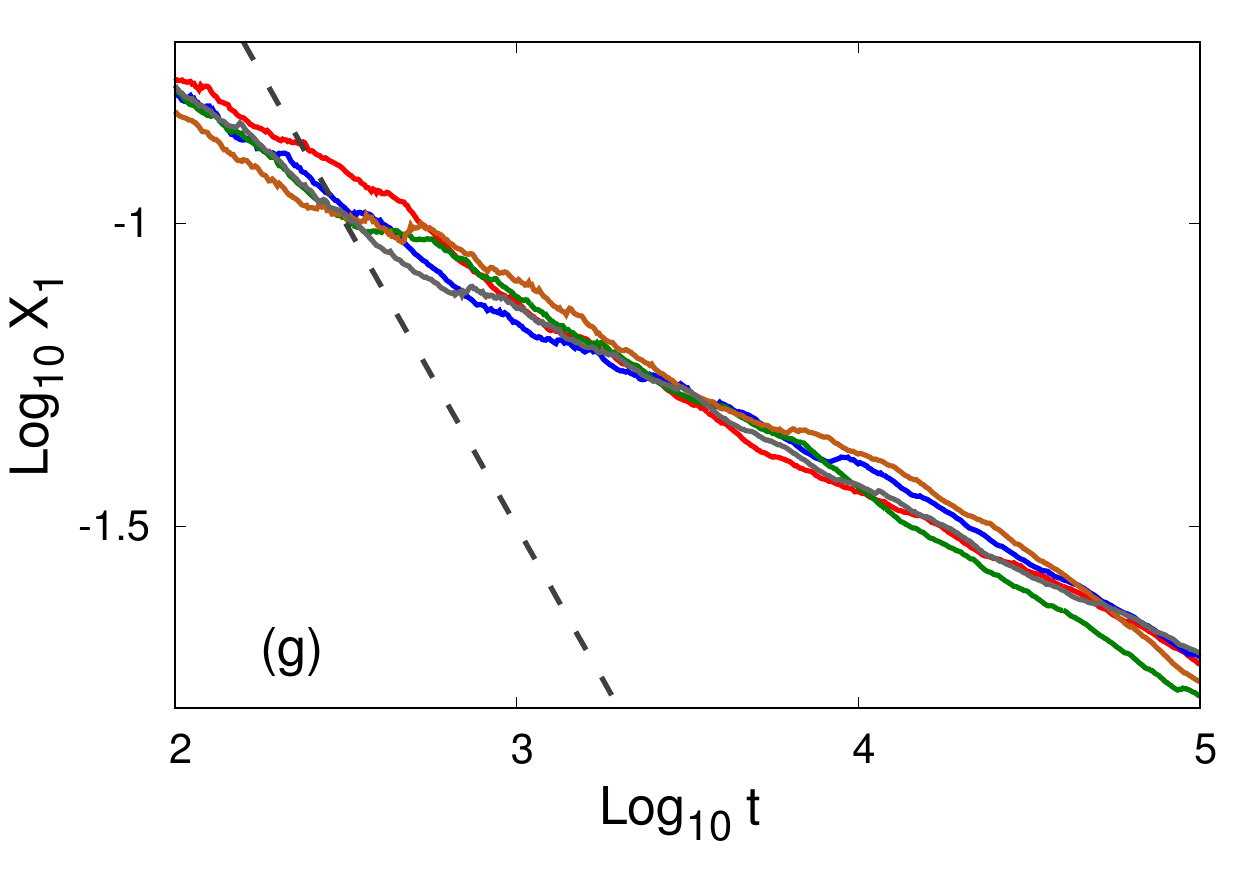}
\includegraphics[scale=0.5]{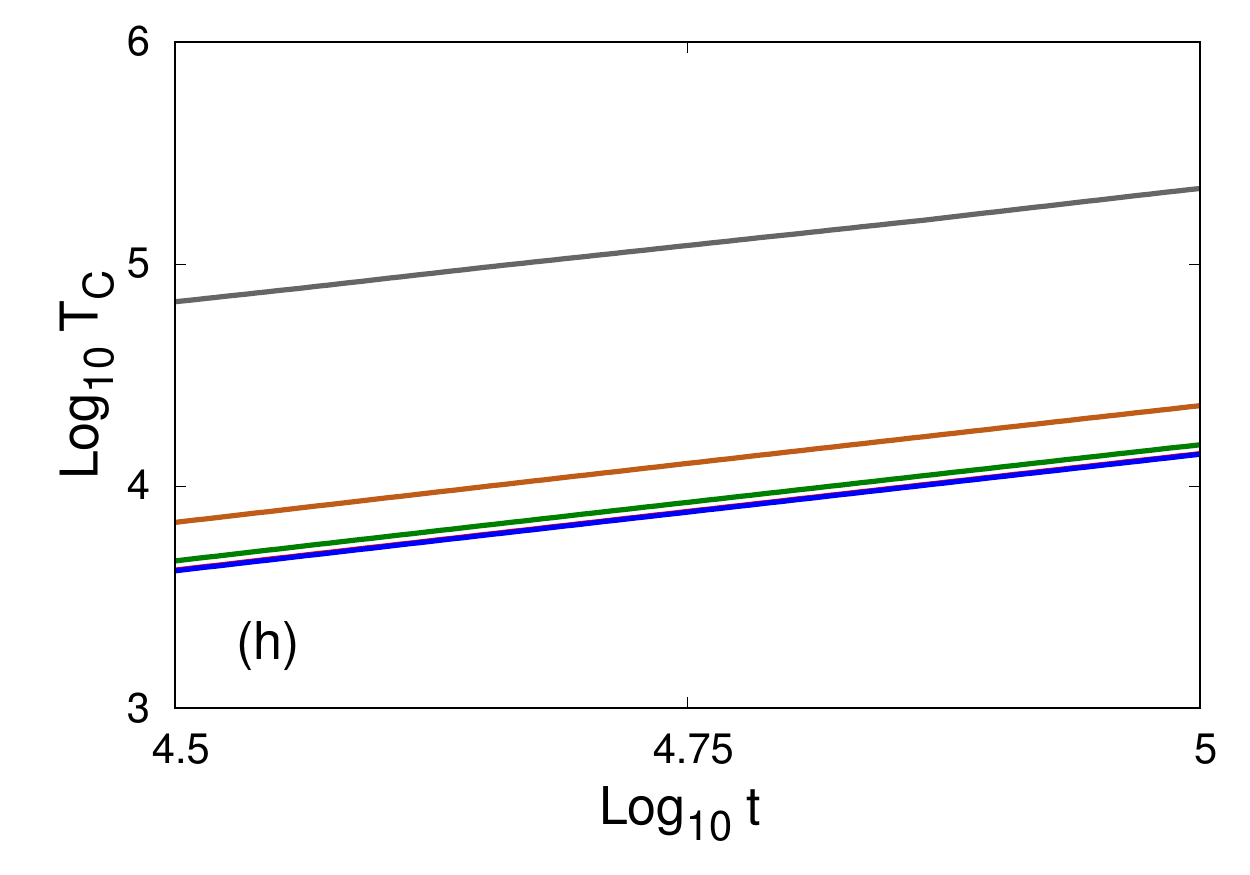}
\caption{Results for the integration of case $\text{I}_{2D}$ (see text for details) of the 2D
  DDNLS Hamiltonian [Eq.~\eqref{eq:DNLS_2D_real}] by the fourth order
  SI $\mathcal{ABC}Y4$ for $\tau = 0.025$, the sixth order SIs
  $s11\mathcal{ABC}6$ for $\tau = 0.125$, $s9\mathcal{ABC}6$ for $\tau
  = 0.105$ and $\mathcal{ABC}Y6\_A$ for $\tau = 0.08$, along with the
  non-symplectic scheme $DOP853$ for $\tau = 0.05$ [brown, red,
  blue, green and grey curves respectively]. Time evolution of (a)
  $E_r(t)$, (b) $S_r(t)$, (e) $m_2(t)$, (f) $P(t)$, (g) $X_1(t)$ and
  (f) $T_C(t)$. In (c) and (d) we plot the  norm density 
  distributions along the lines $i=100$ and $j=100$ respectively, at time $t_f=10^5$. The
  dashed line in panels (e) and (f) guides the eye for slope $1/3$,
  while in panel (g) denotes slope -1.}
\label{fig:weakchaos2Dequationsmotion}
\end{figure}
%%%%%%%%%%%%%%%%%%%%%%%%%%%%%

%%%%%%%%%%%%%%%%%%%%%%%%%%%%%
%\begin{table}[!hhhtb]
\begin{table}
		\centering
                \caption{Similar to Table \ref{table:performance1DDNLSstrong} but for  case $\text{I}_{\text{2D}}$ of  the 2D DDNLS model [Eq.~\eqref{eq:DNLS_2D_real}].
     See \cite{sim_details_dnls} practical information details on the simulations.}
	\begin{tabular}{lcclr|lcclr}
		\toprule
		\multicolumn{5}{c}{$E_r \approx 10^{-5}$} &
		\multicolumn{5}{c}{$E_r \approx 10^{-9}$} \\
		\toprule
		Integrator           & $n$ & Steps & $\tau$  & $T_{\text{C}}$ & Integrator           & $n$ & Steps             & $\tau$   & $T_{\text{C}}$ \\
		\midrule
		$s9\mathcal{ABC}6$   & $6$ & $45$  & $0.105$ & $13914$        & $s17\mathcal{ABC}8$  & $8$ & $77$              & $0.075$  & $36528$        \\
		$s11\mathcal{ABC}6$  & $6$ & $37$  & $0.125$ & $14000$        & $s19\mathcal{ABC}8$  & $8$ & $69$              & $0.08$   & $38270$        \\
		$\mathcal{ABC}Y6\_A$ & $6$ & $29$  & $0.08$  & $15344$        & $s9\mathcal{ABC}6$   & $6$ & $45$              & $0.0235$ & $65287$        \\
		$\mathcal{ABC}Y4$    & $4$ & $13$  & $0.025$ & $23030$        & $s11\mathcal{ABC}6$  & $6$ & $37$              & $0.0275$ & $67314$        \\
		$SS864S$             & $4$ & $17$  & $0.085$ & $23887$       & $\mathcal{ABC}Y8\_D$ & $8$ & $61$              & $0.008$  & $140506$       \\
		$\mathcal{ABC}S4Y6$  & $6$ & $49$  & $0.03$  & $77424$        & $DOP853$             & $8$ & $\delta=10^{-16}$ & $0.05$   & $218704$       \\
		$\mathcal{ABC}Y4Y6$&$6$&$37$&$0.0165$&$87902$&  & & & \\
		$\mathcal{ABC}S4$&$4$&$21$&$0.0065$&$132713$& & & & \\
		$\mathcal{ABC}2$&$2$&$5$&$0.005$&$157694$&  & & &			\\
		\\
		\bottomrule
	\end{tabular}
		\label{table:performance2DDNLSstrong_1new}
	\end{table}
%%%%%%%%%%%%%%%%%%%%%%%%%%%%%

The norm density distributions at the final integration time $t_f =10^5$ along the axis $i=100$ [Fig.~\ref{fig:weakchaos2Dequationsmotion}(c)] and $j=100$ [Fig.~\ref{fig:weakchaos2Dequationsmotion}(d)] obtained by the various integrators practically overlap indicating the ability of all numerical schemes to correctly capture the system's dynamics, as well as the fact that the initial excitations expand along all directions of the 2D lattice. From Fig.~\ref{fig:weakchaos2Dequationsmotion}(e) [Fig.~\ref{fig:weakchaos2Dequationsmotion}(f)] we see that $m_2 (t)$ [$P(t)$] is increasing according to the power law $m_2 = t^{1/3}$ [$P= t^{1/3}$] as expected from the analysis presented in~\cite{laptyeva2012subdiffusion}, indicating that the 2D lattice is being thermalized. The results of Figs.~\ref{fig:weakchaos2Dequationsmotion}(e) and (f) provide additional numerical evidences that all numerical methods reproduce correctly the dynamics. This is also seen by the similar behavior of the  finite time mLE curves in  Fig.~\ref{fig:weakchaos2Dequationsmotion}(g). From the results of this figure we see that $X_1$ exhibits a tendency to decrease following a completely different decay from the $X_1\propto t^{-1}$ power law observed for regular motion. This behavior was also observed for the 2D DKG model \cite{senyange2017symplectic}, as well as for the 1D DKG and DDNLS systems in \cite{skokos2013nonequilibrium,senyange2018characteristics} where a power law $X_1(t) \propto t^{\alpha_L}$ with $\alpha _L \approx -0.25$ and
$\alpha _L \approx -0.3$ for, respectively, the weak and strong chaos dynamical regimes was established. Further investigations of the behavior of the finite mLE in 2D disordered systems are required in order to determine a potentially global behavior of $X_1$, since here and in  \cite{senyange2017symplectic} only some isolated cases were discussed. Such studies will require the statistical analysis of results obtained for many different disorder realizations, parameter sets and initial conditions. Thus, the utilization of efficient and accurate numerical integrators, like the ones presented in this study, will be of utmost importance for the realization of this goal.

From Tables \ref{table:performance2DDNLSstrong_1new} and \ref{table:performance2DDNLSweak_2new}, where the CPU times $T_C$ required by the tested integrators are reported, we see that, as in the case of the 1D DDNLS model, the SIs $s11\mathcal{ABC}6$   and $s9\mathcal{ABC}6$ have the best performance for $E_r\approx 10^{-5}$ and the SIs $s19\mathcal{ABC}8$ and $s17\mathcal{ABC}8$ for
$E_r\approx 10^{-9}$.
%%%%%%%%%%%%%%%%%%%%%%%%%%%%%
%\begin{table}[!hhhtb]
\begin{table}
		\centering
                \caption{Similar to Table \ref{table:performance1DDNLSstrong} but for  case $\text{II}_{\text{2D}}$ of  the 2D DDNLS model [Eq.~\eqref{eq:DNLS_2D_real}]. See \cite{sim_details_dnls} for practical information on the simulations.
                }
	\begin{tabular}{lcclr|lcclr}
		\toprule
		\multicolumn{5}{c}{$E_r \approx 10^{-5}$} &
		\multicolumn{5}{c}{$E_r \approx 10^{-9}$} \\
		\toprule
		Integrator           & $n$ & Steps & $\tau$    & $T_{\text{C}}$ & Integrator           & $n$ & Steps             & $\tau$   & $T_{\text{C}}$ \\
		\midrule
		$s11\mathcal{ABC}6$  & $6$ & $45$  & $0.1515$  & $11443$        & $s19\mathcal{ABC}8$  & $8$ & $77$              & $0.135$  & $20952$        \\
		$s9\mathcal{ABC}6$   & $6$ & $37$  & $0.11$    & $13408$        & $s17\mathcal{ABC}8$  & $8$ & $69$              & $0.0875$ & $28966$        \\
		$\mathcal{ABC}Y6\_A$ & $6$ & $29$  & $0.0775$  & $14607$        & $s11\mathcal{ABC}6$  & $6$ & $45$              & $0.0335$ & $50301$        \\
		$SS864S$             & $4$ & $17$  & $0.0915$  & $15564$        & $s9\mathcal{ABC}6$   & $6$ & $37$              & $0.024$  & $58187$        \\
		$\mathcal{ABC}Y4$    & $4$ & $13$  & $0.0215$  & $25898$        & $\mathcal{ABC}Y8\_D$ & $8$ & $61$              & $0.009$  & $150045$       \\
		$\mathcal{ABC}S4Y6$  & $6$ & $49$  & $0.035$   & $64423$        & $DOP853$             & $8$ & $\delta=10^{-16}$ & $0.05$   & $166145$       \\
		$\mathcal{ABC}Y4Y6$  & $6$ & $37$  & $0.01375$ & $102580$       &                      &     &                   &          &                \\
		$\mathcal{ABC}S4$    & $4$ & $21$  & $0.005$   & $185615$       &                      &     &                   &          &                \\
		$\mathcal{ABC}2$     & $2$ & $5$   & $0.00155$ & $198534$       &                      &     &                   &          &                \\
		\bottomrule
	\end{tabular}
		\label{table:performance2DDNLSweak_2new}
	\end{table}
%%%%%%%%%%%%%%%%%%%%%%%%%%%%%

%-----------------------------------------------------
\section{Conclusions} \label{sec:conclusion}

In this work we carried out a methodical and detailed analysis of the performance of several symplectic and non-symplectic integrators, which were used to integrate the equations of motion and the variational equations of some important  many-body classical Hamiltonian systems in one and two spatial dimensions: the $\alpha$-FPUT chain, as well as the 1D and 2D DDNLS models.
In the  case of the $\alpha$-FPUT system we used two part split SIs, while for the integration of the DDNLS models we implemented several three part split SIs. In order to evaluate the efficiency of  all these integrators we evolved in time  different sets of initial conditions and evaluated quantities related to (a) the dynamical evolution of the studied systems (e.g.~the second moment of norm density distributions for the DDNLS models), (b) the quantification of the systems' chaotic behavior (i.e.~the finite time mLE), and (c) the accurate computation of the systems' integrals of motion (relative energy and norm errors), along with the CPU times needed to perform the simulations.

For the $\alpha$-FPUT system several two part split SIs
showed very good performances, among which we mention the  $ABA864$ and $ABAH864$  SIs of order four to be the best schemes for moderate energy accuracies ($E_r \approx 10^{-5}$), while the  $SRKN^a_{14}$  and $SRKN^b_{11}$ SIs of order six  were the best integration schemes for higher accuracies ($E_r \approx 10^{-9}$).
In particular, the $ABA864$ scheme appears to be an efficient, general choice as it showed  a quite good behavior  also for $E_r \approx 10^{-9}$.
Concerning the 1D and the 2D DDNLS models our simulations showed that the SIs  $s9\mathcal{ABC}6$ and $s11\mathcal{ABC}6$ (order six),  along with the SIs $s17\mathcal{ABC}8$ and  $s19\mathcal{ABC}8$ (order eight) are the best integrators for moderate ($E_r \approx 10^{-5}$) and high ($E_r \approx 10^{-9}$) accuracy levels respectively.

The $DOP853$ and $TIDES$ non-symplectic integrators required, in general, much longer CPU times to carry out the simulations, although they produced more accurate results (i.e~smaller $E_r$ and $S_r$ values) than the symplectic schemes. Apart from the drawback of the high CPU times, the fact that  $E_r$ (and $S_r$) values exhibit a constant increase in time signifies that such schemes should not be preferred over SIs when very long time simulations are needed.

It is worth noting that two part split SIs of order six and higher often do now not produce reliable results for relative low energy accuracies like $E_r \approx 10^{-5}$ for the $\alpha$-FPUT system (similar behaviors were reported in \cite{senyange2017symplectic} for the DKG model). This happens because the required integration time step $\tau$ needed to keep the relative energy error at $E_r \approx 10^{-5}$ is typically large, resulting to an unstable behavior of the integrator i.e.~the produced $E_r$ values do not remain bounded. Thus, SIs of order $n \geq 6$ are more suitable for calculations that require higher accuracies (e.g.~$ E_r \approx 10^{-9}$ or lower).

We note that we presented here a detailed comparison of the performance of several two and three part split SIs for the integration of the variational equations through the tangent map method and consequently for the computation of a chaos indicator (the mLE), generalizing, and completing in some sense, some sporadic previous investigation of the subject \cite{senyange2017symplectic, skokos2010numerical,
gerlach2011comparing,
gerlach2012efficient}, which were only focused on two part split SIs.

We hope that the clear description of the construction of several two and three part split SIs in Sec.~\ref{sec:integration_schemes}, along with the explicit presentation in the Appendix of the related differential operators for many commonly used classical, many-body Hamiltonians will be useful to researchers working on lattice dynamics. The numerical techniques presented here can be used for the computation of several chaos indicators, apart from the mLE (e.g.~the  SALI and the GALI methods \cite{skokos2016smaller}) and for the dynamical study of various lattice models, like for example of arrays  of Josephson junctions in regimes of weak non-integrability~\cite{mithun2018:rotor},  granular chains~\cite{achilleos2016energy} and DNA models~\cite{hillebrand2018chaos}, to name a few.

\clearpage
\appendix

\section*{Appendix}

%-----------------------------------------------------
\section{Explicit forms of tangent map method operators}\label{sec:app_A}

We present here the exact expressions of the  operators needed by the various SIs we implemented in our study to simultaneously solve the Hamilton equations of motion and the variational equations, or in order words to  solve the system of  Eq.~\eqref{eq:genaralX}
\begin{equation}
\dot{\boldsymbol{ X}} = (\dot{\boldsymbol{ x}}(t),\dot{\delta \boldsymbol{ x}}(t))=\boldsymbol{ f}(\boldsymbol{ X})=
  \begin{bmatrix}

\boldsymbol{ J}_{2N} \cdot \boldsymbol{ D}_H(\boldsymbol{ x}(t))         \\
   \big[  \boldsymbol{ J}_{2N} \cdot \boldsymbol{ D}_{ H}^2 (\boldsymbol{ x}(t) ) \big] \cdot \delta \boldsymbol{ x}(t)
\end{bmatrix}.
\end{equation}

%-----------------------------------------------------
\subsection{The $\alpha$-Fermi-Pasta-Ulam-Tsingou model}\label{sec:FPUT_app}

The Hamiltonian of the $\alpha$-FPUT chain  [Eq.~\eqref{eq:alphaFPUT}] can be split into two integrable parts as
\begin{equation}
A\left(\boldsymbol{ p}\right) = \sum_{ i=0}^N\ \frac{p_i^2}{2}, \qquad B\left(\boldsymbol{ q}\right) = \sum_{ i=0}^N\ \frac{1}{2}(q_{i+1} - q_{ i})^2  + \frac{\alpha}{3}(q_{i+1} - q_{ i})^3.
\label{eq:split2partFPUT}
\end{equation}
As we have already stated, the split into two integrable parts is not necessarily unique. In this particular case another possible choice of integrable splits for the $\alpha$-FPUT chain is to group together the quadratic terms of the Hamiltonian [i.e.~$A({\bf p},{\bf q})= \sum_{ i=0}^N\ \frac{p_i^2}{2} + \frac{1}{2}(q_{i+1} - q_{ i})^2$] and keep separately the nonlinear terms [i.e.~$B({\bf q})= \sum_{ i=0}^N\ \frac{\alpha}{3}(q_{i+1} - q_{ i})^3$].
The set of equations of motion and variational equations for the  Hamiltonian function $A\left(\boldsymbol{ p}\right)$ is
\begin{equation}
\frac{d\boldsymbol{X}}{dt} = L_{AZ}\boldsymbol{X} \colon \begin{cases}
\dot{q}_i &= p_i \\
\dot{p}_i &= 0 \\
\dot{\delta q}_i &= \delta p_i \\
\dot{\delta p}_i &= 0
\end{cases}, \qquad \text{for}\qquad  1\leq i\leq N,
\end{equation}
and the corresponding operator $e^{\tau L_{AZ}}$, which propagates the values of $q_i$, $p_i$, $\delta q_i$ and $\delta p_i$ for $\tau$ time units in the future, obtaining $q_i'$, $p_i'$, $\delta q_i'$ and $\delta p_i'$, takes the form
\begin{equation}
\label{eq:res_A_FPU}
e^{ \tau L_{AZ}} \colon
\begin{cases}
    q_i' &= q_i + \tau p_i        \\
  p_i' &=  p_i \\
  \delta q_i' &=\delta q_i +   \tau \delta p_i   \\
    \delta p_i' &= \delta  p_i
\end{cases}, \qquad \text{for}\qquad 1\leq i\leq N.
\end{equation}
In a similar way for the $B\left(\boldsymbol{ q}\right)$ Hamiltonian of Eq.~\eqref{eq:split2partFPUT} we get
\begin{equation}
\frac{d\boldsymbol{X}}{dt} = L_{BZ} \boldsymbol{X} \colon
\begin{cases}
    \dot{q}_i &= 0        \\
   \dot{p}_i &=   (q_{i+1} + q_{i-1} -2 q_i) +\alpha \big[ (q_{i+1} - q_i)^2 - (q_i - q_{i-1})^2\big] \\
    \dot{\delta q}_i &=  0   \\
   \dot{ \delta p}_i &=    [ 2\alpha  (q_{i-1} - q_{i+1} ) - 2 ]\delta  q_i   + [ 1 +2\alpha (q_{i+1} - q_i) ]\delta  q_{i+1} + [1+ 2\alpha (q_{i} - q_{i-1}) ]\delta  q_{i-1}
\end{cases},
\label{eq:eqBFPUT}
\end{equation}
and
\begin{equation}
e^{ \tau L_{BZ}}\colon
\begin{cases}
    q_i' &=    q_i         \\
   p_i' &=  p_i +\tau \big\{   (q_{i+1} + q_{i-1} -2 q_i) +\alpha \big[ (q_{i+1} - q_i)^2 - (q_i - q_{i-1})^2\big] \big\}  \\
    \delta q_i' &=  \delta  q_i   \\
    \delta p_i' &=  \delta p_i + \tau \big\{  [ 2\alpha  (q_{i-1} - q_{i+1} ) - 2 ]\delta  q_i + [ 1 +2\alpha (q_{i+1} - q_i) ]\delta  q_{i+1} + [1+ 2\alpha (q_{i} - q_{i-1}) ]\delta  q_{i-1}     \big\}
\end{cases}.
\label{eq:opBFPUT}
\end{equation}

According to  Eq.~\eqref{eq:2st_int_corre}
the accuracy of the $SABA_n$ and $SBAB_n$ integrators can be improved by using a  corrector Hamiltonian $K$ \cite{laskar2001high}. In the case of a separable Hamiltonian $H(\boldsymbol{q},\boldsymbol{p})  = A(\boldsymbol{p})+B(\boldsymbol{q})$ with $A\left(\boldsymbol{ p}\right) = \sum_{i=1}^N p_i^2/2$, the corrector $K$ becomes
\begin{equation}
\label{eq:corrector_app}
 K(\boldsymbol{ q})  = \{B\{B, A\}\} = \sum_{ i=1}^N\ \left(\frac{\partial B}{\partial q_i} \right)^2.
\end{equation}
For the $\alpha$-FPUT chain, the corrector Hamiltonian $K$ is
\begin{equation}
\begin{split}
K(\boldsymbol{ q}) &= \sum_{i=1}^{N}\Big[ \big(2q_i - q_{i+1} - q_{i-1}  \big) \big(1+\alpha (q_{i+1} - q_{i-1})\big)  \Big]^2\ .\\
\end{split}
\label{eq:C_alphaFPU}
\end{equation}
As the equations of motion and variational equations associated to the corrector Hamiltonian $K$ are cumbersome,
we report here  only the form of the operator  $e^{ \tau L_{KZ}}$
\begin{equation}
e^{ \tau L_{KZ}}\colon
\begin{cases}
    q_i' &=    q_i         \\
  p_i' &= p_i + 2 \tau \Big\{2 \big(  q_{i+1} +  q_{i-1}  -  2 q_i   \big) \big[  1+\alpha\big( q_{i+1} -   q_{i-1}  \big) \big]^2  \\
         & - \big(  q_{i+2} +  q_{i}  -  2 q_{i+1}   \big)  \big[  1 +\alpha\big( q_{i+2} -   q_{i}  \big)  \big]  \big[  1 - 2\alpha\big( q_{i} -   q_{i+1}  \big)  \big]\\
          & - \big(  q_{i-2} +  q_{i}  -  2 q_{i-1}   \big)   \big[  1 +\alpha\big( q_{i} -   q_{i-2}  \big)  \big]  \big[  1 - 2\alpha\big( q_{i-1} -   q_{i}  \big)\big]  \Big\}\\
    \delta q_i' &=  \delta  q_i   \\
    \delta p_i' &= \delta p_i +\tau  \big\{ \gamma_i \delta  q_{i}  +  \gamma_{i+1} \delta  q_{i+1} +  \gamma_{i+2} \delta  q_{i+2} +  \gamma_{i-1} \delta  q_{i-1} +  \gamma_{i-2} \delta  q_{i-2}   \big\}
\end{cases},
\label{eq:e_LAW_LBW_LKW_gen_FPUT}
\end{equation}
where
\begin{equation}
\begin{split}
\gamma_{i}&=  - 2 \Big\{4\big[  1+\alpha\big( q_{i+1} -   q_{i-1}  \big) \big]^2 \\
 &  \qquad  \quad+ \big[  1 +\alpha\big( q_{i+2} -   q_{i}  \big)  \big]  \big[  1 - 2\alpha\big( q_{i} -   q_{i+1}  \big)  \big] \\
   &  \qquad \quad+   \big[  1 +\alpha\big( q_{i} -   q_{i-2}  \big)  \big]  \big[  1 - 2\alpha\big( q_{i-1} -   q_{i}  \big)\big] \\
        &  \qquad\quad+\alpha \big(2 q_{i+1} -  q_{i+2} -  q_{i}  \big)  \big[  3 - 4\alpha q_{i} +2\alpha q_{i+1}  +2\alpha q_{i+2} \big]\\
          &  \qquad\quad -  \alpha \big(2 q_{i-1} -  q_{i-2} -  q_{i}    \big) \big[ 3 + 4 \alpha q_{i} - 2\alpha q_{i-1} - 2\alpha q_{i-2}   \big]   \Big\}\\
\gamma_{i+1}&=  4 \Big\{\big[  1+\alpha\big( q_{i+1} - q_{i-1} \big) \big] \big[ 1 - \alpha \big(4 q_i   -3 q_{i+1}  -  q_{i-1} \big) \big] \\
          & \qquad\quad + \big[  1 +\alpha\big( q_{i+2} -   q_{i}  \big)  \big]  \Big[  1 + \alpha\big( 4 q_{i+1} -  3 q_{i} - q_{i+2}  \big) \Big] \Big\}\\
\gamma_{i-1}& =  4\Big\{\big[  1+\alpha\big( q_{i+1} - q_{i-1}  \big) \big] \big[ 1 + \alpha \big(4 q_i   -3 q_{i-1}  - q_{i+1} \big) \big] \\
          & \qquad\quad+ \big[  1 +\alpha\big( q_{i} -   q_{i-2}  \big)  \big]  \Big[  1 - \alpha\big( 4 q_{i-1} -  3 q_{i} - q_{i-2}  \big) \Big] \Big\} \\
\gamma_{i+2}&=2  \big[1 - 2\alpha\big( q_{i} - q_{i+1}  \big)\big]  \big[ 2 \alpha\big( q_{i+1} - q_{i+2}\big)  -1 \big]    \\
\gamma_{i-2} &=2\big[  1 - 2\alpha\big( q_{i-1} -   q_{i}  \big)\big]   \big[  2\alpha\big( q_{i-2} -q_{i-1}  \big) -1 \big]
\end{split}.
\label{eq:e_LC_alphaFPUT_coeff}
\end{equation}
We did not specify the range of index $i$  in Eqs.~\eqref{eq:eqBFPUT}, \eqref{eq:opBFPUT} and \eqref{eq:e_LAW_LBW_LKW_gen_FPUT} intentionally, because it depends on the type of the used boundary conditions. In particular, the expression of Eqs.~\eqref{eq:eqBFPUT}, \eqref{eq:opBFPUT} and \eqref{eq:e_LAW_LBW_LKW_gen_FPUT} are accurate for the case of periodic boundary conditions, i.e.~$q_0=q_N$, $p_0=p_N$, $\delta q_0=\delta q_N$,  $\delta p_0=\delta p_N$,
$q_{N+1}=q_1$, $p_{N+1}=p_1$, $\delta q_{N+1}=\delta q_1$,  $\delta p_{N+1}=\delta p_1$. In the case of fixed boundary conditions we considered in our numerical simulations, some adjustments have to be done for the $i=1$ and $i=N$ equations, like the ones reported in  the Appendix of~\cite{senyange2017symplectic} where the operators for the 1D and 2D DKG models were reported (with the exception of the corrector term).

For completeness sake in Sec.~\ref{sec:app_other} we provide the explicit expression of the $e^{ \tau L_{BZ}}$ and $e^{ \tau L_{KZ}}$ operators for some commonly used Hamiltonians, which can be split into two integrable parts, one of which is the usual kinetic energy $A\left(\boldsymbol{ p}\right) = \sum_{i=1}^N p_i^2/2$.

%-----------------------------------------------------
\subsection{The 1D disordered discrete nonlinear Schr\"{o}dinger equation} \label{sec:1D_DNLS_app}

Here we focus on the 1D DDNLS system, whose Hamiltonian [Eq.~\eqref{eq:DNLS_1D_real}] can be split  into three integrable parts as
\begin{equation}
\begin{split}
 \mathcal{A}_1 =  \sum_{i = 1}^{N} \frac{\epsilon _i}{2}(q_i ^2 + p_i ^2) + \frac{\beta}{8}(q_i ^2 + p_i ^2)^2,   \qquad
\mathcal{B}_1  = - \sum_{i = 1}^{N} p_{i+1}p_i\ ,  \qquad \mathcal{C }_1 = -  \sum_{i = 1}^{N}q_{i+1}q_i
\end{split}.
\end{equation}
The set of equations of motion and variational equations associated with the   Hamiltonian function $\mathcal{A}_1$ is
\begin{equation}
\frac{d\boldsymbol{X}}{dt} =
L_{\mathcal{A}_1Z}\boldsymbol{X}
\colon
\begin{cases}
\dot{q}_{i} &= p_{i} \theta_{i}  \\
\dot{p}_{i} & = -q_{i} \theta_{i} , \\
\dot{\delta q}_{i} &= \left[ \theta_{i}  +  \beta p_{i} ^2\right] \delta p_{i} + \beta q_{i} p_{i} \delta q_{i} \\
\dot{\delta p}_{i} &= - \left[ \theta_{i}  + \beta q_{i}^2\right] \delta q_{i} - \beta q_{i} p_{i} \delta p_{i}
\end{cases}, \qquad \text{for}\qquad 1\leq i\leq N
\label{eq:DNLS_1D_3ps_A}
\end{equation}
with $\theta_{i} = \epsilon _{i} + \beta (q_{i} ^ 2 + p_{i} ^2 )/2$ for $i=1, 2, \ldots, N$ being constants of the motion. The corresponding operator  $e^{\tau L_{\mathcal{A}_1Z}}$ takes the form
\begin{equation}
e^{\tau L_{\mathcal{A}_1Z}}
\colon
\begin{cases}
q_i' &= q_i \cos (\tau \alpha _i) +  p_i \sin (\tau \alpha _i) \\
p_i' &= p_i \cos (\tau \alpha _i) -  q_i \sin (\tau \alpha _i) \\
\delta q_i' &= \frac{q_i  \cos (\tau \alpha _i ) + p_i  \sin (\tau \alpha _i )}{ 2 J_i} \delta J_i  + \left(p_i  \cos (\tau \alpha _i ) - q_i  \sin (\tau \alpha _i )\right) \left(\beta \delta J_i  \tau + \delta \theta  _i   \right) \\
\delta p_i'  & = \frac{p_i  \cos (\tau \alpha _i ) - q_i \sin (\tau \alpha _i )}{2J_i } \delta J_i  - \left( q_i  \cos ( \tau \alpha _i ) + p_i \sin (\tau \alpha _i ) \right) \left(\beta \delta J_i  \tau + \delta \theta  _i \right)
\end{cases}, \quad \text{for}\quad 1\leq i \leq N
\label{eq:etA_DNLS_1D}
\end{equation}
with $J_i \neq 0$ and
\begin{equation}
\begin{split}
J_i  = \frac{1}{2} (q_i  ^2 + p_i ^2)\ , \qquad
\alpha _i = \epsilon _i + \beta  J_i  \ ,\qquad
\delta J_i   = q_i   \delta q_i   + p_i   \delta p_i  \ , \qquad
\delta \theta _i  = \frac{p_i  }{2J_i }\delta q_i  - \frac{q_i  }{2J_i } \delta p_i
\end{split}.
\label{eq:DNLS2st_coeff}
\end{equation}
We note that in the special case of $J_i=0$ we have $q_i=p_i=0$. Then the system of Eq.~\eqref{eq:DNLS_1D_3ps_A} takes the simple form $\dot{q_i}=0$, $\dot{p_i}=0$, $\dot{\delta q}_{i}=  \epsilon_i \delta p_i$, $\dot{\delta p}_{i}= - \epsilon_i \delta q_i$, leading to
$q_i' = q_i$, $p_i' = p_i$,
$\delta q_i'=\delta q_i \cos (\epsilon_i \tau) + \delta p_i\sin (\epsilon_i \tau)$,
$\delta p_i'=\delta p_i \cos (\epsilon_i \tau) - \delta q_i\sin (\epsilon_i \tau)$.

The set of equations of motion and variational equations associated to the intermediate Hamiltonian functions  $\mathcal{B}_1$ and $\mathcal{C}_1$  are respectively
\begin{equation}
\begin{split}
\frac{d\boldsymbol{X}}{dt} = L_{\mathcal{B}_1Z}\boldsymbol{X} : \left\{
\begin{array}{rl}
\dot{q}_i  &=  - p_{i-1} - p_{i+1} \\
\dot{p}_i  &= 0\\
\dot{\delta  q }_i &=  - \delta  p_{i-1} - \delta  p_{i+1} \\
\dot{\delta  p }_i &= 0
\end{array} \right.,
\quad \text{and}\quad \frac{d\boldsymbol{X}}{dt} = L_{\mathcal{C}_1Z}\boldsymbol{X} : \left\{
\begin{array}{rl}
\dot{q_i}  &=  0 \\
\dot{p_i}  &= q_{i-1} + q_{i+1} \\
\dot{\delta  q_i } &= 0\\
\dot{\delta  p_i } &= \delta  q_{i-1} + \delta  q_{i+1}
\end{array} \right..
\end{split}
\label{eq:DNLS_1D_3ps_B_3ps_C}
\end{equation}
These yield to the  operators $e^{L_{\mathcal{B}_1Z}}$ and $e^{L_{\mathcal{C}_1Z}}$ given by
\begin{equation}
\begin{split}
e^{\tau L_{\mathcal{B}_1Z}} \colon  \left \{
\begin{array}{rl}
q_i' &= q_i  - \tau (p_{i-1} + p_{i+1})  \\
p_i' &= p_i \\
\delta  q_i' &= \delta  q_i  - \tau (\delta  p_{i-1} + \delta  p_{i+1}) \\
\delta  p_i' &= \delta  p_i
\end{array} \right.
\qquad\qquad
e^{\tau L_{\mathcal{C}_1Z}} \colon  \left \{
\begin{array}{rl}
q_i' & = q_i  \\
p_i' &= p_i +   \tau (q_{i-1} + q_{i+1}) \\
\delta q_i' &= \delta q_i   \\
\delta p_i' &= \delta p_i +   \tau (\delta q_{i-1} +\delta  q_{i+1})\\
\end{array} \right..
\end{split}
\label{eq: 3ps var eq of b}
\end{equation}
As in the case of the $\alpha$-FPUT model,  Eqs.~\eqref{eq:DNLS_1D_3ps_B_3ps_C} and \eqref{eq: 3ps var eq of b} correspond to the case of periodic boundary conditions and adjustments similar to the ones presented in the Appendix of~\cite{senyange2017symplectic} for the 1D DKG model, should be implemented when fixed boundary conditions are imposed.

%-----------------------------------------------------
\subsection{The 2D disordered discrete Nonlinear Schr\"{o}dinger equation} \label{sec:2D_DNLS_app}

The Hamiltonian $H_{\text{2D}}$ of Eq.~\eqref{eq:DNLS_2D_real} can be split into three integrable parts $\mathcal{A}_2$, $\mathcal{B}_2$ and $\mathcal{C}_2$ as
\begin{equation}
\begin{split}
\mathcal{A}_2 = \sum _{i = 1}^{N} \sum _{j=1}^{M}  \frac{\epsilon _{i, j} }{2} \left[q_{i, j} ^2 + p_{i, j}^2\right] + \frac{\beta}{8} \left[q_{i, j} ^2 + p_{i, j}^2\right] ^2,
\qquad\quad
\mathcal{B}_2 & =  \sum _{i = 1}^{N} \sum _{j=1}^{M} - p _{i, j+1} p _{i, j} - p _{i + 1, j}p _{i, j},  \\  \mathcal{C}_2 &=  \sum _{i = 1}^{N} \sum _{j=1}^{M} -q_{i, j+1} q _{i, j} - q_{i + 1, j} q _{i, j}.
\end{split}
\label{eq:2D_DNLS_3split_app}
\end{equation}
The equations of motion and the variational equations associated with the $\mathcal{A}_2$  Hamiltonian are
\begin{equation}
\begin{split}
\frac{d\boldsymbol{X}}{dt} =
L_{\mathcal{A}_2Z}\boldsymbol{X} & : \left\{
\begin{array}{rl}
\dot{q}_{i,j} &= p_{i,j} \theta_{i,j}  \\
\dot{p}_{i,j} & = -q_{i,j} \theta_{i,j} , \\
\dot{\delta q}_{i,j} &= \left[ \theta_{i,j}  +  \beta p_{i,j} ^2\right] \delta p_{i,j} + \beta q_{i,j} p_{i,j} \delta q_{i,j} \\
\dot{\delta p}_{i,j} &= - \left[ \theta_{i,j}  + \beta q_{i,j}^2\right] \delta q_{i,j} - \beta q_{i,j} p_{i,j} \delta p_{i,j}
\end{array} \right., \\
\end{split}
\label{eq:DNLS_2D_3ps_A}
\end{equation}
for $1 \leq i \leq N$, $1 \leq j \leq M$, with $\theta_{i,j} = \epsilon _{i,j} + \beta (q_{i,j} ^ 2 + p_{i,j} ^2 )/2$ being constants of motion for the Hamiltonian $\mathcal{A}_2$. Then, the
operator $e^{\tau L_{\mathcal{A}_2Z}}$ is
 \begin{equation}
  \begin{split}
e^{\tau L_{\mathcal{A}_2Z}} :\left \{
\begin{array}{rl}
q_{i, j}' &= q_{i, j} \cos (\tau \alpha _{i, j}) +  p_{i, j} \sin (\tau \alpha _{i, j}) \\
p_{i, j}' &= p_{i, j} \cos (\tau \alpha _{i, j}) -  q_{i, j} \sin (\tau \alpha _{i, j}) \\
\delta q_{i, j}' &= \frac{q_{i, j}  \cos (\tau \alpha _{i, j}) + p_{i, j}  \sin (\tau \alpha _{i, j} )}{ 2 J_{i, j}} \delta J_{i, j}  + \left(p_{i, j}  \cos (\tau \alpha _{i, j} ) - q_{i, j}  \sin (\tau \alpha _{i, j} )\right) \left(\beta \delta J_{i, j}  \tau + \delta \theta  _{i, j}  \right) \\
\delta p_{i, j}'  & = \frac{p_{i, j}  \cos (\tau \alpha _{i, j} ) - q_{i, j} \sin (\tau \alpha _{i, j} )}{2J_{i, j} } \delta J_{i, j}  - \left( q_{i, j}  \cos ( \tau \alpha _{i, j} ) + p_{i, j} \sin (\tau \alpha _{i, j} ) \right) \left(\beta \delta J_{i, j}  \tau + \delta \theta  _{i, j} \right)
\end{array} \right.
\end{split}
\end{equation}
with  $J_{i, j}\neq 0$ and
\begin{equation}
\begin{split}
&\qquad \quad J_{i, j}  = \frac{1}{2} (q_{i, j}  ^2 + p_{i, j} ^2)\ ,\qquad \alpha _{i, j} = \epsilon _{i, j} + \beta  J_{i, j} \ , \\
&\delta J_{i, j}   = q_{i, j}   \delta q_{i, j}   + p_{i, j}   \delta p_{i, j}  \ , \qquad
\delta \theta _{i, j}  = \frac{p_{i, j}  }{2J_{i, j} }\delta q_{i, j}  - \frac{q_{i, j}  }{2J_{i, j} } \delta p_{i, j}
\end{split}.
\label{eq:2D_DNLS2st_coeff}
\end{equation}
Again, in the special case of $J_{i, j}=0$, the system of Eq.~\eqref{eq:DNLS_2D_3ps_A} takes the form $\dot{q_{i, j}}=0$, $\dot{p_{i, j}}=0$, $\dot{\delta q}_{i, j}=  \epsilon_{i, j} \delta p_{i, j}$, $\dot{\delta p}_{i, j}= - \epsilon_{i, j} \delta q_{i, j}$, leading to
$q_{i, j}' = q_{i, j}$, $p_{i, j}' = p_{i, j}$,
$\delta q_{i, j}'=\delta q_{i, j} \cos (\epsilon_{i, j}\tau) + \delta p_{i, j}\sin (\epsilon_{i, j}\tau)$,
$\delta p_{i, j}'=\delta p_{i, j}\cos (\epsilon_{i, j}\tau) - \delta q_{i, j}\sin (\epsilon_{i, j}\tau)$.

The  equations of motion and the variational equations for Hamiltonians $\mathcal{B}_2$ and  $\mathcal{C}_2$  are
\begin{equation}
\begin{split}
\frac{d\boldsymbol{X}}{dt}=L_{\mathcal{B}_2Z}\boldsymbol{X} &: \left\{
\begin{array}{rl}
\dot{q}_{i, j} &= - p_{i-1, j} - p_{i, j-1} - p_{i, j+1} - p_{i+1, j} \\
\dot{p}_{i, j} &= 0 \\
\dot{\delta q}_{i, j} &= - \delta p_{i-1, j} - \delta p_{i, j-1} - \delta p_{i, j+1} - \delta p_{i+1, j} \\
\dot{\delta p}_{i, j}  &= 0
\end{array} \right.
\end{split},
\label{eq:2D_DDNLS_eq_B}
\end{equation}
and
\begin{equation}
\begin{split}
\frac{d\boldsymbol{X}}{dt}=L_{\mathcal{C}_2Z}\boldsymbol{X} & : \left\{
\begin{array}{rl}
\dot{p}_{i, j} &= q_{i-1, j} +q_{i, j-1}+q_{i, j+1} +q_{i+1, j}  \\
\dot{q}_{i, j} &= 0 \\
\dot{\delta p}_{i, j}  &=  \delta q_{i-1, j} +\delta q_{i, j-1} +\delta q_{i, j+1} +\delta q_{i+1, j} \\
\dot{\delta q}_{i, j}  &= 0 \\
\end{array} \right.
\end{split},
\label{eq:2D_DDNLS_eq_C}
\end{equation}
while the corresponding operators $e^{L_{\mathcal{B}_2Z}}$ and $e^{L_{\mathcal{C}_2Z}}$ are respectively
 \begin{equation}
  \begin{split}
e^{\tau L_{\mathcal{B}_2Z}} &:\left \{
\begin{array}{rl}
q_{i, j}' &= q_{i, j} - \tau \left( p_{i-1, j} + p_{i, j-1} + p_{i, j+1} + p_{i+1, j} \right) \\
p_{i, j}'  &= p_{i, j} \\
\delta q_{i, j}' &= \delta q_{i, j} - \tau \left( \delta p_{i-1, j} + \delta p_{i, j-1} + \delta p_{i, j+1} + \delta p_{i+1, j} \right) \\
\delta p_{i, j}' &= \delta p_{i, j} \\
\end{array} \right. \\
\end{split},
\label{eq:e_LBZ_DNLS_2D_app}
\end{equation}
and
\begin{equation}
\begin{split}
e^{\tau L_{\mathcal{C}_2Z}} & : \left \{
\begin{array}{rl}
q_{i, j}' &= q_{i, j} \\
p_{i, j}' &=p_{i, j} + \tau \left( q_{i-1, j} +q_{i, j-1}+q_{i, j+1} +q_{i+1, j} \right) \\
\delta q_{i, j}' &= \delta  q_{i, j} \\
\delta p_{i, j}' &=\delta p_{i, j} + \tau \left( \delta q_{i-1, j} +\delta q_{i, j-1}+\delta q_{i, j+1} +\delta q_{i+1, j} \right)
\end{array} \right..
\end{split}
\label{eq:e_LCZ_DNLS_2D_app}
\end{equation}
Here again Eqs.~\eqref{eq:2D_DDNLS_eq_B}, \eqref{eq:2D_DDNLS_eq_C}, \eqref{eq:e_LBZ_DNLS_2D_app} and \eqref{eq:e_LCZ_DNLS_2D_app} correspond to the case of periodic boundary conditions. For fixed boundary conditions adjustments similar to the ones report in   the Appendix of~\cite{senyange2017symplectic} for the 2D DKG lattice should be performed.

%-----------------------------------------------------
\subsection{Other Hamiltonian models which can be split into two integrable parts}\label{sec:app_other}

We present here the exact expressions of the  tangent map operators needed in symplectic integration schemes which can be used to numerically integrate some important models in the field of classical many-body systems:
the $\beta$-FPUT chain, the KG model, and the classical XY model (a JJC system)~\cite{mithun2018:rotor,Livi2987:chaotic,Binder2000:observation, Blackburn2016:survey}. Similarly to the $\alpha$-FPUT chain of Eq.~\eqref{eq:alphaFPUT}, the Hamiltonians $H(\boldsymbol{q},\boldsymbol{p})$ of  each of these systems can be split as
\begin{equation}
H(\boldsymbol{q},\boldsymbol{p})  = A(\boldsymbol{p})+B(\boldsymbol{q})
= \sum _{i=1}^N \frac{p_i^2}{2} +  B\left(\boldsymbol{ q}\right) %= \sum _{i} \frac{1}{2}(q_{i+1} - q_{ i})^2  + \frac{\alpha}{3}(q_{i+1} - q_{ i})^3.
\label{eq:split2part_others}
\end{equation}
with appropriately defined  potential terms $B\left(\boldsymbol{ q}\right)$:
\begin{align}
\label{eq:beta}
\text{$\beta$-FPUT:}\   \qquad B_\beta\left(\boldsymbol{ q}\right)  &  =  \sum_{i=0}^{N}  \frac{1}{2}(q_{i+1} - q_i)^2  + \frac{\beta}{4}(q_{i+1} - q_i)^4,  \\
\label{eq:KG}
\text{KG:}\  \qquad B_{K}\left(\boldsymbol{ q}\right)  & = \sum_{i=1}^{N}  \frac{q_i^2}{2}  +  \frac{q_i^4 }{4}  +  \frac{k}{2}(q_{i+1} - q_i)^2 ,  \\
\label{eq:rotor}
\text{JJC:}\  \qquad  B_R\left(\boldsymbol{ q}\right) & =  \sum_{i=1}^{N} E_J \left[1-\cos(q_{i+1}-q_i) \right],
\end{align}
where $\beta$, $k$ and $E_J$ are real coefficients.
Obviously for all these systems the  operator $e^{\tau L_{AZ}}$ of the kinetic energy part is the same as for the $\alpha$-FPUT chain in Eq.(\ref{eq:res_A_FPU}).  Thus, we report  below only the expressions of the operators $e^{\tau L_{BZ}}$ and $e^{\tau L_{KZ}}$ when periodic boundary conditions are imposed.

\begin{enumerate}
\item \textbf{The $\beta$-Fermi-Pasta-Ulam-Tsingou chain} \\
The operator $e^{\tau L_{BZ}}$ of the $\beta$-FPUT chain of Eq.~\eqref{eq:beta} is
\begin{equation}
\begin{split}
e^{ \tau L_{BZ}}: \left\{
\begin{array}{rl}
    q_i' &=    q_i         \\
   p_i' &=  \big\{ q_{i-1} + q_{i+1} - 2q_i + \beta \big[ (q_{i+1} - q_i)^3 - (q_i - q_{i-1})^3\big] \big\} \tau +  p_i\\
   % q_{i+1} +  q_{i-1}  -  2 q_i   \big]    \big[  1+\alpha\big( q_{i+1} -   q_{i-1}  \big)
    \delta q_i' &=  \delta  q_i   \\
    \delta p_i' &=  \big\{  \big[- 3 \beta \big[ (q_i - q_{i-1})^2 + (q_{i+1} - q_i)^2\big]- 2 \big]\delta  q_i  \\
    &\qquad \quad + \big[ 1 + 3\beta (q_{i+1} - q_i)^2 \big]\delta  q_{i+1} + \big[1 + 3\beta (q_i - q_{i-1})^2\big]\delta  q_{i-1}     \big\}\tau + \delta p_i
\end{array} \right.
\end{split}.
\label{eq:e_LB_betaFPU}
\end{equation}
The corrector Hamiltonian $K$ of Eq.(\ref{eq:corrector_app}) becomes
%. Since $a^3 - b^3 = (a-b)(a^2+ab+b^2)$, it follows
\begin{equation}
\begin{split}
K(\boldsymbol{ q}) &= \sum_{i=1}^{N}\Big\{ 2q_i - q_{i-1} - q_{i+1} + \beta \big[ (q_i - q_{i-1})^3 - (q_{i+1} - q_i)^3\big]  \Big\}^2,\ \\
\end{split}
\label{eq:C_betaFPU}
\end{equation}
while the corresponding operator is
\begin{equation}
\begin{split}
e^{ \tau L_{KZ}}: \left\{
\begin{array}{rl}
    q_i' &=    q_i         \\
  p_i' &=
  2\Big\{ \Big[ q_{i-1} +  q_{i+1} - 2q_i  + \beta \big[  (q_{i+1} - q_i)^3 - (q_i - q_{i-1})^3\big]  \Big] \Big[2 + 3\beta \big[ (q_i - q_{i-1})^2 + (q_{i+1} - q_i)^2 \big]  \Big]    \\
&- \Big[q_{i} + q_{i+2} - 2q_{i+1} + \beta \big[(q_{i+2} - q_{i+1})^3 - (q_{i+1} - q_{i})^3\big]  \Big] \Big[ 1 +  3 \beta ( q_{i+1} - q_{i})^2 \Big]  \\
&- \Big[ q_{i} + q_{i-2} - 2q_{i-1} + \beta \big[ (q_{i} - q_{i-1})^3 - (q_{i-1} - q_{i-2})^3 \big]  \Big] \Big[ 1 +  3 \beta ( q_{i} - q_{i-1})^2 \Big] \Big\}\tau + p_i \\
    \delta q_i &=  \delta  q_i   \\
    \delta p_i &= \big\{ \gamma_i \delta  q_{i}  +  \gamma_{i+1} \delta  q_{i+1} +  \gamma_{i+2} \delta  q_{i+2} +  \gamma_{i-1} \delta  q_{i-1} +  \gamma_{i-2} \delta  q_{i-2}   \big\}\tau + \delta p_i
\end{array} \right.
\end{split},
\label{eq:e_LC_betaFPU}
\end{equation}
with
\begin{equation}
\begin{split}
\gamma_{i}&=
-2 \Big\{ \Big[2 + 3 \beta \big[  (q_i - q_{i-1})^2 +(q_{i+1} - q_i)^2 \big]  \Big]^2 + \Big[ 1 +  3 \beta ( q_{i+1} - q_{i})^2 \Big]^2 +  \Big[ 1 +  3 \beta ( q_{i} - q_{i-1})^2 \Big]^2  \\
&\qquad\qquad \quad +6\beta \big(2q_i - q_{i-1} - q_{i+1} \big)  \Big[2q_i - q_{i-1} - q_{i+1} + \beta \big[ (q_i - q_{i-1})^3 - (q_{i+1} - q_i)^3\big]  \Big] \\
&\qquad\qquad \quad-  6\beta ( q_{i} - q_{i+1}) \Big[2q_{i+1} - q_{i} - q_{i+2} + \beta \big[ (q_{i+1} - q_{i})^3 - (q_{i+2} - q_{i+1})^3\big]  \Big]  \\
&\qquad\qquad \quad-6\beta ( q_{i} - q_{i-1})  \Big[2q_{i-1} - q_{i} - q_{i-2} + \beta \big[ (q_{i-1} - q_{i-2})^3 - (q_{i} - q_{i-1})^3\big]  \Big]  \Big\}
 \\
\gamma_{i+1}&=
 2\Big\{ \Big[ 1 + 3 \beta (q_{i+1} - q_i)^2   \Big]   \Big[4 +3 \beta \big[  (q_i - q_{i-1})^2 + 2(q_{i+1} - q_i)^2 + (q_{i+2} - q_{i+1})^2 \big] \Big]  \\
&\qquad\qquad \quad- 6 \beta(q_{i+1} - q_i) \Big[3q_i - q_{i-1} - 3q_{i+1} + q_{i+2} + \beta \big[(q_i - q_{i-1})^3 - 2(q_{i+1} - q_i)^3 + (q_{i+2} - q_{i+1})^3\big]  \Big] \Big\}
  \\
\gamma_{i-1}&=
2 \Big\{  \Big[  1 + 3 \beta (q_i - q_{i-1})^2  \Big]  \Big[4 +3 \beta \big[ (q_{i+1} - q_i)^2 + 2 (q_i - q_{i-1})^2 + (q_{i-1} - q_{i-2})^2 \big]  \\
&\qquad\qquad \quad- 6\beta (q_{i-1} - q_{i}) \Big[3q_i - 3q_{i-1} - q_{i+1} + q_{i-2} + \beta \big[ (q_{i} - q_{i+1})^3 - 2 (q_{i-1} - q_{i})^3 + (q_{i-2} - q_{i-1})^3 \big] \Big\}
 \\
\gamma_{i+2}&=  - 2  \Big[ 1 + 3\beta (q_{i+2} - q_{i+1})^2  \Big] \Big[ 1 +  3 \beta ( q_{i+1} - q_{i})^2 \Big]   \\
\gamma_{i-2}&=  - 2 \Big[ 1 + 3\beta(q_{i-1} - q_{i-2})^2 \Big] \Big[ 1 +  3 \beta ( q_{i} - q_{i-1})^2 \Big]
\end{split}.
\label{eq:e_LC_betaFPU_coeff}
\end{equation}

\item
\textbf{The 1D Klein-Gordon chain model}\\
The operator $e^{\tau L_{BZ}}$ of the Klein-Gordon chain [Eq.~\eqref{eq:KG}] is
\begin{equation}
\begin{split}
e^{ \tau L_{BZ}}: \left\{
\begin{array}{rl}
    q_i' &=    q_i         \\
   p_i' &=  \big\{  - q_i(1 + q_i^2) +   k(q_{i+1} + q_{i-1} -2 q_i )  \big\} \tau +  p_i\\
    \delta q_i' &=  \delta  q_i   \\
    \delta p_i'&=  \big\{  - [ 1 + 3q_i^2 + 2k ]\delta  q_i  + k \delta  q_{i+1} +  k \delta  q_{i-1}     \big\}\tau + \delta p_i
\end{array} \right.
\end{split}.
\label{eq:e_LB_KG}
\end{equation}
The corresponding corrector Hamiltonian $K$ [Eq.(\ref{eq:corrector_app})] is written as
\begin{equation}
K(\boldsymbol{ q}) = \sum_{i=1}^{N}\left[ q_i(1 + q_i^2) +   k(2 q_i - q_{i+1} - q_{i-1} )   \right]^2\ ,
\label{eq:C_KG}
\end{equation}
while $e^{\tau L_{KZ}}$ takes the form
\begin{equation}
\begin{split}
e^{ \tau L_{KZ}}: \left\{
\begin{array}{rl}
    q_i' &=    q_i         \\
  p_i' &=   2 \Big\{  \big[ - q_i(1 + q_i^2)  +  k(  q_{i+1} + q_{i-1} - 2 q_i) \big]\big[ 1 + 3 q_i^2  +   2k  \big]   \\
         &\quad  + k\big[ q_{i-1}(1 + q_{i-1}^2)  -   k( q_{i} + q_{i-2} - 2 q_{i-1}) \big]\\
          & \quad + k\big[  q_{i+1}(1 + q_{i+1}^2)   -  k( q_{i+2} + q_{i} -  2 q_{i+1})\big] \Big\}\tau + p_i \\
    \delta q_i' &=  \delta  q_i   \\
    \delta p_i' &= \big\{ \gamma_i \delta  q_{i}  +  \gamma_{i+1} \delta  q_{i+1} +  \gamma_{i+2} \delta  q_{i+2} +  \gamma_{i-1} \delta  q_{i-1} +  \gamma_{i-2} \delta  q_{i-2}   \big\}\tau + \delta p_i
\end{array} \right.
\end{split},
\label{eq:e_LC_KG}
\end{equation}
with
\begin{equation}
\begin{split}
\gamma_{i}& = -2 \Big\{ \big[ 1 + 3 q_i^2 +   2k  \big]^2 + 6 q_i \big[ q_i(1 + q_i^2) +   k(2 q_i - q_{i+1} - q_{i-1} ) \big] +2k^2 \Big\} \\
\gamma_{i+1}&= 2k \big[ 2 + 4k + 3 q_i^2 + 3q_{i+1}^2  \big]    \\
\gamma_{i-1}&=  2k \big[ 2 + 4k + 3 q_i^2 + 3q_{i-1}^2  \big]  \\
\gamma_{i+2}&= -2  k^2   \\
\gamma_{i-2}&= -2  k^2
\end{split}.
\label{eq:e_LC_KG_coeff}
\end{equation}

\item
\textbf{The XY model of a Josephson junctions array}\\
The operator $e^{\tau L_{BZ}}$ for the potential of  Eq.~\eqref{eq:rotor} is
\begin{equation}
\begin{split}
e^{ \tau L_{BZ}}: \left\{
\begin{array}{rl}
    q_i' &=    q_i         \\
   p_i' &=  E_J \big[  \sin(q_{i+1}-q_i)+\sin(q_{i-1} - q_i)  \big] \tau+  p_i\\
    \delta q_i' &=  \delta  q_i   \\
    \delta p_i' &=  \big\{  - E_J \big[\cos(q_{i+1}-q_i)+\cos(q_i-q_{i-1}) \big] \delta q_i \\
    &\qquad  E_J \big[ \cos(q_{i+1}-q_i) \big]\delta q_{i+1} + E_J \big[ \cos(q_i-q_{i-1}) \big] \delta q_{i-1} \big\}\tau + \delta p_i
\end{array} \right.
\end{split}.
\label{eq:e_LB_rot}
\end{equation}
In this case the  corrector Hamiltonian $K$ of Eq.(\ref{eq:corrector_app}) becomes
\begin{equation}
K(\boldsymbol{ q}) =   \sum_{i=1}^{N} E_J^2 \left[ \sin(q_{i+1}-q_i)+\sin(q_{i-1} - q_i)  \right]^2,
\label{eq:C_rot}
\end{equation}
and the operator $e^{\tau L_{KZ}}$ is given by the following set of equations
\begin{equation}
\begin{split}
e^{ \tau L_{KZ}}: \left\{
\begin{array}{rl}
    q_i' &=    q_i         \\
      p_i' &=   E_J^2 \Big\{ 2 \big[  \sin(q_{i+1}-q_i)+\sin(q_{i-1} - q_i)  \big] \cdot \big[ \cos(q_{i+1}-q_i) + \cos(q_{i-1} - q_i)  \big] \\
& -2 \big[  \sin(q_{i+2}-q_{i+1})+\sin(q_{i} - q_{i+1})  \big] \cdot  \cos(q_{i} - q_{i+1})   \\
& - 2 \big[  \sin(q_{i}-q_{i-1})+\sin(q_{i-2} - q_{i-1})  \big] \cdot   \cos(q_{i} - q_{i-1})  \Big\}\tau + p_i \\
    \delta q_i' &=  \delta  q_i   \\
    \delta p_i' &= \big\{ \gamma_i \delta  q_{i}  +  \gamma_{i+1} \delta  q_{i+1} +  \gamma_{i+2} \delta  q_{i+2} +  \gamma_{i-1} \delta  q_{i-1} +  \gamma_{i-2} \delta  q_{i-2}   \big\}\tau + \delta p_i
\end{array} \right.
\end{split}
\label{eq:e_LC_rot}
\end{equation}
with
\begin{equation}
\begin{split}
\gamma_{i}&=  E_J^2 \Big\{  - 4 \cos(2 (q_{i+1}-q_i)) - 4 \cos(q_{i-1} -2 q_i+ q_{i+1}) - 4 \cos(2 (q_{i-1}-q_i))  \\
& \qquad + 2 \sin(q_{i+2}-q_{i+1}) \sin(q_{i} -  q_{i+1}) + 2 \sin(q_{i-2} - q_{i-1})  \sin(q_{i} - q_{i-1})  \Big\} \\
\gamma_{i+1}&=  E_J^2 \Big\{   2 \cos(q_{i-1} -2 q_i+ q_{i+1}) + 4 \cos(2 (q_{i+1}-q_i))  + 2 \cos(q_{i+2}-2 q_{i+1}+q_{i})     \Big\} \\
\gamma_{i-1}&=   E_J^2 \Big\{  2 \cos(q_{i-1} -2 q_i+ q_{i+1}) + 4\cos(2 (q_{i}-q_{i-1})) + 2 \cos(q_{i-2} - 2 q_{i-1}+ q_{i})   \Big\}     \\
\gamma_{i+2}&=  E_J^2  2 \cos(q_{i+2}-q_{i+1}) \cos(q_{i} - q_{i+1})     \\
\gamma_{i-2}&=  E_J^2  2 \cos(q_{i-2} - q_{i-1})  \cos(q_{i} - q_{i-1})
\end{split}.
\label{eq:e_LC_rot_coeff}
\end{equation}
\end{enumerate}

%-----------------------------------------------------
\section*{Acknowledgments}

We thank S.~Flach for useful discussions. C.D.~and M.T.~acknowledge financial support from the IBS (Project Code No.~IBS-R024-D1). Ch.S.~and B.M.M.~were supported by the National Research Foundation of South Africa (Incentive Funding for Rated Researchers, IFFR and Competitive Programme for Rated Researchers, CPRR). B.M.M.~would also like to thank the High Performance Computing facility of the University of Cape Town (\url{http://hpc.uct.ac.za}) and the Center for High Performance Computing (\url{https://www.chpc.ac.za}) for the provided computational resources needed for the study of the two  DDNLS models, as well as their user-support teams for their help on many practical issues.

%-----------------------------------------------------
\section*{Conflict of interest}

All authors declare  no conflict of interest in this paper.

%-----------------------------------------------------

%For more questions regarding reference style, please refer to the \href{http://www.ncbi.nlm.nih.gov/books/NBK7256/}{Citing Medicine}.
%
%\section*{Supplementary (if necessary)}

\end{document}